\documentclass[12pt]{article}
\usepackage{graphics}
\usepackage{amsmath}
\usepackage{amssymb}

\numberwithin{equation}{section} 
\def\G{\Gamma}
\def\D{\Delta}
\def\L{\Lambda}

\def\O{\Omega}
\def\a{\alpha}
\def\b{\beta}
\def\g{\gamma}
\def\d{\delta}
\def\e{\varepsilon}
\def\m{\mu}
\def\n{\nu}
\def\s{\sigma}
\def\r{\rho}
\def\l{\lambda}
\def\t{\tau}
\def\o{\omega}

\def\vt{\vartheta}
\def\mc{\mathcal}
\def\mf{\mathfrak}
\def\N{\nabla}
\def\p{\partial}
\def\Tr{\text{Tr}}
\def\tr{\text{tr}}
\def\la{\langle}
\def\ra{\rangle}
\def\lra{\longrightarrow}
\def\lr{\longleftrightarrow}
\def\dg{\dagger}
\def\wt{\widetilde}
\def\da{\dot{a}}
\def\db{\dot{b}}
\def\Zop{\bbbz}

\def\bbbz {{\sf Z\!\!Z}}

\def\RR{R-R }

\begin{document}
\thispagestyle{empty}
\addtocounter{page}{-1}
\def\thefootnote{\alph{footnote}}
\begin{flushright}
  hep-th/0307027 \\
  AEI-2003-054 \\
  KCL-MTH-03-10
\end{flushright}

\vskip 0.5cm

\begin{center}\LARGE
{\bf Strings in plane wave backgrounds}\footnote{Based on the author's 
PhD thesis, Humboldt University, Berlin.}
\end{center}

\vskip 1.0cm

\begin{center}
{\large A. Pankiewicz\,\footnote{E-mail address: {\tt apankie@aei.mpg.de}}${}^{,c}$}

\vskip 0.5cm

{\it $^b$ Max-Planck-Institut f\"ur Gravitationsphysik, Albert-Einstein-Institut \\ 
Am M\"uhlenberg 1, D-14476 Golm, Germany}

\vskip 0.2cm

{\it $^c$ Department of Mathematics, King's College London \\
Strand, London WC2R 2LS, United Kingdom}
\end{center}

\vskip 1.0cm

\begin{center}
July 2003
\end{center}

\vskip 1.0cm

\begin{abstract}
\noindent 
I review aspects of string theory on plane wave backgrounds emphasising the connection to gauge theory
given by the BMN correspondence. Topics covered include the Penrose limit and its role in deriving the
BMN duality from AdS/CFT, light-cone string field theory in the maximally supersymmetric plane wave
and extensions of the correspondence to less supersymmetric backgrounds. 
\end{abstract}

\vfill

\setcounter{footnote}{0}
\def\thefootnote{\arabic{footnote}}
\newpage

\renewcommand{\theequation}{\thesection.\arabic{equation}}
\tableofcontents
\pagebreak

\section{Introduction}

\subsection{Motivation}

The intimate connection between string and gauge theories has been one of the dominant
themes in theoretical high energy physics over the last years. A famous example is
the equivalence (duality) of string theory on Anti-de Sitter (AdS) spaces with conformal
field theories, the AdS/CFT
correspondence~\cite{Maldacena:1998re,Gubser:1998bc,Witten:1998qj},
see e.g.~\cite{Aharony:1999ti} for a review.

Several arguments support the expectation of a duality between
string and gauge theories or, even more generally, gravitational and non-gravitational
theories.
For example, a qualitative one comes from the fact that QCD, the $SU(3)$
gauge theory of strong interactions, confines chromoelectric flux to flux tubes -- the
QCD string -- at low energies. After all, 
string theory was originally discovered in attempts to describe the spectrum of hadronic resonances.
A quantitative argument is
't Hooft's analysis of the large $N$ limit of $SU(N)$ gauge theories~\cite{'tHooft:1974jz}:
for large $N$ and fixed 't Hooft coupling $\l=g^2_{\text{YM}}N$,
the Feynman diagram expansion can be rearranged according to the genus $g$ of the
Riemann surface
which the diagram can be drawn on and every amplitude can be written in an
expansion of the form $\sum_{g=0}^{\infty}N^{2-2g}f_g(\l)$, i.e.\ $1/N^2$ is
the effective genus counting parameter. This is like the
perturbation series of a string theory, where the string coupling $g_{\text{s}}$ is
identified with $1/N$ and $\l$ corresponds to the loop-counting parameter of the string
non-linear $\s$-model. This a very general argument for the large $N$ duality between gauge
theories and certain string theories, but it does not give an answer to what kind of string
theory one should look for.

Further hints come from the study of black holes. The simplest example is the
Schwarzschild solution of general relativity depending on a single parameter, the mass
$M$ of the black hole.
They have a horizon and are black classically, everything
crossing the horizon is inevitably pulled into the black hole singularity.
However, semi-classical analysis
shows that due to quantum processes black holes start to emit Hawking
radiation: the emission spectrum is roughly that of a blackbody
with temperature $T\sim 1/M$; the deviation of the pure blackbody spectrum is
encoded in the so called `greybody factor'.
As radiating systems black holes are expected to obey the laws of thermodynamics.
If one defines the black hole entropy,
as first proposed by Bekenstein and Hawking by $S=\frac{1}{4}A\sim M^2$, $A$ the area of the
black hole horizon, these laws are in fact satisfied.
A quantum theory of gravity should e.g.\ provide the framework for a microscopic derivation of the
black hole entropy via a counting of states and predict its greybody factor.
As the Bekenstein-Hawking entropy involves the area instead of the volume, as is
the case for statistical mechanics and local quantum field theories,
one may wonder if one can find a holographic description in terms of
local quantum field theories `living' on the horizon, such that
$S_{\text{QFT}}\sim A$. More generally, the holographic principle~\cite{'tHooft:1993gx,Susskind:1995vu}
asserts that the number of degrees of freedom of quantum gravity on some manifold
scales as the area of its boundary: this suggests  that a field theory on the boundary
of space-time might capture the physics of gravity in the bulk. For reviews of the
holographic principle see~\cite{Bigatti:1999dp,Bousso:2002ju};
for an introduction on black holes in string theory see e.g.~\cite{Peet:2000hn}.

The AdS/CFT correspondence explicitly realizes the general principles of
large $N$ duality and holography.
The simplest and best understood example is the equivalence of
string theory on $AdS_5\times S^5$ and the maximally supersymmetric gauge theory in
four dimensions, ${\mc N}=4$ $SU(N)$ super Yang-Mills (SYM).
The latter arises as the low-energy (i.e.\ energies much smaller than the string scale
$1/\sqrt{\a'}$) effective theory on the world-volume of $N$ D3-branes.
As these are charged under the \RR four-form potential~\cite{Polchinski:1995mt}, their
presence generates a five-form flux in the (flat) transverse six-dimensional space. This flux
contributes to the energy-momentum tensor, so the geometry backreacts and curves.
The backreaction is negligible if the effective gravitational coupling is small,
which is the case if $g_{\text{s}}N\sim g^2_{\text{YM}}N\ll 1$. In this regime
the gauge theory is weakly coupled. In the regime of strong coupling, the
large $N$ limit, the backreaction is no longer small and the geometry will change
significantly. To be more precise, for $1\ll g_{\text{s}}N<N$ we can use the dual
description of D3-branes in terms
of extremal three-branes in type IIB supergravity~\cite{Polchinski:1995mt}: in this
picture,
considering low-energy excitations on the D3-brane, translates to going to the
near-horizon region of the three-brane since energies are red-shifted for an asymptotic
observer~\cite{Maldacena:1998re}. The near-horizon region has the geometry
of $AdS_5\times S^5$ with radii $R^4/{\a'}^2=g^2_{\text{YM}}N$ and the five-form flux
on the $S^5$ equals $N$, the number of colors in the gauge theory. Strongly coupled
${\mc N}=4$ SYM is identified with supergravity (since the curvature $\a'/R^2\ll 1$) on
$AdS_5\times S^5$. It is believed that this duality is true for all values of parameters and
extends to the full string theory; this however is difficult to verify with the present
technology, though there are some exceptions, see~\cite{Aharony:1999ti}.
For reviews of attempts to use AdS/CFT as
a starting point to obtain a string description of QCD or at least of pure ${\mc N}=1$
SYM, see e.g.~\cite{Aharony:2002up,Bigazzi:2003ui}.

It was realized by Berenstein, Maldacena and Nastase (BMN)~\cite{Berenstein:2002jq}
that plane (or pp) wave backgrounds provide an interesting example where the string/gauge
correspondence can be studied beyond the supergravity approximation.
As will be
explained in detail in what follows, on the geometric side this involves the Penrose
limit~\cite{Penrose1976} applied for example to $AdS_5\times S^5$;
roughly speaking, one focuses on the neighborhood of the geodesic of a massless particle,
in the center of $AdS_5$ and rotating on the $S^5$.
String theory in the
resulting plane wave background can be exactly quantized in light-cone
gauge~\cite{Metsaev:2001bj}. On the other hand, in the gauge theory this limit singles out
composite operators carrying a large charge~\cite{Berenstein:2002jq}.
Though I will not discuss this here, let me mention that one can also consider
macroscopic rotating strings vs. large spin operators~\cite{Gubser:2002tv}.

\subsection{Outline}

This work is organized as follows:
section~\ref{chapter2} starts with a fairly general introduction to
pp-wave backgrounds in ten/eleven-dimensional supergravities.
I discuss various basic aspects of these backgrounds, in particular their
(super)symmetries, emphasizing the emergence of special maximally supersymmetric
solutions that will play a major role in the rest of this work. Then I introduce
the notion and properties of the Penrose limit of a space-time and show that
this connects maximally supersymmetric pp-waves to the $AdS\times S$
spaces. Having introduced the necessary background material, the correspondence
between IIB string theory on the maximally supersymmetric plane wave and a
double scaling limit of ${\mc N}=4$ $SU(N)$ super Yang-Mills will be derived from the
AdS/CFT correspondence.
Several features of this novel BMN correspondence, for example symmetries, the
comparison of states and spectra, and holography, will be discussed in detail both
from the (free) string theory and the gauge theory point of view.

Section~\ref{chapter3} presents extensions of the BMN duality. First an
overview over various possible approaches is given to provide
a feeling for the general picture that emerges. The ingredients are then used to
describe in detail the specific example of the duality between strings on
supersymmetric orbifolds of the plane wave background and ${\mc N}=2$ quiver
gauge theories.
In addition to these generalizations, further issues to be discussed include
D-branes on the plane wave and more
complicated pp-wave backgrounds leading to interacting world-sheet theories.

We return to string theory on the plane wave background in section~\ref{chapter4}, where
string interactions are introduced. These correspond to non-planar
corrections in the (interacting) dual gauge theory. I explain why it is natural to
describe string interactions in the setup of light-cone string field theory and
discuss its principles, in particular additional complications arising in the superstring
as compared to its bosonic version.
To make the presentation self-contained a review of the
free string is included.
In the following, the full construction of the cubic interaction vertex as well
as the dynamical supercharges is presented; the focus is mostly on the general methods and
technical details are relegated to two appendices. The results thus obtained are applied
to compute the mass shift of certain string states induced by interactions. In an
approximation to be specified, the leading non-planar corrections to the anomalous dimension
of the dual gauge theory operators are exactly recovered within string theory.

Finally, I conclude in section~\ref{chapter5} and discuss some open problems.

\section{Strings on the plane wave from gauge theory}\label{chapter2}

\subsection{pp-waves in supergravity}\label{sec21}

It is known that maximally supersymmetric backgrounds of 11-dimen\-sional supergravity include flat Minkowski space,
$AdS_4\times S^7$ and $AdS_7\times S^4$~\cite{Freund:1980xh}.
In addition to these three spaces there is another maximally supersymmetric solution
discovered by Kowalski-Glikman~\cite{Kowalski-Glikman:1984wv}. This solution -- which will be referred to as the KG space -- arises as a special case of the
more general pp-wave\footnote{pp-wave geometries are space-times admitting a covariantly constant null vector field.}
solutions~\cite{Hull:1984vh} of the form
\begin{equation}\label{pp11}
\begin{split}
ds^2 & = 2dx^+dx^-+H(x^I,x^+)\bigl(dx^+\bigr)^2+dx^Idx^I\,,\\
F_4 & = dx^+\wedge\varphi(x^I,x^+)\,,
\end{split}
\end{equation}
where $I$ labels the transverse nine-dimensional space, $F_4$ is the four-form field
strength of 11d supergravity and $H$ obeys
\begin{equation}\label{condition}
\Delta H = -\varphi^2\,,\qquad \varphi^2\equiv\frac{1}{3!}\varphi_{IJK}\varphi^{IJK}\,.
\end{equation}
$\D$ is the Laplacian in the transverse space $\mathbb{E}^9$ and $\varphi$ is
closed and co-closed in $\mathbb{E}^9$. $\p/\p x^-$ is a covariantly constant null vector.
For constant $\varphi$ this solution
preserves at least 16 supercharges~\cite{Hull:1984vh,Figueroa-O'Farrill:2001nz}.
An important subclass of solutions are the homogeneous plane wave space-times, where the field strength is constant and $H$ is
independent of $x^+$ and quadratic in the $x^I$
\begin{equation}\label{cw}
H(x^I) = A_{IJ}x^Ix^J\,,
\end{equation}
with $A$ a constant, symmetric matrix. In this case the metric describes a Lorentzian symmetric space $G/K$
with $K=\mathbb{R}^9$ and $G$ a (solvable) Lie group depending on $A$~\cite{Cahen1970,Figueroa-O'Farrill:2001nz}.
Solutions of this kind are space-times with a null ($F_4^2=0$) homogeneous flux and were referred to as Hpp-waves
in~\cite{Figueroa-O'Farrill:2001nz}. Up to the overall scale and permutations
these solutions are parameterized by the eigenvalues of $A$.
Modulo diffeomorphisms, there is precisely one choice for $A$ for which the solution is maximally supersymmetric. This is the KG solution
\begin{equation}\label{kg}
A_{IJ} =
\begin{cases} -\frac{1}{9}\d_{IJ}\,,\quad I,J=1,2,3 \\ -\frac{1}{36}\d_{IJ}\,,\quad I,J=4,\ldots,9
\end{cases}\qquad
\varphi = dx^1\wedge dx^2\wedge dx^3\,.
\end{equation}
Let me briefly sketch the derivation of some of the statements that I made above. It is possible to verify that the
pp-wave geometry in equation~\eqref{pp11} is a solution of the supergravity equations of motion provided the conditions on $\varphi$
and $H$ are satisfied.
To analyze the number of preserved supersymmetries one has to consider the Killing spinor equation.
A solution to the supergravity equations of
motion is supersymmetric if it is left invariant under some non-trivial supersymmetry transformation. If the fermions have been put to zero
in the solution non-trivial conditions following from the requirement of unbroken supersymmetry only arise in the transformation of
the fermions themselves. The gravitino transformation law gives rise to the Killing spinor equation
\begin{equation}\label{killing}
\d_{\e}\psi_M = {\mc D}_M\e = 0\,,
\end{equation}
where the supercovariant derivative is
\begin{equation}
{\mc D}_M\e=\N_M\e-\frac{1}{288}\left({\G^{PQRS}}_M+8\G^{PQR}\d^S_M\right)F_{PQRS}\e\,.
\end{equation}
Iterating the first order Killing equation implies the second order supergravity equations
of motion. In other supergravities containing additional bosonic and fermionic fields
the number of unbroken supersymmetries may be further constrained by algebraic equations
arising from the variations of other fermions, such as for example the dilatino in type IIB
supergravity. Computing the supercovariant derivative in the background equation~\eqref{pp11}
and solving the Killing equation leads to~\cite{Hull:1984vh}
\begin{equation}
\p_+\e = \frac{1}{24}\varphi_{IJK}\G^{IJK}\e\,,\qquad \G_-\e=0\,,
\end{equation}
where $\e=\e(x^+)$ is only a function of $x^+$ and $\varphi$ is assumed to be constant. This
is a first order ordinary differential
equation, which has a unique solution for each initial value. Hence, for constant field strength, the background in
equation~\eqref{pp11} generically preserves 16 supersymmetries. If one chooses the three-form
$\varphi$ and the matrix $A$ to be of the form given in equation~\eqref{kg} spinors
satisfying $\G_+\e=0$ solve the Killing equation as well~\cite{Chrusciel:1984gr,Figueroa-O'Farrill:2001nz} and hence the KG
solution is maximally supersymmetric. The fact that the Hpp-wave geometry is a
Lorentzian symmetric space can be seen as follows~\cite{Figueroa-O'Farrill:2001nz}: consider the 20-dimensional
Lie algebra
\begin{equation}\label{alg}
[e_+,e_I]=e_I^*\,,\qquad [e_+,e_I^*]=A_{IJ}e_J\,,\qquad [e_I^*,e_J]=A_{IJ}e_-\,,
\end{equation}
which is isomorphic to ${\mf h}(9)\rtimes\mathbb{R}$, ${\mf h}(9)$ the Heisenberg algebra generated by $\{e_I,e_I^*,e_-\}$,
$e_-$ being the central element, and $e_+$ an outer automorphism which
rotates coordinates $\{e_I\}$ and momenta $\{e_I^*\}$. The Hpp-wave space-time can then be constructed as the coset $G/K$, where $G$ is the
Lie group with the algebra in~\eqref{alg} and $K$ is generated by $\{e_I^*\}$~\cite{Figueroa-O'Farrill:2001nz}.
To verify this one proceeds in the standard way by choosing a representative of the coset and solving the Cartan-Maurer equations.
Notice that the inclusion of the form flux respects these symmetries as $F_4$ is parallel. For a generic Hpp-wave background these are all
the isometries, in special cases however, the number of isometries is enlarged due to symmetries of $A$ and $F_4$. For example, for the
KG solution the isometry is enhanced to a semi-direct product
\begin{equation}
{\mf h}(9)\rtimes\bigl({\mf s}{\mf o}(3)\oplus{\mf s}{\mf o}(6)\oplus\mathbb{R}\bigr)\,,
\end{equation}
due to the degeneracy of the eigenvalues of $A$.
Notice that the dimension of the isometry algebra of the KG solution is 38, which equals
the dimension of the isometry algebras
of the two other non-trivial maximally supersymmetric solutions $AdS_4\times S^7$ and
$AdS_7\times S^4$ (${\mf s}{\mf o}(3,2)\oplus {\mf s}{\mf o}(8)$ and ${\mf s}{\mf o}(6,2)\oplus {\mf s}{\mf o}(5)$, respectively).
One suspects that this is not merely a coincidence. Recall that flat space and $AdS_4\times S^7$ ($AdS_7\times S^4$) play
the role of asymptotic and near-horizon limits of the M2-brane (M5-brane) and as such are related to each other. Is there a connection
to the KG solution as well? I will say more about this in the next section.
The full superalgebra can be obtained by utilizing the fact that for $\e_1$, $\e_2$ Killing spinors, $\bar{\e}_1\G^M\e_2$ is a Killing
vector and by analyzing the transformations of Killing spinors induced by the action of the Killing vectors. This has been done
in~\cite{Figueroa-O'Farrill:2001nz} to which I refer for details.

The story is similar for type IIB supergravity~\cite{Blau:2001ne}. The analogue of
equation~\eqref{pp11} is
\begin{equation}\label{pp10}
\begin{split}
ds^2 & = 2dx^+dx^-+H(x^I,x^+)\bigl(dx^+\bigr)^2+dx^Idx^I\,,\\
F_5 & = dx^+\wedge\varphi(x^I,x^+)\,,
\end{split}
\end{equation}
with the dilaton being constant and all other supergravity fields set to zero.
The equations of motion for $F_5$ require that
the four-form $\varphi$ is self-dual and closed in $\mathbb{E}^8$ and hence also co-closed. Again, $H$ has to satisfy the
Poisson equation in transverse space
\begin{equation}\label{condition1}
\Delta H = -\frac{1}{2}\varphi^2\,,\qquad \varphi^2\equiv\frac{1}{4!}\varphi_{IJKL}\varphi^{IJKL}\,.
\end{equation}
For $\varphi$ constant, this solution preserves as least 16 supersymmetries~\cite{Blau:2001ne}.
In analogy with the 11d case, the subclass of solutions in which $H$ is of the
form~\eqref{cw}, describe Lorentzian symmetric space-times
with homogeneous five-form flux. There is again one exceptional, maximally supersymmetric
solution~\cite{Blau:2001ne}
\begin{equation}
A_{IJ}=-\m^2\d_{IJ}\,,\qquad \varphi=4\m \bigl(dx^1\wedge dx^2\wedge dx^3\wedge dx^4+dx^5\wedge dx^6\wedge dx^7\wedge dx^8\bigr)\,.
\end{equation}
Here $\m$ is a parameter with dimension of mass, which by a rescaling of $x^+$ and $x^-$ can be set to any non-zero value. It has
become common in the literature to refer to this solution as the {\em plane wave background}. The isometry algebra of the plane wave
solution is
\begin{equation}
{\mf h}(8)\rtimes\bigl({\mf s}{\mf o}(4)\oplus{\mf s}{\mf o}(4)\oplus\mathbb{R}\bigr)\,.
\end{equation}
Notice that the metric by itself has an ${\mf s}{\mf o}(8)$ symmetry, which however, is broken by the \RR field strength to
${\mf s}{\mf o}(4)\oplus{\mf s}{\mf o}(4)$. The isometry group also contains a discrete ${\mathbb Z}_2$ exchanging the two transverse
${\mathbb R}^4$'s. The dimension of the isometry algebra is 30 -- again the same as of the
${\mf s}{\mf o}(4,2)\oplus {\mf s}{\mf o}(6)$ of $AdS_5\times S^5$. Let me be more explicit about the
Killing vectors of the plane wave solution generating the algebra ${\mf h}(8)\rtimes\mathbb{R}$.
A convenient parametrization is~\cite{Blau:2001ne}
\footnote{Strictly speaking one should write $P_+$ instead of $P^-$ since indices are raised
and lowered with the plane wave metric and $g_{++}$ is non-zero. So $P^-\equiv P_+$ by
definition.}
\begin{equation}\label{generators}
\begin{split}
P^- & = -i\p_+\,,\qquad P^+=-i\p_-\,,\\
P^I & = -i\cos(\m x^+)\p_I-i\m\sin(\m x^+)x_I\p_-\,,\\
J^{+I} & = -i\m^{-1}\sin(\m x^+)\p_I+i\cos(\m x^+)x_I\p_-\,.
\end{split}
\end{equation}
They obey the algebra
\begin{equation}\label{ppalg}
[P^-,P^I]  = i\m^2J^{+I}\,,\qquad [P^I,J^{+J}]=i\d_{IJ}P^+\,,\qquad [P^-,J^{+I}]=-iP^I\,,
\end{equation}
and transform in the obvious way under the transverse ${\mf so}(4)\oplus{\mf so}(4)$.
The generators $\{P^I,J^{+I}$, $P^+,P^-\}$ are hermitian and related to
$\{e_I,e_I^*,e_-,e_+\}$ by trivial rescaling. It is convenient to work with the former to
make the flat space limit $\m \to 0$ manifest.
I will present some of the remaining (anti)commutation relations of the plane wave
superalgebra in section~\ref{chapter4} when I need them, see~\cite{Blau:2001ne} for the full algebra.

One might wonder if there are any further maximally supersymmetric solutions of
ten/eleven-dimensional supergravities, however, as was proved
in~\cite{Figueroa-O'Farrill:2002ft} by careful analysis of the constraints
arising from the supersymmetry variations, this is not the case.
It is instructive to discuss the
issue of supersymmetry in Hpp-wave backgrounds in more detail, in particular the dependence
of the Killing spinors on the coordinate $x^+$.
For $\varphi$ constant and hence $H$ independent of $x^+$,
the Killing spinors of the background~\eqref{pp10} are independent of $x^-$
and can be expressed as~\cite{Blau:2001ne}
\begin{equation}\label{spinor}
\e = \left(1+\frac{i}{2}x^I\G_-[\G_I,W]\right)\chi\,,\qquad
W\equiv\frac{1}{4!}\varphi_{IJKL}\G^{IJKL}\,,
\end{equation}
where $\chi$ has only $x^+$ dependence determined by
\begin{equation}
\bigl(\p_++iW\bigr)\chi  = 0\,.
\end{equation}
Additionally one has the requirement that
\begin{equation}
\bigl(x^IW^2+32\p^IH\bigr)\G_I\G_-\chi = 0\,.
\end{equation}
This equation determines the number of Killing spinors. As $\chi=\G_-\chi_0$ is a solution for any $H$ satisfying
equation~\eqref{condition1}, the generic Hpp-wave background has 16
{\em standard} Killing spinors~\cite{Cvetic:2002hi}. By equation~\eqref{spinor} these are also
independent of the $x^I$. Generically the standard spinors depend on the
coordinate $x^+$ but they are independent of it if $W\chi=0$. This equation may or may not
have solutions depending on the explicit form of the four-form $\varphi$. If $H$ is quadratic
in $x^I$ the above equation may admit additional Killing spinors
$\chi$ that are annihilated by $\G_+$. These {\em supernumerary} spinors are always
independent of $x^+$~\cite{Cvetic:2002hi} but depend on the $x^I$ via equation~\eqref{spinor}.
Performing a T-duality along $x^+$, those Killing spinors which are
independent of $x^+$ will survive at the level of the low-energy effective field theory and
the resulting type IIA solution will also be
supersymmetric.\footnote{In the full string theory including winding states, all
supersymmetries must survive as T-duality is an exact symmetry.}
So in the generic case (only standard Killing spinors, all depending on $x^+$), performing a
T-duality along $x^+$ results in a non-supersymmetric
solution of type IIA supergravity. In special cases like the plane wave background
(16 supernumerary spinors), the IIA solution will be
supersymmetric.
Lifting this solution to 11 dimensions gives rise to a supersymmetric deformed
M2-brane with additional four-form flux~\cite{Cvetic:2002hi}.
One can also revert this logic~\cite{Cvetic:2002si} and analyze the Killing spinors of the
11d Hpp-waves. In this case the supernumerary Killing
spinors generically also depend on $x^+$. Dimensionally reducing the Hpp-wave on $x^+$ or $x^I$ (provided the latter is a
Killing direction) one gets a D0-brane or IIA pp-wave, respectively and the number of preserved supersymmetries is again determined by
the coordinate dependence of the Killing spinors in 11 dimensions.

\subsection{The Penrose-G\"{u}ven limit}

We have seen in the previous subsection that ten/eleven-dimensional supergravities
admit maximally supersymmetric solutions of the
pp-wave type, the plane-wave background and the KG solution, respectively. These are on equal footing with the other more standard maximally
supersymmetric backgrounds, that is flat space and the $AdS\times S$ solutions. But whereas the latter are connected being the asymptotic
and near-horizon regions respectively of fundamental branes, no such connection was known for the pp-waves. I have already mentioned
that the dimensions of the superalgebras of the KG and plane wave solutions agree with those
of $AdS\times S$, so one might expect that
there exists a connection between the two. In fact it does~\cite{Blau:2002dy} and the connection is the Penrose-G\"{u}ven limit as defined
originally by Penrose~\cite{Penrose1976} and extended to supergravity by G\"{u}ven~\cite{Gueven:2000ru}. I review this limit below.

Consider a Lorentzian space-time and a null geodesic $\g$ in it. According to~\cite{Penrose1976,Gueven:2000ru} for a sufficiently
well-behaved geodesic
one can introduce local coordinates $U$, $V$ and $Y^I$ such that the metric in the neighborhood of $\g$ takes the form
\begin{equation}\label{g}
ds^2 = dV\left(dU+\a dV+\b_IdY^I\right)+C_{IJ}dY^IdY^J\,,
\end{equation}
where $\a$, $\b_I$ and $C_{IJ}$ are functions of the coordinates. The coordinate $U$ is the affine parameter of $\g$ and for $\g$ to be
well-behaved $C$ must be invertible, otherwise the coordinate system breaks down. Supergravities contain additional fields besides the
metric, such as the dilaton $\Phi$ and $p$-form potentials $A_p$. In particular the $p$-forms have a gauge symmetry and this gauge freedom
can be used to eliminate some of the components of $A_p$. Indeed, one can choose locally~\cite{Gueven:2000ru}
\begin{equation}\label{A}
A_{UVI_1\cdots I_{p-2}}=0=A_{UI_1\cdots I_{p-1}}\,.
\end{equation}
This is the starting point of the Penrose-G\"{u}ven limit: a null geodesic $\g$ which locally is described by the metric in
equation~\eqref{g} plus (possibly) additional background fields which are gauge fixed to have the local form in equation~\eqref{A}.
The next step consists in introducing a real, positive constant $\O$ and rescaling the coordinates as
\begin{equation}\label{resc}
U=u\,,\qquad V=\O^2 v\,,\qquad Y^I=\O y^I\,.
\end{equation}
This diffeomorphism results in a $\O$-dependent family of fields $g(\O)$, $A_p(\O)$ and $\Phi(\O)$ and the coordinate choices in
equations~\eqref{g} and~\eqref{A} ensure that the following Penrose limit~\cite{Penrose1976},
extended by G\"{u}ven~\cite{Gueven:2000ru} to fields other than
the metric, is well-defined:
\begin{equation}\label{scaling}
\bar{g}=\lim_{\O\to0}\O^{-2}g(\O)\,,\qquad
\bar{A}_p=\lim_{\O\to0}\O^{-p}A_p(\O)\,,\qquad
\bar{\Phi}=\lim_{\O\to0}\Phi(\O)\,.
\end{equation}
Due to the rescaling of coordinates in~\eqref{resc} the limiting fields only depend on $u$ and the background takes the form
\begin{equation}
\begin{split}
ds^2 & = dudv+\bar{C}_{IJ}(u)dy^Idy^J\,,\\
\bar{F}_{p+1} & = du \wedge \bar{A}_p'(u)\,.
\end{split}
\end{equation}
Here $\bar{F}_{p+1}$ is the $(p+1)$-form field strength of $\bar{A}_p$ and $'$ denotes $d/du$. This background describes a pp-wave with
null flux in {\em Rosen} coordinates~\cite{Blau:2002dy}. It is possible to change to {\em Brinkmann} coordinates, where
the resulting metric takes the form
\begin{equation}
ds^2 =  2dx^+dx^-+A_{IJ}(x^+)x^Ix^J\bigl(dx^+\bigr)^2+dx^Idx^I\,,
\end{equation}
considered in the previous subsection. For more details, see~\cite{Blau:2002dy}.
Before I explicitly show that this mechanism connects the KG and plane wave solutions with the $AdS$ ones,
it is instructive to discuss some important {\em hereditary} properties of the Penrose limit~\cite{Blau:2002mw}. As we have seen, the
Penrose limit basically consists of two steps, performing a diffeomorphism and gauge-fixing with a subsequent rescaling of the supergravity
fields. It is a general property of supergravity actions that they transform homogeneously under the rescaling of fields in
equation~\eqref{scaling}. Hence, if the original background is a solution to the supergravity equations of motion, so is the
new $\O$-dependent one for any $\O>0$ and by continuity the limiting configuration~\eqref{scaling} is a valid supergravity
background~\cite{Penrose1976,Gueven:2000ru}. The Penrose limit inherits further properties of its parent solution, involving for example the
curvature tensor; the Penrose limit of a conformally flat space-time is conformally flat, that of an Einstein space is
Ricci-flat and another hereditary property is that of being locally symmetric, see for example~\cite{Blau:2002mw}.
One may also wonder about the fate of isometries and supersymmetries; these
are hereditary in the sense that the resulting background has {\em at least} as many isometries and supersymmetries as the
parent background~\cite{Blau:2002mw}. Let me show that this is the case. Consider a Killing vector $\xi$ of the metric $g$. Performing the
rescaling of coordinates and fields in equations~\eqref{resc} and~\eqref{scaling}, $\xi\to\xi(\O)$ and $\xi(\O)$ is a Killing vector
for the transformed metric $\O^{-2}g(\O)$ for non-zero $\O$. The question is if a weight $\D_{\xi}$ exists such that the limit
\begin{equation}
\bar{\xi} = \lim_{\O\to0}\O^{\D_{\xi}}\xi(\O)\,,
\end{equation}
is both non-singular and non-zero. In the local coordinates adapted to the null geodesic $\xi$ can be written as
\begin{equation}
\xi = \a(U,V,Y^I)\p_U+\b(U,V,Y^I)\p_V+\g^I(U,V,Y^I)\p_{Y^I}\,.
\end{equation}
Performing the rescaling of coordinates one can expand $\xi(\O)$ around $\O=0$ as
\begin{equation}
\O^2\xi(\O) = \bar{\b}(u)\p_v+\O\bigl(\bar{\g}^I(u)\p_{y^I}+y^I\p_{y^I}\bar{\b}(u)\p_v\bigr)+\cdots
\end{equation}
Then for $\O^{k_{\xi}}$ being the coefficient of the first non-vanishing term in this expansion
\begin{equation}
\bar{\xi} = \lim_{\O\to0}\O^{2-k_{\xi}}\xi(\O)
\end{equation}
is finite and non-zero. Now suppose we have two linearly independent Killing vectors $\xi_1$ and $\xi_2$. Then it might happen that
their leading order terms in a small-$\O$ expansion are linearly dependent, for definiteness assume they are equal. Do we loose
a Killing vector here? Consider the difference
\begin{equation}
\xi_-(\O) = \xi_1(\O)-\xi_2(\O)\,.
\end{equation}
By construction the leading order term is zero. The next to leading term defines a new Killing vector $\bar{\xi}_-$. If
$\bar{\xi}_-$ and  $\bar{\xi}_1$ are linearly independent
we are done, if not one has to iterate the procedure. One can show~\cite{Blau:2002mw} that eventually one ends up
with two linearly independent Killing vectors of the limiting space-time. Hence the number of Killing vectors never decreases in
the Penrose-G\"{u}ven limit. Notice however that it may very well happen that it {\em increases}. This is because we have seen that
the resulting space-time is of the Hpp-wave form and as we know from the previous section this space-time has always an isometry algebra
isomorphic to a $(2D-3)$-dimensional Heisenberg algebra plus outer automorphism (in D dimensions).
So some isometries need not have a counterpart in the original
space-time and can arise only in the limit $\O\to 0$. It is also important to realize that because different Killing vectors $\xi$
may have to be rescaled with different weights $\D_{\xi}$ the original isometry algebra may get contracted in the limit.
The discussion of the hereditary properties of Killing spinors is similar. Again, no supersymmetries are lost
in the limit, though the number of Killing spinors may increase (as we have seen Hpp-waves preserve at least 16 supersymmetries).
For a more detailed and rigorous discussion see~\cite{Blau:2002mw}.

The information acquired above is already quite powerful. Consider for example the  Penrose limit of $AdS$. Anti de-Sitter
is a conformally flat, locally symmetric, Einstein space. The limiting space-time is Ricci-flat, conformally flat and locally
symmetric and hence isometric to flat Minkowski space. We are primarily interested in the maximally supersymmetric
$AdS\times S$ backgrounds. Now the result depends on the geodesic: if it lies purely in $AdS$ we get Minkowski space
(the sphere is blown up to flat space in the limit as well); if not it follows from the hereditary properties that we have to get
the KG solution and the plane wave background as limiting space-times~\cite{Blau:2002dy,Blau:2002mw}. I will also show this explicitly below for the
case of $AdS_5\times S^5$. For $AdS_4\times S^7$ and $AdS_7\times S^4$ the Penrose-G\"{u}ven limits are isomorphic to
each other and result in the KG solution~\cite{Blau:2002dy}.

The spaces $AdS_{p+2}\times S^{D-p-2}$ with radii of curvature related by $R_{AdS}/R_{S}=\r$
provide an example which
illustrates the above behavior of isometries~\cite{Blau:2002mw}. The original isometry algebra
is ${\mf s}{\mf o}(2,p+1)\oplus{\mf s}{\mf o}(D-p-1)$. The ${\mf s}{\mf o}(2,p+1)$
factor is contracted to ${\mf h}(p+1)\rtimes {\mf s}{\mf o}(p+1)$. The $p+1$ creation- and $p+1$ annihilation operators transform as vectors
under ${\mf s}{\mf o}(p+1)$. Similarly ${\mf s}{\mf o}(D-p-1)$ contracts to ${\mf h}(D-p-3)\rtimes{\mf s}{\mf o}(D-p-3)$. The central
elements of the two Heisenberg algebras coincide; this is due to the fact that two Killing vectors of the parent space-time agree to
leading order in small $\O$.  Thus the two Heisenberg algebras combine into ${\mf h}(D-2)$. The remaining Killing vector $\bar{\xi}_-$
becomes an outer automorphism and the resulting contracted algebra is~\cite{Blau:2002mw}
\begin{equation}
{\mf h}(D-2)\rtimes\bigl({\mf s}{\mf o}(p+1)\oplus{\mf s}{\mf o}(D-p-3)\oplus\mathbb{R}\bigr)\,.
\end{equation}
If the radii of curvature are equal (as is the case for $p=3$) the subalgebra ${\mf s}{\mf o}(p+~1)\oplus{\mf s}{\mf o}(D-p-3)$ is enlarged
to the full ${\mf s}{\mf o}(D-2)$. This has no counterpart in the original background.

{}Finally, consider the Penrose-G\"{u}ven limit of $AdS_5\times S^5$ explicitly.
The dilaton is constant and in global coordinates the metric and five-form flux is
\begin{equation}
\begin{split}
ds^2 & = R^2\bigl[-\cosh^2\r dt^2+d\r^2+\sinh^2\r d\O_3^2+\cos^2\theta d\psi^2+d\theta^2+\sin^2\theta d{\O'}_3^2\bigr]\,,\\
F_5 & = 4R^4\bigl[\cosh\r\sinh^3\r dt\wedge d\r\wedge d\O_3+
\cos\theta\sin^3\theta d\psi\wedge d\theta \wedge d\O'_3\bigr],
\end{split}
\end{equation}
where $R^4 \equiv 4\pi g_{\text{s}}{\a'}^2N$ and
$\r\ge0$, $t\in\mathbb{R}$, $\psi\in[0,2\pi]$ and $\theta\in[0,\frac{\pi}{2}]$.
As alluded to above, in order that the limiting space-time will be non-trivial the
null geodesic must not lie purely within $AdS_5$; so consider a
massless particle sitting at the origin of $AdS_5$ ($\r=0$)
and rotating around the circle of the $S^5$ parameterized by $\psi$ and
$\theta=0$~\cite{Blau:2002dy,Berenstein:2002jq}. To focus
on the geometry in the neighborhood of this geodesic the coordinates are rescaled such that
a tube around the geodesic is blown up. Explicitly, introduce
light-cone coordinates $x^{\pm}$ and perform a rescaling
\begin{equation}
x^+ = \frac{1}{2\m}(t+\psi)\,,\qquad x^- = -\m R^2(t-\psi)\,,\qquad \r=\frac{r}{R}\,,\qquad \theta=\frac{y}{R}\,,
\end{equation}
where $\m$ is an arbitrary (non-zero) mass parameter. Blowing up the neighborhood of the
geodesic is equivalent to taking $R\to\infty$
and the metric and five-form flux become
\begin{equation}\label{planewave}
\begin{split}
ds^2 & = 2dx^+dx^--\m^2\vec{x}^2\bigl(dx^+\bigr)^2+d\vec{x}^2\,,\\
F_5 & = 4\m dx^+\wedge\bigl(dx^1\wedge dx^2\wedge dx^3\wedge dx^4+dx^5\wedge dx^6\wedge dx^7\wedge dx^8\bigr)\,.
\end{split}
\end{equation}
This is the plane wave solution of type IIB supergravity~\cite{Blau:2002dy}.

\subsection{The BMN correspondence}\label{sec23}

In the previous subsection I reviewed the connection of $AdS_5\times S^5$ and the plane wave
background via the Penrose-G\"{u}ven limit. 
As IIB string theory on $AdS_5\times S^5$ is dual to ${\mc N}=4$ $SU(N)$ super Yang-Mills
by the AdS/CFT correspondence~\cite{Maldacena:1998re,Gubser:1998bc,Witten:1998qj,Aharony:1999ti} the
implications of the Penrose-G\"{u}ven limit on the dual CFT can be studied. 

It has been known for some time that strings on pp-wave NS-NS backgrounds are exactly solvable, 
see e.g.~\cite{Amati:1988ww,Horowitz:1990bv,deVega:1990nr,deVega:1990kk,deVega:1992ke,deVega:1993rr}.
In light-cone gauge this is also true for a large class of pp-wave \RR backgrounds, in particular 
the maximally supersymmetric plane wave, in spite of the presence of the constant \RR flux~\cite{Metsaev:2001bj}.
So one may hope
that this simpler setup allows to extend our understanding of the AdS/CFT duality beyond the
supergravity approximation by the inclusion of string states on the plane wave. This is
indeed the case as was demonstrated by Berenstein, Maldacena and Nastase in~\cite{Berenstein:2002jq}.
The formulation of the BMN correspondence is the subject of this subsection.

Following the construction of the type IIB superstring action on $AdS_5\times S^5$ using superspace
coset methods~\cite{Metsaev:1998it}, the action on the plane wave background was constructed
by Metsaev in~\cite{Metsaev:2001bj}.
Let me briefly sketch this construction.
The action has to obey the following conditions: its bosonic part is the $\s$-model
with the plane wave geometry being the target space; it is globally supersymmetric with respect
to the plane wave superalgebra and locally $\kappa$-symmetric; it reduces to the standard
Green-Schwarz action in the flat space limit. As shown in~\cite{Metsaev:2001bj} this
conditions uniquely specify the action, which as in flat space can be
written as a sum of a `kinetic' $\s$-model term and a Wess-Zumino term. The latter is needed
to obey the condition of $\kappa$-symmetry.
To find the explicit form of the superstring action in terms of the coordinate (super)fields
a parametrization of the coset representative has to be specified and the
Cartan-Maurer equations have to solved. Not surprisingly, the resulting covariant action
is non-polynomial~\cite{Metsaev:2001bj}.
The simplest way to proceed is to study the action in light-cone gauge.
As in flat space the light-cone gauge-fixing procedure consists of two steps, first
$\kappa$-symmetry is fixed by the fermionic light-cone gauge choice
$\G^+S=0$, then the diffeomorphism and Weyl-symmetry on the world-sheet is fixed by the
bosonic light-cone gauge $\sqrt{-g}g^{ab}=\eta^{ab}$ and $x^+(\s,\t)=\t$.
The resulting action is quadratic in both bosonic and
fermionic superstring $2d$ fields, and hence can be quantized explicitly~\cite{Metsaev:2001bj}.
In fact, from the form of the metric in equation~\eqref{planewave}, it is obvious that
the action for the eight transverse directions in light-cone gauge is just that for eight
bosons of mass $\m$. Similarly the fermions acquire masses due to the coupling to
the \RR background~\cite{Metsaev:2002re}. Masses of bosons and fermions are equal due to world-sheet supersymmetry:
after imposing the light-cone gauge conditions the world-sheet $\kappa$-symmetry and space-time
supersymmetries transmute into rigid world-sheet supersymmetries. As in flat space
16 of the 32 supersymmetries are linearly realized in light-cone gauge and commute
with the Hamiltonian~\cite{Metsaev:2001bj}. It was shown in~\cite{Cvetic:2002nh} that the linearly realized supersymmetries
correspond to the supernumerary Killing spinors of the pp-wave
backgrounds. This is in agreement with their independence of $x^+$~\cite{Cvetic:2002hi}
(cf.\ section~\ref{sec21}).

After gauge-fixing the light-cone action becomes~\cite{Metsaev:2001bj,Metsaev:2002re}
\begin{equation}\label{lcaction}
S_{\text{l.c.}}=\frac{1}{2\pi\a'}\int\,d\t\int_0^{2\pi\a' p^+}\,d\s
\left[\frac{1}{2}\dot{x}^2-\frac{1}{2}x^{\prime\,2}-\frac{1}{2}\m^2x^2
+i\bar{S}\bigl(\p\!\!\!/+\m\Pi\bigr)S\right]\,,
\end{equation}
where $\Pi=\G^1\G^2\G^3\G^4$ and $S$ is a Majorana spinor on the world-sheet and a positive chirality
$SO(8)$ spinor under rotations in the eight transverse directions. It is not difficult
to quantize this action and the resulting light-cone Hamiltonian is~\cite{Metsaev:2001bj,Metsaev:2002re}
\begin{equation}\label{spec}
H=\m\sum_{n\in\mathbb{Z}}N_n\sqrt{1+\frac{n^2}{\bigl(\m\a' p^+\bigr)^2}}\,.
\end{equation}
Here $n$ is a label for the Fourier mode and $N_n$ is the occupation number of that mode
including bosons and fermions. The ground state energy is cancelled between bosons and
fermions. In contrast to flat space, modes with $n=0$ are also harmonic oscillators due to
the mass terms on the world-sheet and string theory on the plane wave has a unique ground
state $|v,p^+\ra$,  $p^+$ the light-cone momentum.
The single string Hilbert space is built
by acting with the bosonic and fermionic creation oscillators (for all $n$) on
$|v,p^+\ra$ subject to the level-matching condition for physical states
\begin{equation}\label{lm}
\sum_{n\in\mathbb{Z}}nN_n=0\,.
\end{equation}
Truncation to the zero-mode sector gives rise to the spectrum of
IIB supergravity on the plane wave~\cite{Metsaev:2002re}. I will provide more details on the
quantization of strings on the plane wave in section~\ref{free}, where I need them.

To understand the effect of the Penrose-G\"{u}ven limit on the dual CFT,
consider the scaling behavior of the energy $E=i\p_t$ and angular momentum
$J=-i\p_{\psi}$ of a state in $AdS_5\times S^5$. Recall that the AdS/CFT correspondence
relates the energy of a string state in $AdS_5\times S^5$ to the energy of a state in
${\mc N}=4$ SYM living on $\mathbb{R}\times S^3$~\cite{Gubser:1998bc,Witten:1998qj},
which is the (conformal) boundary of $AdS_5\times S^5$ in global
coordinates. By the operator-state map, the energy of a state on $\mathbb{R}\times S^3$, where the $S^3$ has unit radius, translates
to the conformal dimension $\D$ of an operator on $\mathbb{R}^4$.
Likewise, the angular momentum $J$ on the $S^5$ translates to the R-charge
under a $U(1)_R$ subgroup of the full $SU(4)_R\simeq SO(6)_R$ R-symmetry of ${\mc N}=4$ SYM.
Then we have the following relations
\begin{equation}\label{Hp+}
\begin{split}
H & =-p_+=i\p_+=i\m (\p_t+\p_{\psi})=\m(\D-J)\,,\\
p^+ & = p_-=-i\p_-=\frac{i}{2\m R^2}(\p_t-\p_{\psi})=\frac{\D+J}{2\m R^2}\,.
\end{split}
\end{equation}
Now what happens if we apply the limit $R\to\infty$? Firstly, $R\to\infty$ means
$N\to\infty$, the string coupling $g_{\text{s}}$ and hence also $g^2_{\text{YM}}=4\pi g_{\text{s}}$ should be kept fixed. Then a
configuration with fixed, non-zero $p^+$ requires to scale $\D$, $J\sim\sqrt{N}$. In fact, the plane
wave superalgebra implies that $H$ and $p^+$ are non-negative or equivalently $\D\ge |J|$;
this also follows from the representation theory of the 4d superconformal algebra.
So the Penrose-G\"{u}ven limit induces the following double-scaling or BMN limit in
${\mc N}=4$ $SU(N)$ SYM~\cite{Berenstein:2002jq}
\begin{equation}
N\to\infty\quad\text{and}\quad J\to\infty\quad\text{with}\quad\frac{J^2}{N}\quad
\text{fixed}\,,\quad g_{\text{YM}}\quad\text{fixed}\,.
\end{equation}
As a first check consider how the bosonic part of the plane wave superalgebra
${\mf h}(8)\rtimes({\mf s}{\mf o}(4)\oplus{\mf s}{\mf o}(4)\oplus\mathbb{R})$ is realized
in the gauge theory on $\mathbb{R}\times S^3$.  The conformal group $SO(4,2)$ is
generated by the seven Killing vectors of $\mathbb{R}\times SO(4)$ and eight
additional conformal Killing vectors.
By singling out a $U(1)_R$ subgroup with generator $J$ the
$SO(6)_R$ symmetry is broken to $SO(4)_R\times U(1)_R$. So we see that the
transverse symmetry corresponds to $SO(4)_R$ and the isometry group of the
$S^3$~\cite{Berenstein:2002jq,Berenstein:2002sa}.
In the BMN limit, the eight conformal Killing vectors together with the
eight broken generators of R-symmetry
give rise to a Heisenberg algebra ${\mf h}(8)$ with central element $J$ and
outer automorphism $E-J$, see for example~\cite{Okuyama:2002zn,Das:2002cw}.
In other words the ${\mc N}=4$ superalgebra contracts
to the plane wave superalgebra in the Penrose-G\"{u}ven limit. In the previous subsection I have argued
that this is the case, see also~\cite{Hatsuda:2002xp} for an explicit demonstration.
It is an important question how the unitary irreducible
representations -- e.g.\ composite operators in ${\mc N}=4$ SYM -- behave under the
contraction~\cite{Arutyunov:2002xd}. In the limit they should
form representations of the plane wave superalgebra.
In particular, as the conformal dimension diverges in the BMN limit, the
space-time dependence of their correlation functions is ill-defined and hence
requires special treatment. One way to achieve this was proposed in~\cite{Arutyunov:2002xd}
and requires to combine space-time with an auxiliary R-symmetry space much in the same
way that $\D$ and $J$ combine into the finite quantity $\D-J$.
The manifestation of the discrete $\mathbb{Z}_2$ exchanging the two
transverse $\mathbb{R}^4$'s in the gauge theory is somewhat mysterious.

The BMN limit is different from the 't Hooft limit of $SU(N)$ gauge theories and at first
sight puzzling. To see why this is so, recall that the 't Hooft limit takes
$N\to\infty$, $g_{\text{YM}}\to 0$, such that the 't Hooft coupling
$\l\equiv g_{\text{YM}}^2N$ is fixed. Away from the strict
$N\to\infty$ limit all Feynman diagrams of a given order in $1/N$ can be
drawn on a Riemann surface whose Euler number is precisely the power of $N$ to which these
diagrams contribute~\cite{'tHooft:1974jz}. So $1/N^2$ is identified with the genus counting parameter and
the perturbation series of the gauge theory may then be organized in a
double series expansion in the effective coupling $\l$ and the genus counting parameter
$1/N^2$. This is the standard lore why large $N$ gauge theories are expected to be dual
to some weakly coupled string theory with coupling $1/N$. The AdS/CFT
correspondence provides a concrete example where this is realized.
The above reasoning breaks down because operators in
the field theory are not held fixed in the limit but acquire an infinite charge as $N\to\infty$. Indeed, using
equation~\eqref{Hp+} and $(\D-J) \ll J$, in the BMN limit
\begin{equation}\label{dict}
\frac{1}{\bigl(\m\a' p^+\bigr)^2}=\frac{g_{\text{YM}}^2N}{J^2}\equiv\l'\,,\qquad
4\pi g_{\text{s}}\bigl(\m\a' p^+\bigr)^2 = \frac{J^2}{N}\equiv g_2\,.
\end{equation}
These relations are quite suggestive. It looks like a new effective coupling $\l'$ and a new effective genus counting parameter $g_2^2$
might develop as a consequence of the simultaneous infinite scaling of $N$ and $J$. This is in some sense correct as I will explain
in more detail below.

While most of the (unprotected) operators acquire infinite anomalous dimension and decouple
in the BMN limit, it is conceivable that some (BMN) operators with a suitable scaling of charge
survive and be dual to string states in the plane wave background (for a general discussion, see ~\cite{Polyakov:2001af}).
At the planar level this class of operators has been identified in~\cite{Berenstein:2002jq}.
Recall that ${\mc N}=4$ SYM contains six scalar fields $\phi^r$ of conformal dimension one transforming
in the ${\bf 6}$ of $SO(6)_R$. Take $J$ to be the $U(1)_R$ generator rotating
the 5-6-plane and define $Z=\frac{1}{\sqrt{2}}\bigl(\phi^5+i\phi^6\bigr)$. $Z$ carries
unit $J$-charge and the remaining four scalars $\phi^i$, $i=1,\ldots,4$  are invariant under
$U(1)_R$. For simplicity, consider only single-trace operators for the moment.
The operator corresponding to the string ground state should carry large $J$ charge and have
$\D-J=0$. There is a unique single-trace operator satisfying this requirement which
subsequently is identified with $|v,p^+\ra$~\cite{Berenstein:2002jq}
\begin{equation}\label{ground}
\frac{1}{\sqrt{JN^J}}\Tr\bigl[Z^J\bigr] \lr |v,p^+\ra\,,
\end{equation}
where the trace is over color indices.
At weak coupling the dimension of this operator is
$J$ since each $Z$ field has dimension one. As the operator is a chiral primary~\cite{Berenstein:2002jq}
it is protected by supersymmetry and $\D-J=0$ for all values of the coupling.
The normalization is chosen such that the operator has normalized two-point function
when we restrict ourselves to planar diagrams.
However, non-planar diagrams do give a non-vanishing contribution in the BMN limit and the two-point function of $\Tr\bigl[Z^J\bigr]$
can be computed exactly for all genera~\cite{Kristjansen:2002bb,Constable:2002hw}.
This can be understood by noting that at genus $h$ diagrams are weighted by $1/N^{2h}$ as expected, but
at the same time the number of diagrams grows as $J^{4h}$, see also~\cite{Berenstein:2002sa,Gross:2002mh}. So we see
the quantity $g_2^2$ emerging
as the effective genus counting parameter for the above operator.
This will also be true for more general BMN
operators, to be described below. There is an additional complication:
at finite $g_2$ single-trace operators are no longer orthogonal to multi-trace operators and it is therefore
no longer justified to restrict attention to single-trace operators only.
To simplify matters let me assume $g_2=0$ in what follows; then equation~\eqref{ground}
is a precise identification. I will return to the issue of operator mixing below.

Next consider the supergravity states obtained by acting with the eight bosonic and fermionic
zero-mode oscillators $a_0^{I\,\dg}$ and $S_0^{a\,\dg}$ on the plane wave
vacuum. Each oscillator raises the energy by $\m$.
In the gauge theory these are
obtained by the action of the broken symmetries on the trace of $Z$'s~\cite{Berenstein:2002jq}.
For example we can rotate $Z$ into $\phi^i$ by a broken $SO(6)_R$ transformation.
Applying this to $\Tr\bigl[Z^{J+1}\bigr]$ one obtains~\cite{Berenstein:2002jq}
\begin{equation}
\frac{1}{\sqrt{N^{J+1}}}\Tr\bigl[\phi^iZ^J\bigr] \lr a_0^{i\,\dg}|v,p^+\ra\,,
\end{equation}
where the cyclicity of the trace was used. Acting a second time with such a transformation
changes another $Z$ to $\phi^j$ or, if $i=j$, $\phi^i$ to $\bar{Z}$. For $i\neq j$
\begin{equation}
\frac{1}{\sqrt{JN^{J+2}}}\sum_{l=0}^{J}\Tr\bigl[\phi^iZ^l\phi^jZ^{J-l}\bigr]
\lr a_0^{i\,\dg}a_0^{j\,\dg}|v,p^+\ra\,.
\end{equation}
Similarly the action of broken superconformal symmetries give rise to insertions of
$D_{i}Z=\p_{i}Z+[A_{i},Z]$ and the components of the gaugino with $J=1/2$, $\chi^a_{J=1/2}$, in the trace of
$Z$'s~\cite{Berenstein:2002jq}. In this way one obtains a precise
correspondence between supergravity states on the plane wave and (at the planar level)
single-trace chiral primary operators. This is already known from the AdS/CFT
correspondence~\cite{Gubser:1998bc,Witten:1998qj}.
One of the crucial insights of~\cite{Berenstein:2002jq} was to extend this
identification to `massive' string states. These are constructed similarly to the above
but now each insertion is accompanied with a phase. For example, the operator
\begin{equation}
\sum_{l=0}^{J}e^{\frac{2\pi i nl}{J+1}}\Tr\bigl[Z^l\phi^i Z^{J-l}\bigr]
\end{equation}
reduces to the supergravity state considered above for $n=0$, but it vanishes for nonzero $n$
due to the cyclicity of the trace. This is precisely how it should be: a single non-zero-mode acting on the vacuum does not satisfy the level-matching condition~\eqref{lm}.
So the next-simplest possibility is to consider~\cite{Berenstein:2002jq}
\begin{equation}\label{On}
\frac{1}{\sqrt{JN^{J+2}}}\sum_{l=0}^{J}e^{\frac{2\pi i nl}{J}}\Tr\bigl[\phi^iZ^l\phi^j Z^{J-l}\bigr]
\lr a_n^{i\,\dg}a_{-n}^{j\,\dg}|v,p^+\ra\,,
\end{equation}
where $i\neq j$, the cyclicity of the trace was used to put one operator at the first position
and $1/J$ contributions have been neglected in the power of the phase factor.
The general rule is quite simple, each insertion of an `impurity' is accompanied with a
phase depending on the world-sheet momentum; those operators where the momenta do not sum to
zero vanish due to cyclicity of the trace, in this way implementing the level matching
condition; the dictionary between impurity insertions and string oscillators is thus
roughly (cf.\ the discussion below) as follows~\cite{Berenstein:2002jq}
\begin{equation}
\begin{split}
a^{i\,\dg} & \lr \phi^i\,,\quad i=1,2,3,4\,,\\
a^{i'\,\dg} & \lr D_{i'-4}Z\,,\quad i'=5,6,7,8\,,\\
S^{a\,\dg} & \lr \chi^a_{J=\frac{1}{2}}\,.
\end{split}
\end{equation}
To check this identification it is useful to expand the string theory Hamiltonian~\eqref{spec}
for large $\m\a' p^+$ or equivalently for small $\l'$ (cf.\ equation~\eqref{dict})
\begin{equation}\label{expand}
\frac{1}{\m}H \simeq
\sum_{n\in\mathbb{Z}} N_n\left(1+\frac{1}{2}\frac{n^2}{\bigl(\m\a'p^+\bigr)^2}\right)
= \sum_{n\in\mathbb{Z}} N_n\left(1+\frac{1}{2}\frac{\l}{J^2}n^2\right)\,.
\end{equation}
We see that for $\m\a' p^+\gg 1$ all string states have approximately the same energy; this
is reproduced by the construction of the BMN operators: in free field theory the inclusion of
the phases does not make a difference, it is only in the interacting theory that this gets
important because these operators are no longer protected. Notice however, that
the BMN operators proposed to be dual to string states are built by sewing together protected operators with
varying phases.
One might imagine that these operators are nearly BPS in the sense
that a delicate cancellation of renormalization and large $J$ effects protects them
from leaving the spectrum in the BMN limit. This is exactly what happens~\cite{Berenstein:2002jq}.
Remarkably it turns out that the anomalous dimensions of these operators are not just finite in the BMN limit, but as
has been argued in~\cite{Berenstein:2002jq}, they are  perturbatively computable with $\l'$ playing the role of the effective coupling.
Indeed, notice that the first correction in~\eqref{expand} involves the 't Hooft coupling
$\l$ so it seems one might reproduce this from a {\em perturbative} (in $g^2_{\text{YM}}$ or $\l$) field theory
computation. Consider for example the operator in~\eqref{On}. Taking into account
interactions the relevant diagrams arise from the quartic vertex
\begin{equation}\label{intq}
\sim g^2_{\text{YM}}\Tr\bigl([Z,\phi^i][\bar{Z},\phi^i]\bigr)\,.
\end{equation}
The effect of this vertex can be analyzed as follows. The above interaction can be split
into two parts, depending on whether the position of the operator $\phi$ in
the `string' of $Z$'s is effectively moved to a neighboring position or not.
Since at the planar level operators with $\phi$'s sitting at different positions
are orthogonal to each other, contracting all the fields
gives a result which, for the first class, does not depend on the insertion of the phases,
whereas for the second class it does. Combining the relevant contributions,
utilizing the fact that other interactions involving gauge bosons and scalar loops
cancel due to supersymmetry and taking the large $N$ and $J$
limit one precisely reproduces the first non-trivial correction
in~\eqref{expand}~\cite{Berenstein:2002jq}.
For a careful treatment see for example~\cite{Kristjansen:2002bb}.
Notice that
the computation was done perturbatively in $\l$, but to take the BMN limit requires to
send $\l\to\infty$. But the result for small $\l$ equals the one for large $\l$ obtained from
the string Hamiltonian and it is tempting to assume that it is correct for all $\l$ at the planar level.
Further support to this conjecture comes from~\cite{Gross:2002su} which extended the above
computation to two loops and presented arguments for higher loops, again matching the expectation
coming from the expansion of the
square root in~\eqref{expand}. In~\cite{Santambrogio:2002sb} superconformal representation theory was used to argue that the full square
root is reproduced; alternatively this was seen to be the case in~\cite{Berenstein:2002jq} by exponentiating the quartic vertex;
let me sketch how this works.
SYM on $\mathbb{R}\times S^3$ can be expanded in spherical harmonics on the $S^3$.
In particular the zero-modes
of scalar fields on the $S^3$ have unit energy and the `string' of oscillators corresponding to the zero-mode of $Z$
carries $\D-J=0$. To raise the energy we insert for example the zero-mode of $\phi\sim b+b^{\dg}$ at some position along the
string of $Z$ oscillators. In the free theory the position of $\phi$ is unchanged and operators with $\phi$ inserted at different positions
are orthogonal in the planar approximation. So we can think of the $J$ $Z$'s as defining a lattice with $J+1$ sites
and an insertion of $\phi$ at different positions corresponds to the excitations $b^{\dg}_l$ at the $l$-th site of the lattice.
As alluded to above, the interaction in~\eqref{intq} can move an operator $\phi$ to a neighboring position, so when acting on the
string of $Z$ oscillators the effective Hamiltonian for $\phi$ consisting of the free and interacting parts is~\cite{Berenstein:2002jq}
\begin{equation}\label{discr}
H \sim \sum_{l}\left(b^{\dg}_lb_l+\frac{\l}{4\pi^2}\bigl[(b_{l+1}+b_{l+1}^{\dg})-(b_{l}+b_{l}^{\dg})\bigr]^2\right)\,.
\end{equation}
In the  large $N$ and $J$ continuum limit the discretized Hamiltonian reduces to
\begin{equation}\label{lcham}
H \sim \int_0^Ld\s\,\bigl[\dot{\phi}^2+{\phi'}^2+\m^2\phi^2\bigr]\,,\qquad L=\frac{2\pi}{\sqrt{\l}}\frac{J}{\m}=2\pi\a'p^+\,.
\end{equation}
This {\em is} the bosonic part of the string light-cone Hamiltonian on the plane wave. Consequently
the full square root is reproduced from planar gauge theory in the BMN limit and the `string' of $Z$'s plus insertion
of impurities becomes equivalent to the physical string~\cite{Berenstein:2002jq}.
So there is evidence that $\l'$ emerges as a new effective coupling
in the BMN limit and one might think that the perturbation series of SYM in the
BMN limit can be reorganized as a double series expansion in the effective coupling
$\l'$ and the effective genus counting parameter $g_2^2$.
If true, the BMN duality has the interesting property that regimes in string theory on the plane wave and SYM in the BMN limit
are {\em simultaneously}
perturbatively accessible. This is in contrast to the usual AdS/CFT correspondence,
where due to our limited ability to perform calculations for finite $\l$ in SYM -- or
equivalently in the full string theory on $AdS_5\times S^5$ -- the relation is a strong/weak
coupling duality. Note however, while perturbative calculations in $\l$ of BMN operator two- and three-point functions
can be reorganized in $\l'$~\cite{Berenstein:2002jq,Kristjansen:2002bb,Constable:2002hw,Gross:2002su,Santambrogio:2002sb} --
and hence an extrapolation to large $\l$ seems viable --
this is no longer the case for higher point functions:
computing for example the 4-point function of $\tr\bigl[Z^J\bigr]$ perturbatively in $\l$,
a naive extrapolation to large $\l$ leads to divergences~\cite{Beisert:2002bb}.

The above heuristic discussion is in fact oversimplifying. Consider for example the BMN operators with $\D-J=2$, that is a
{\em defect charge} of two.
Instead of inserting two impurities (defects) into the trace of $Z$'s we could also insert one $\bar{Z}$, $D\phi^i$, $D^2Z$ etc., that
is fields carrying multiple defect charge.
Indeed, all of these are present, even at the planar level~\cite{Beisert:2002tn}. However, they do not give rise to additional string
states (there are none) but are {\em hidden} within the ordinary operators with single charge defects by
{\em operator mixing}~\cite{Beisert:2002tn}. One example where
this happens is the $SO(4)$ singlet~\cite{Beisert:2002bb,Beisert:2002tn}
\begin{equation}\label{long}
{\mc O}_n^J \sim \sum_{l=0}^J\cos\frac{\pi n(2l+3)}{J+3}\Tr\bigl[\phi^iZ^l\phi^iZ^{J-l}\bigr]
-4\cos\frac{\pi n}{J+3}\Tr\bigl[\bar{Z}Z^{J+1}\bigr]\,.
\end{equation}
Written like this it is in fact an exact one-loop eigenstate of $\D$ even for
finite $J$~\cite{Beisert:2002tn}. Roughly speaking the above mixing is needed to cancel
singularities that occur when the two $\phi$ impurities collide~\cite{Beisert:2002bb}.
For non-zero $n$ the above operator is the primary of a long ${\mc N}=4$ superconformal multiplet and
all the other defect charge two operators dual to string states in the BMN limit are contained in this multiplet as
descendants~\cite{Beisert:2002tn}.
All fields with defect charge two {\em do} appear in these generalized BMN operators. Analogously, for $n=0$ the operator
in equation~\eqref{long} is the primary of a half BPS multiplet; all operators dual to supergravity states with up to two oscillators
are descendants. One might conjecture that this pattern generalizes to higher defect charge~\cite{Beisert:2002tn}.

At finite $g_2$ mixing of single-trace with multi-trace operators has to be taken
into account~\cite{Kristjansen:2002bb,Constable:2002hw}.
For
example, to compute the anomalous dimension on the torus single- and double-trace operators have to be redefined (mixed) in order to
normalize and diagonalize their two-point functions. For the (redefined) operator in~\eqref{long} one finds
at order ${\mc O}(g_2^2\l')$~\cite{Beisert:2002bb,Constable:2002vq}
\begin{equation}\label{anom}
\bigl(\D-J\bigr)_n=2+\l'\left[n^2+\frac{g_2^2}{4\pi^2}\left(\frac{1}{12}+\frac{35}{32\pi^2n^2}\right)\right]\,.
\end{equation}
In fact, the above result holds for all BMN operators with defect charge two transforming in the various irreducible representations
of $SO(4)\times SO(4)$; this is a consequence of superconformal symmetry~\cite{Beisert:2002tn}. For the explicit form of some of the
redefined operators at this order see~\cite{Beisert:2002bb,Constable:2002vq}. It is actually simpler to
consider directly the dilatation operator,
work with the `bare' operators and diagonalize
the resulting anomalous dimension matrix. This approach was followed in~\cite{Beisert:2002ff,Beisert:2003tq}
and results in a simple derivation
of equation~\eqref{anom}. Further results on higher genus correlators
include~\cite{Minahan:2002ve,Eynard:2002df}, scalar/vector, vector/vector and multi-trace BMN operators
have also been considered in~\cite{Gursoy:2002yy,Klose:2003tw,Gursoy:2002fj,Beisert:2002ff}.
For an extension of equation~\eqref{anom} to order ${\mc O}(g_2^4\l')$ see~\cite{Beisert:2002ff}. The
contribution of
higher genus corrections to the anomalous dimension is
related to a mass-shift of the dual string states due to interactions.
A detailed study of string interactions will be deferred to section~\ref{chapter4}. Let me however
mention a route -- which will not be pursued in what follows --
to study interacting strings on the plane wave, the string-bit formalism~\cite{Verlinde:2002ig}.
Inspired by the emergence of the free string, discretized into $J$ bits along the
string coordinate $\s$ as in~\eqref{discr} and from matrix string theory~\cite{Motl:1997th,Banks:1997my,Dijkgraaf:1997vv},
one interprets the $J$ small strings as describing the quantization of the $J$-th symmetric product of the plane wave
target space. This leads to a quantum-mechanical orbifold model. In a
spirit reminding of the matrix string, string splitting and joining is then
realized by an operator that roughly speaking
exchanges two string bits; see~\cite{Verlinde:2002ig} for details.
This approach was further studied in~\cite{Zhou:2002mi,Vaman:2002ka,Pearson:2002zs} and led to results in agreement with field theory.
Very recently, doubts on the consistency of this model have been voiced in~\cite{Bellucci:2003qi}. The reason for this is the so-called
fermion doubling problem, which leads to the loss of supersymmetry -- inevitably broken by the discretization -- even in the
continuum limit. For a posible resolution of this puzzle, see~\cite{Danielsson:2003yc}. 
Moreover, repeating the above derivation of the string
Hamiltonian~\eqref{lcham}
by truncation to the lowest modes corresponding to the operators $DZ$ and the fermions,
apparently does not lead to the correct string Hamiltonian~\cite{Okuyama:2002zn}.

Finally, let me briefly discuss the issue of holography on the plane wave.
As already mentioned, the conformal boundary of $AdS_5\times S^5$ in global coordinates
is $\mathbb{R}\times S^3$ on which the dual SYM theory lives. However, in the
Penrose-G\"{u}ven limit one focuses on the neighborhood of a null geodesic located
at the origin of $AdS_5$ and rotating around a great circle of the $S^5$.
It was shown in~\cite{Berenstein:2002sa} that the
conformal boundary of the plane wave is a one-dimensional null line.
This can be seen by a conformal mapping of the plane wave to the Einstein static universe
$\mathbb{R}\times S^9$. Since the Einstein static universe is regular, the boundary consists
of the space-time region for which the Weyl factor is divergent.
This is the case for a null line,
a $S^7$ inside the $S^9$ shrinks to zero size and the spatial projection of the null line
is a circle on the $S^9$~\cite{Berenstein:2002sa}. One can picture this as a line winding in time on
the Einstein cylinder, see~\cite{Berenstein:2002sa}.
For a thorough discussion of the causal structure of more general pp-wave geometries, which are
not conformally flat and hence the above trick of identifying the boundary by a conformal
mapping does not work, see~\cite{Marolf:2002ye,Hubeny:2002zr}. For a large class of pp-waves satisfying certain
conditions, the boundary is again one-dimensional.
The conformal boundaries and geodesics of
$AdS_5\times S^5$ and the plane wave and how the former approach the latter in the Penrose limit
have been analyzed in~\cite{Dorn:2003ct}.

So the boundary of the plane wave
is a null line, whereas SYM lives on $\mathbb{R}\times S^3$ before the limit is taken.
Here one should recall again that the geodesic is rotating on the $S^5$, so when projected
on the boundary it is time-like and can be identified with $t$. As the $S^3$ has disappeared
in the process this supports the expectation~\cite{Berenstein:2002sa} (see also~\cite{Kiritsis:2002kz}) that
the holographic dual of string theory on the
plane wave is a quantum mechanical matrix model obtained by a truncation of SYM on the $S^3$.
It would be nice to gain a precise understanding in which sense such a truncation can be
consistently performed, see also~\cite{Kim:2003rz}. An alternative approach, the construction of
a holographic screen consisting of a four-dimensional hypersurface in the plane wave,
was followed in~\cite{Siopsis:2002vw,Siopsis:2002nx}. It would be interesting to understand if
this has some connection to~\cite{Metsaev:2002sg}, where supersymmetric D3-branes and ${\mc N}=4$
SYM on a four-dimensional plane wave, arising from a Penrose limit of $\mathbb{R}\times S^3$,
was studied. For further remarks on holography in the plane wave see~\cite{Yoneya:2003mu}. 
One would also like
to go beyond the comparison of masses vs. anomalous dimensions in both theories. Some ideas
in this respect have been formulated in~\cite{Berenstein:2002sa} (see however, also~\cite{Spradlin:2003bw,Mann:2003qp}), 
a consistent truncation
of SYM in the BMN limit would suggest to compare finite time transition amplitudes in this
model to string amplitudes on the plane wave.

\section{Extensions of the BMN duality}\label{chapter3}

\subsection{Various approaches}

It is an interesting
question whether the BMN proposal is applicable to other less
trivial backgrounds. Can the string spectrum in less supersymmetric situations again be deduced
from a subsector of a dual gauge theory with reduced, possibly even no supersymmetry? This question was addressed in several
publications~\cite{Itzhaki:2002kh,Gomis:2002km,PandoZayas:2002rx,Alishahiha:2002ev,Kim:2002fp}
appearing shortly after~\cite{Berenstein:2002jq}. Recall that orbifolds of type IIB string theory on $AdS_5\times S^5$~\cite{Kachru:1998ys}
provide a simple way to reduce the amount of supersymmetry in the
AdS/CFT correspondence. For example, the world-volume theory of $kN$ D3-branes
located at the $\mathbb{Z}_k$ orbifold singularity of an ALE space is a ${\mc N}=2$ $[U(N)]^k$ quiver gauge theory~\cite{Douglas:1996sw} which is
dual to string theory on $AdS_5 \times (S^5/{\mathbb Z}_k)$~\cite{Kachru:1998ys}.
${\mc N}=1$ field theories can arise from D3-branes on orbifold singularities of the form $\mathbb{C}^3/\G$, with $\G$ a discrete
proper subgroup
of $SU(3)$. These are dual to strings on $AdS_5 \times (S^5/\G)$~\cite{Kachru:1998ys}. One can also consider
$N$ D3-branes located at a conifold singularity of a Calabi-Yau three-fold. In this case the world-volume theory is a
${\mc N}=1$ $SU(N)\times SU(N)$ field theory coupled to four bi-fundamental chiral multiplets with a IR fixed point and
an exactly marginal superpotential~\cite{Klebanov:1998hh}. This theory is dual to string theory on $AdS_5\times T^{1,1}$,
$T^{1,1}$ being the base of the conifold.

What happens if we apply the Penrose-G\"{u}ven limit to these
situations?\footnote{In
general there exist distinct classes of geodesics which
give rise to different space-times in the limit. The statements I
make usually refer to the generic case if not stated otherwise.}
Let me sketch the case of $AdS_5\times T^{1,1}$ which was studied
in~\cite{Itzhaki:2002kh,Gomis:2002km,PandoZayas:2002rx}. Topologically $T^{1,1}$ is a $U(1)$ bundle
over $S^2\times S^2$ and its $SU(2)\times SU(2)\times U(1)$
isometry is identified with a $SU(2)\times SU(2)$ global symmetry
and $U(1)_R$ symmetry of the dual superconformal field
theory~\cite{Klebanov:1998hh}. The surprising result found
in~\cite{Itzhaki:2002kh,Gomis:2002km,PandoZayas:2002rx} is that blowing up the neighborhood of a
null geodesic rotating around the $U(1)$ fiber one ends up with
the maximally supersymmetric plane wave background again.
Consequently a subsector of the gauge theory with enhancement from ${\mc N}=1$ to
${\mc N}=4$ supersymmetry should emerge in the BMN limit. Indeed,
one finds that the string Hamiltonian in this case is related to
that of the plane wave by a twisting~\cite{Itzhaki:2002kh,Gomis:2002km,PandoZayas:2002rx}
\begin{equation}
H_{ T^{1,1}} = H_{S^5}+J_1+J_2\,,
\end{equation}
where $J_1$ and $J_2$ are rotation generators of a $\mathbb{R}^2\times \mathbb{R}^2$ subspace of the plane wave transverse geometry.
From the gauge theory perspective $H_{ T^{1,1}}$ is identified with $\D-\frac{3}{2}R$, where $R$ is the generator of the $U(1)_R$
symmetry and $J_a=Q_a-\frac{1}{2}R$, where $Q_a$ are the Cartan generators of the  $SU(2)\times SU(2)$ global symmetry.
All these combinations remain fixed in the limit, similarly to $\D-J$ in the ${\mc N}=4$ case. In particular the sector in the ${\mc N}=1$
theory with supersymmetry enhancement is specified by~\cite{Itzhaki:2002kh,Gomis:2002km,PandoZayas:2002rx}
\begin{equation}
H_{S^5} = \D-\frac{1}{2}R-Q_1-Q_2\,.
\end{equation}
One can explicitly identify these operators in the gauge theory. The matter content
consists of chiral multiplets $A_i$ and $B_i$ with $R$-charge 1/2 and conformal dimension $3/4$ transforming as $({\bf 2},{\bf 1})$ and
$({\bf 1},{\bf 2})$ under the global symmetry. Then the unique operator corresponding to the string ground state
is $\tr (A_1B_1)^R$, analogous to $\tr Z^J$ in ${\mc N}=4$.
Oscillators in the $\mathbb{R}^2\times \mathbb{R}^2$ direction are roughly speaking identified with the action of the raising operators
of $SU(2)\times SU(2)$ on the ground state and a possible addition of phases. For more details, see~\cite{Itzhaki:2002kh,Gomis:2002km,PandoZayas:2002rx}. Another
example where ${\mc N}=1$ is enhanced to ${\mc N}=4$ arises from the Penrose-G\"{u}ven limit of the dual pair obtained from
$N$ D3 branes on a $\mathbb{C}^3/\mathbb{Z}_3$ orbifold singularity~\cite{Gomis:2002km}. Further discussion of supersymmetry enhancement
in ${\mc N}=1$ theories arising from various orbifolds of $S^5$ and $T^{1,1}$ can be found in~\cite{Oh:2002sv}.

However, supersymmetry enhancement is not a generic feature, as can be seen from the examples involving ${\mc N}=2$ $[U(N)]^k$ quiver
gauge theory~\cite{Alishahiha:2002ev,Kim:2002fp,Takayanagi:2002hv} (the case $k=2$ has also been discussed
in~\cite{Itzhaki:2002kh}).
The reason for this is that in the generic case the Penrose-G\"{u}ven limit of $AdS_5 \times (S^5/{\mathbb Z}_k)$ yields
the $\mathbb{Z}_k$ orbifold of the plane wave background and hence breaks half of the supersymmetry.
This example will be discussed in more detail in the next subsection.
Penrose-G\"{u}ven limits of various orbifolds and orientifolds of $AdS\times S$ spaces have also been
considered in~\cite{Floratos:2002uh}.
I have said above that generically supersymmetry is not enhanced in the Penrose limit of $AdS_5 \times (S^5/{\mathbb Z}_k)$.
A precise statement is the following: if the null geodesic is fixed by the group action, the resulting space-time will be an
orbifold of the plane wave; if this is not the case one recovers the pure plane wave
again~\cite{Alishahiha:2002ev}. Following the logic
above this means that strings on plane waves can also arise in a sector of ${\mc N}=2$ theory
with enhancement to ${\mc N}=4$.
This observation leads to a further interesting development. Suppose we have $N_1$ D3-branes placed on a $\mathbb{C}^2/\mathbb{Z}_{N_2}$
singularity. Blowing up the region around a null geodesic not fixed by the group action
one can also take $N_1$, $N_2\to\infty$ and keep the $R$ charge
finite~\cite{Mukhi:2002ck,Alishahiha:2002jj}. How does this affect the resulting geometry?
Again, introduce light-cone coordinates
\begin{equation}\label{period}
x^+ = \frac{1}{2\m}(t+\psi)\,,\qquad x^-=-\m R^2(t-\psi)\,,\qquad R^4\equiv 4\pi g_{\text{s}}\a'^2 N_1N_2\,,
\end{equation}
however, this time $\psi\sim\psi+\frac{2\pi}{N_2}$ since the geodesic is not fixed by $\mathbb{Z}_{N_2}$.
Taking $N_1\sim N_2\to\infty$
yields the standard plane wave geometry with the difference that due to~\eqref{period} the light-like
coordinate $x^-$ becomes {\em compact} with period
\begin{equation}
x^-\sim x^-+2\pi R^-\,,\qquad R^-\equiv \m\a'\sqrt{4\pi g_{\text{s}}\frac{N_1}{N_2}}\,.
\end{equation}
Consequently the light-cone momentum $p^+$ is quantized in units of $1/R^-$ and we have a description of discrete light-cone quantization
of strings on the plane wave in terms of a quiver gauge theory~\cite{Mukhi:2002ck,Alishahiha:2002jj}. An interesting new feature is for
example the appearance of momentum and winding states along the compact direction. These are also realized in the gauge
theory~\cite{Mukhi:2002ck,Alishahiha:2002jj}: the dual gauge theory is a $[U(N_1)]^{N_2}$ quiver gauge theory, in particular it contains
$N_2$ bi-fundamental hypermultiplets~\cite{Douglas:1996sw} or, in ${\mc N}=1$ language, $2N_2$ chiral multiplets in the bi-fundamental.
Denote their scalar components by $(A_I,B_I)$. The operator
$\tr(A_1\cdots A_{N_2})$ has precisely the correct quantum numbers to describe a state with one unit of light-cone momentum and zero
winding. This looks like a `string' winding once around the quiver diagram (which is a circle). Similarly an operator with $k$ units
of momentum winds $k$ times around the quiver. Winding states are shown to be dual to operators with insertions of adjoint scalars from the
vector multiplet together with a phase. The picture that emerges is quite suggestive: strings carrying momentum are described by
operators winding around a large quiver circle, whereas strings with non-zero winding are dual to operators which carry `momentum'
(the phase). Indeed it was argued in~\cite{Mukhi:2002ck,Alishahiha:2002jj}, using T-duality, that the `strings' winding the
quiver circle {\em are} so called non-relativistic winding strings in the T-dual description. I refer the reader
to~\cite{Mukhi:2002ck,Alishahiha:2002jj} for more details. One can also study compactifications of string theory on the plane wave along
space-like circles~\cite{Michelson:2002wa}.
The plane wave with a manifest space-like isometry is related to the standard one by a coordinate transformation, resulting in a shift
of the Hamiltonian by a rotation generator. For a classification of the preserved supersymmetry under toroidal
compactifications see~\cite{Michelson:2002wa}. Plane waves with space-like isometries can also arise from non-standard Penrose limits of
$AdS_5\times S^5$ and $AdS_5\times S^5/\mathbb{Z}_k$ and are dual to triple scaling limits of ${\mc N}=4$ or ${\mc N}=2$ gauge
theories~\cite{Bertolini:2002nr}. The identification of momentum and winding states along the space-like circle with operators in the dual
gauge theory is similar in spirit to~\cite{Mukhi:2002ck,Alishahiha:2002jj}, see~\cite{Bertolini:2002nr} for the details.

A further interesting direction is the generalization of the BMN correspondence to non-conformal
backgrounds~\cite{PandoZayas:2002rx}. In particular
one can consider examples known to be dual to RG flows from ${\mc N}=4$ in the UV to ${\mc N}=1$ IR fixed points and
take the Penrose-G\"{u}ven limit `along the flow'~\cite{Gimon:2002sf,Brecher:2002ar}.
Non-conformal backgrounds do, however, not lead
to solvable string theories, rather they share the generic feature that the Penrose limit leads to time-dependent mass terms for
the world-sheet theory in light-cone gauge~\cite{PandoZayas:2002rx}.
Despite of this fact it has been argued in~\cite{Gimon:2002sf} that some features of the RG flow,
such as the branching of a given operator in the UV into operators of the IR, can be captured by
studying the corresponding problem of a point particle propagating in this time dependent background.
This system is exactly solvable~\cite{Gimon:2002sf}.
One may also focus on the geometry in the IR~\cite{Brecher:2002ar,Casero:2002ks,Gimon:2002nr} and
the resulting background will be one of a deformed Hpp-wave containing additional
constant three-form fluxes. This leads
again to a solvable string theory, see also~\cite{Russo:2002rq}. By choosing a
non-standard geodesic, one can use the resulting string theory to study heavy hadrons with
mass proportional to a large global charge in the confining dual IR gauge theory~\cite{Gimon:2002nr}.
An interesting {\em solvable} example of a time-dependent plane wave background
supported by a non-constant dilaton was considered in~\cite{Papadopoulos:2002bg}.
Finally, a non-supersymmetrix example has been discussed in~\cite{Bigazzi:2002gw}. In many models 
the light-cone zero-point energy turns out to be negative; this raises the question about their stability 
which was addressed (and, at least classically, answered affirmatively) in~\cite{Bigazzi:2003jk}.

\subsection{Strings on orbifolded plane waves from quiver gauge theory}\label{pporbifold}

In the previous subsection I tried to give a flavor of the possible extensions of the
BMN duality. In this subsection the case of the plane wave
orbifold~\cite{Alishahiha:2002ev,Kim:2002fp,Takayanagi:2002hv} will be discussed
in more detail.
Specifically, I will consider a ${\mathbb Z}_k$ orbifold of one of the two ${{\mathbb R}}^4$
subspaces transverse to the propagation null vector
and show that first-quantized free string theory is described correctly by the large $N$,
fixed gauge coupling limit of ${\mc N}=2$ $[U(N)]^k$ quiver gauge theory.
Apart from being an interesting example with less supersymmetry, a further motivation comes from the fact that, as
shown in~\cite{Berenstein:2002jq,Metsaev:2001bj,Metsaev:2002re}, the plane wave background acts as a 
harmonic oscillator potential to the string, and hence the dynamical distinction
between untwisted and twisted states is less clear.
It is thus of intrinsic interest to see if one can find a precise map between type IIB
string oscillation modes and quiver gauge theory operators, both for untwisted and twisted sectors.
Indeed, we will see that operators dual to untwisted and twisted sector states are quite similar.

\subsubsection{IIB superstring on plane wave orbifold}\label{orb2}

As explained in the previous section, the dynamics of superstrings on the maximally supersymmetric
plane wave geometry supported by homogeneous \RR 5-form flux and constant dilaton
\begin{equation}\label{pp1}
\begin{split}
ds^2 & = 2dx^+dx^--\m^2(\vec{x}^2+\vec{y}^2)(dx^+)^2+d\vec{x}^2+d\vec{y}^2\,,\\
F_{+1234} & =F_{+5678} = 4\m\,,
\end{split}
\end{equation}
$(\vec{x},\vec{y})\in{{\mathbb R}}^4 \times {{\mathbb R}}^4$, is governed by an exactly solvable
light-cone world-sheet theory of {\em free}, albeit massive fields~\cite{Metsaev:2001bj}.
The isometry group of the eight-dimensional space transverse to the null propagation
direction is $SO(4)_1\times SO(4)_2$: while the space-time geometry is invariant under $SO(8)$, the 5-form field
strength breaks it to $SO(4)_1\times SO(4)_2$. In the Green-Schwarz action on the plane wave background, the
reduction of the isometry is due to the coupling of spinor fields to the background \RR 5-form field strength.

One is interested in reducing the number of supersymmetries preserved by the background. As alluded to above, one can
break one half of the 32 supersymmetries by taking a ${\mathbb Z}_k$ orbifold of the ${\mathbb R}^4$ subspace
parameterized by $\vec{y}$. The orbifold action is defined by
\begin{equation}
g : \quad (z^1,z^2) \longrightarrow \o(z^1,z^2)\,,\qquad\o=e^{\frac{2\pi i}{k}}\,,
\end{equation}
where
\begin{equation}
z^1\equiv\frac{1}{\sqrt{2}}(y^6+iy^7)\,,\qquad z^2\equiv\frac{1}{\sqrt{2}}(y^8-iy^9)\,,
\end{equation}
and $g$ acts on space-time fields as
\begin{equation}
g=\exp \left( \frac{2\pi i}{k}(J_{67}-J_{89})\right)\,.
\end{equation}
$J_{67}$ and $J_{89}$ are the rotation generators in  the 6-7 and 8-9 planes, respectively. Defined so, the orbifold of the plane wave
background is actually derivable from the Penrose
limit of $AdS_5\times S^5/{{\mathbb Z}_k}$ taken along the great circle of the $S^5$ that is fixed by the ${\mathbb Z_k}$ action.

In the light-cone gauge, the superstring on the background~\eqref{pp1} is described by eight world-sheet
scalars $x^I$ and eight world-sheet fermions $S^a$, all of which are free but massive.
The masses of scalars and fermions are equal by world-sheet supersymmetry (which descends from the light-cone gauge fixing of the
Green-Schwarz action, cf.\ the remark above equation~\eqref{lcaction}) and equal the \RR 5-form field strength $\m$.
$S$ is a positive chirality Majorana-Weyl spinor of $SO(9,1)$, obeying the light-cone gauge condition $\G^+S=0$ and hence
transforming as a positive chirality spinor of $SO(8)$ under rotations in the transverse directions.
Decompose the world-sheet fields into representations of $SO(4)_1\times SO(4)_2$
\begin{equation}
x^I=(\vec{x},\vec{y})\to(\vec{x},z^1, z^2)\,,\qquad
S^a\to (\chi^{\a},\xi^{\dot{\a}})\,,
\end{equation}
where $\a$ and $\dot{\a}$ are spinor indices of $SO(4)_2$, ranging over 1, 2 and I have
suppressed the spinor indices of $SO(4)_1$ under which $\chi^{\a}$
and $\xi^{\dot{\alpha}}$ carry positive and negative chirality, respectively. Then the fields $\vec{x}$ and $\chi^{\a}$
transform trivially under $g$ whereas
\begin{equation}
g:\quad z^m \longrightarrow \o z^m\,,\qquad
\xi^{\dot{\a}} \longrightarrow {\Omega^{\dot{\a}}}_{\dot{\b}}\xi^{\dot{\b}}\,,
\end{equation}
and $\Omega=\text{diag}(\o,\o^{-1})$, that is $\xi^{\dot{1}}$ and $\xi^{\dot{2}}$ transform
oppositely under the ${\mathbb Z}_k$ action.
It is convenient to combine $\xi^{\dot{1}}$, $\bar{\xi}^{\,\dot{2}}$ into a Dirac spinor $\xi$, and
$\bar{\xi}^{\,\dot{1}}$ and $\xi^{\dot{2}}$ into its conjugate $\bar{\xi}$ and analogously
for $\chi$ and $\bar{\chi}$. As the world-sheet theory is free, it is straightforward to quantize the
string in each twisted sector, the only difference among various sectors being the monodromy of the world-sheet fields
sensitive to the orbifolding, that is $z^m$ and $\xi$. The other world-sheet fields remain periodic.
The monodromy conditions in the $q$-th twisted sector, $q=0,\ldots,k-1$, are
\begin{equation}
z^m(\s+2\pi\a'p^+,\t)=\o^qz^m(\s,\t)\,,\qquad \xi (\s+2\pi\a' p^+,\t)=\o^q\xi (\s,\t)\,,
\end{equation}
and the corresponding oscillator modes depend on
$n(q)=n+\frac{q}{k}$  $\left(n\in{\mathbb Z}\right)$.

Physical states are obtained by applying the bosonic and fermionic creation operators to the light-cone vacuum $|v,p^{+}\ra_q$
of each twisted sector. They should satisfy additional constraints ensuring the level-matching condition:
\begin{equation}\label{levelmatching}
\sum_{n\in{\mathbb Z}}nN_n=0\,,\quad
\sum_{n\in{\mathbb Z}}n(q)\left(N_{n(q)}-\bar{N}_{-n(q)}\right)=0\,,
\end{equation}
and ${\mathbb Z}_k$ invariance. The bosonic creation operators are
\begin{equation}
\vec{a}_n^{\dg}\,,\qquad\text{ and }\qquad\a^{\dg\,m}_{n(q)}\,,\quad \bar{\a}^{\dg\,m}_{n(-q)}\,,\qquad
\left(n\in{\mathbb Z}\right)\,.
\end{equation}
Here, $\vec{a}_n$ are the $\vec{x}$ oscillators, whereas $\a^m_{n(q)}$ and $\bar{\a}^m_{n(-q)}$ are $z^m$ and
$\bar{z}^m$ oscillators, respectively. The fermionic creation operators consist, in obvious notation, of
\begin{equation}
\chi^{\dg}_n\,,\quad \bar{\chi}^{\dg}_n\qquad\text{ and }\qquad
\xi^{\dg}_{n(q)}\,,\quad \bar{\xi}^{\dg}_{n(-q)}\,.
\end{equation}
Acting with the fermionic zero-mode oscillators on the light-cone vacua and projecting onto ${\mathbb Z}_k$ invariant states, one fills out
${\mc N}=2$ gravity and tensor supermultiplets of the plane wave background. The action of the bosonic zero-mode
oscillators on these gives rise to a whole tower of multiplets~\cite{Metsaev:2002re}, much as in the $AdS_5 \times S^5$ case.
As an example, we have four invariant states with a single bosonic oscillator
\begin{equation}
\vec{a}_0^{\,\dg}|v,p^+\ra_q\,,
\end{equation}
and states with two bosonic oscillators are
\begin{equation}\label{osc2}
a_n^{\dg\,\m}a_{-n}^{\dg\,\n}|v,p^{+}\ra_q\,,\qquad
\a^{\dg\,l}_{n(q)}\bar{\a}^{\dg\,m}_{-n(q)}|v,p^{+}\ra_q\,.
\end{equation}
In the ${\mathbb Z}_2$ case there are additional invariant states built from two $z^m$ or two $\bar{z}^m$ oscillators.
However, they do not satisfy the level matching condition~\eqref{levelmatching}.
The light-cone Hamiltonian in the $q$-th twisted sector is
\begin{equation}\label{lchamiltonian}
H_q=\sum_{n\in{\mathbb Z}}N_n\sqrt{\m^2+\frac{n^2}{(\a'p^+)^2}}+\sum_{n\in{\mathbb Z}}\left(N_{n(q)}+\bar{N}_{-n(q)}\right)
\sqrt{\m^2+\frac{n(q)^2}{(\a'p^+)^2}}\,.
\end{equation}
The first sum is over those oscillators which are not sensitive to the orbifold and $N_n$ ($N_{n(q)}$ and $\bar{N}_{-n(q)}$) is the
total occupation number of bosons and fermions. The ground state energy is cancelled between bosons and fermions. This corresponds to
a choice of fermionic zero-mode vacuum that explicitly breaks the $SO(8)$ symmetry, which is respected by the metric but
not the field strength background, to $SO(4)_1\times SO(4)_2$~\cite{Metsaev:2002re}.

\subsubsection{Operator analysis in ${\mc N}=2$ quiver gauge theory}\label{orb3}

It is known~\cite{Kachru:1998ys} that type IIB string theory on $AdS_5 \times (S^5/{\mathbb Z}_k)$ is dual to
${\mc N}=2$ $[U(N)]^k$ quiver gauge theory, the world-volume theory of $kN$ D3-branes placed at the orbifold singularity.
In light of the discussion in the previous section, one can anticipate that string theory on the plane wave orbifold is
dual to a new perturbative expansion of the quiver gauge theory at large $N$ and {\em fixed} gauge coupling $g^2_{\text{YM}}=~4\pi g_s k$.
The factor of $k$ in the relation between the string and the gauge coupling
is standard and can be deduced by moving the D3-branes off the tip of the orbifold into the
Higgs branch, see also~\cite{Lawrence:1998ja}.
In the new expansion, one focuses primarily on states with conformal weight $\D$ and $U(1)_R$ charge $J$ which scale as $\D$,
$J\sim\sqrt{N}$, whose difference $(\D-J)$ remains finite in the large $N$ limit. $U(1)_R$ is the subgroup of the original
$SU(4)_R$ symmetry of ${\mc N}=4$ super Yang-Mills theory, which on the gravity side corresponds to the $S^1$ fixed under the orbifolding;
this $U(1)_R$ together with the $SU(2)_1$ subgroup of the remaining $SO(4)\simeq SU(2)_1\times SU(2)_2$ that commutes
with ${\mathbb Z}_k\subset SU(2)_2$ forms the $R$-symmetry group of ${\mc N}=2$ supersymmetric gauge theory.

The reason for the above scaling behavior is that $(\D-J)$ is identified with the light-cone Hamiltonian
on the string theory side, whereas\footnote{Since $\int_{S^5/{\mathbb Z}_k} F_5=N$, the radius of $AdS_5$ is proportional to $(kN)^{1/4}$.}
$\frac{J}{\sqrt{kN}}\sim p^+$, $p^+$ being the longitudinal momentum carried by the string.
When $(\D-J) \ll J$, the light-cone Hamiltonian in~\eqref{lchamiltonian} implies that on the gauge
theory side there are operators
obeying the following relation between the dimension $\D$ and the $U(1)_R$ charge $J$
\begin{equation}\label{lchamiltonian2}
(\D-J)_n=\sqrt{1+\l'n^2}
\qquad\text{ and }\qquad(\D-J)_{n(q)}=\sqrt{1+\l'\left(n(q) \right)^2 }\,.
\end{equation}
In the gauge theory, before orbifolding we have $N\times N$ matrix valued fields, that is the gauge field and three complex scalars
\begin{equation}
A_{\m}\,,\quad Z=\frac{1}{\sqrt{2}}(\phi^4+i\phi^5)\,,\quad \varphi^m=(\varphi^1,\varphi^2)\equiv\frac{1}{\sqrt{2}}
(\phi^6+i\phi^7,\phi^8-i\phi^9)\,,
\end{equation}
and in addition their superpartners,
fermions $\chi$ and $\xi$. The fields $\chi$ and $\xi$ are spinors of $SO(5,1)$, transforming as ${\bf 4}$ and ${\bf 4}'$, respectively.
To define the ${\mathbb Z}_k$ orbifolding in the gauge theory, we promote these fields to $kN\times kN$ matrices ${\mc A}_{\m}$, ${\mc Z}$,
$\Phi^m$, ${\mc X}$ and $\Xi$ and project onto the ${\mathbb Z}_k$ invariant components. The projection is ensured by the conditions
\begin{equation}
S{\mc A}_{\m}S^{-1}={\mc A}_{\m}\,,\qquad S{\mc Z}S^{-1}={\mc Z}\,,\qquad
S{\mc X}S^{-1}={\mc X}
\end{equation}
and
\begin{equation}
S\Phi^mS^{-1}=\o\Phi^m\,,\qquad S\Xi S^{-1}=\o\Xi \,.
\end{equation}
where $S=\text{diag}(1,\o^{-1},\o^{-2},\ldots,\o^{-k+1})$, each block being proportional to the $N\times N$ unit matrix.

The resulting spectrum is that of a four-dimensional ${\mc N}=2$ quiver gauge
theory~\cite{Douglas:1996sw} with $[U(N)]^k$ gauge group,
containing hypermultiplets in the bi-fundamental representations of $U(N)_i\times U(N)_{i+1}$, $i\in{\mathbb Z}$ mod$(k)$. More precisely,
${\mc A}_{\m}$, ${\mc Z}$ and ${\mc X}$ fill out $k$ ${\mc N}=2$ vector multiplets with the fermions transforming as
doublets under $SU(2)_R$ (as its Cartan generator is proportional to $(J_{67}+J_{89})$). The ${\mc Z}$ field has the block-diagonal form
\begin{equation}
{\mc Z}=
\begin{pmatrix}
Z_1 &     &       &       &      & \\
    & Z_2 &       &       &      & \\
    &     &  Z_3  &       &      & \\
    &     &       &\quad\cdot&      & \\
    &     &       &       &\quad\cdot& \\
    &     &       &       &       & Z_k
\end{pmatrix}
\end{equation}
with zeros on the off-diagonal and the diagonal blocks being $N\times N$ matrices of $U(N)_i$'s.
The ${\mc A}_{\m}$ and ${\mc X}$ fields take an analogous form. Likewise, the $\Phi^m$ and $\Xi$ fields fill out $k$ hypermultiplets,
in which the scalars are doublets under $SU(2)_R$, whereas the fermions are neutral.
The $\Phi^m$ fields take the form
\begin{equation}
\Phi^m=
\begin{pmatrix}
0           & \varphi^m_{12} &             &       &       &       & \\
            & 0           & \varphi^m_{23} &       &       &       & \\
            &             & 0           & \quad\cdot &       &       & \\
            &             &             & \quad\cdot & \quad \cdot &     & \\
            &             &             &       &\quad\cdot & \quad\cdot & \\
\varphi^m_{k1} &             &             &       &       &\quad\cdot&\quad\cdot
\end{pmatrix}
\end{equation}
and analogously for $\Xi$.

The light-cone vacua of string theory on the plane wave orbifold ought to be described by $H_q=0$ and in the quiver gauge
theory this translates to operators with $\D-J=0$.
One can build $k$ mutually orthogonal, ${\mathbb Z}_k$ invariant
single-trace operators $\Tr[S^q{\mc Z}^J]$ and associate these operators to the vacuum in the $q$-th twisted sector
\begin{equation}
\frac{1}{\sqrt{kJN^J}}\Tr[S^q{\mc Z}^J]\lr |v,p^+\ra_q\,,\qquad (q=0,\ldots,k-1)\,.
\end{equation}
In what sense is this identification unique? After all, in the quiver gauge theory it appears that the operators
$\Tr[S^q{\mc Z}^J]$ for any $q$ stand on equal footing. However, the orbifold action renders
an additional `quantum' ${\mathbb Z}_k$ symmetry (see for example~\cite{Adams:2001sv}) that acts on fields in the quiver gauge
theory.\footnote{This  ${\mathbb Z}_k$ should not to be confused with the space-time ${\mathbb Z}_k$ used for constructing the orbifold.
By construction, under the orbifold action all the fields are invariant.} Specifically, one can take an element $g$ in this
quantum ${\mathbb Z}_k$ to act
on an arbitrary field $T_{ij}$, $i,j\in{\mathbb Z}$ mod$(k)$, as $g: T_{ij} \lra T_{i+1,j+1}$. In particular, one notes that
$g: \Tr[S^q{\mc Z}^J] \lra \o^{q}\Tr[S^q{\mc Z}^J]$. So one can indeed distinguish classes of operators on the quiver gauge
theory side by their eigenvalues under the quantum ${\mathbb Z}_k$ symmetry.

Next, consider the eight twist invariant operators with $\D-J=1$. They are
\begin{align}
\frac{1}{k\sqrt{N^{J+1}}}\Tr[S^q{\mc Z}^J{\mc D}_{\m}{\mc Z}]
& \lr a_0^{\dg\,\m}|v,p^+\ra_q\,,\\
\frac{1}{k\sqrt{N^{J+1}}}\Tr[S^q{\mc Z}^J{\mc X}_{J=1/2}]
& \lr\chi^{\dg}_0|v,p^+\ra_q\,,\\
\frac{1}{k\sqrt{N^{J+1}}}\Tr[S^q{\mc Z}^J\bar{{\mc X}}_{J=1/2}]
& \lr\bar{\chi}^{\dg}_0|v,p^+\ra_q\,.
\end{align}
These are identified with IIB supergravity modes built out of a single zero-mode oscillator acting
on the $q$-th vacuum. Here, ${\mc D}_{\m}{\mc Z}=\p_{\m}{\mc Z}+[{\mc A}_{\m},{\mc Z}]$ .
Operators corresponding to higher string states on the plane wave orbifold arise as follows. Oscillators of non-zero level $n$
corresponding to the fields not sensitive to the orbifold are identified with insertions of
the operators
${\mc D}_{\m}{\mc Z}$, ${\mc X}_{J=1/2}$ and $\bar{{\mc X}}_{J=1/2}$ with a position dependent phase factor
in the trace $\Tr[S^q{\mc Z}^J]$. For instance, for $\D-J=2$, $\m\neq\n$,
\begin{equation}
\frac{1}{\sqrt{kJN^{J+2}}}
\sum_{l=0}^Je^{\frac{2\pi i ln}{J}}\Tr[S^q{\mc Z}^l{\mc D}_{\m}{\mc Z}{\mc Z}^{J-l}{\mc D}_{\n}{\mc Z}]
\lr a_n^{\m\,\dg}a_{-n}^{\n\,\dg}|v,p^{+}\ra_q\,.
\end{equation}
This is exactly the same as in the unorbifolded case -- the insertion of the position-dependent phase factor ensures
that the level-matching condition is satisfied and that the light-cone energy of the string states is reproduced
correctly~\cite{Berenstein:2002jq}.

The remaining string states involving oscillators with a fractional moding $n(q)$ in the twisted sectors, should be identified
with insertions of the operators $\Phi^m$ and $\Xi_{J=1/2}$ together with position-dependent phase factors of the form
$e^{2\pi i ln(q)/J}$. Similarly, insertions of $\bar{\Phi}^m$ and $\bar{\Xi}_{J=1/2}$ are accompanied with
the phase factor $e^{2\pi i ln(-q)/J}$. Again, the prescription implements the level-matching condition
and yields the correct energy of the
corresponding string states. For $r\neq s$
\begin{equation}\label{op}
\frac{1}{\sqrt{kJN^{J+2}}}\sum_{l=0}^Je^{\frac{2\pi i ln(q)}{J}}\Tr[S^q{\mc Z}^l\Phi^r{\mc Z}^{J-l}\bar{\Phi}^s]
\lr\a^{r\,\dg}_{n(q)}\bar{\a}^{s\,\dg}_{-n(q)}|v,p^{+}\ra_q\,.
\end{equation}
For the ${\mathbb Z}_2$ orbifold, the operator
corresponding to $\a^{r\,\dg}_{n(q)}\a^{s\,\dg}_{m(q)}|v,p^{+}\ra_1$, though being ${\mathbb Z}_2$ invariant,
vanishes for all $m$, $n$ due to the cyclicity of the trace, as it should, cf.\ the remark below equation~\eqref{osc2}.

Finally, operators with insertions such as ${\mc D}^2{\mc Z}$, $\bar{\mc Z}$
or ${\mc X}_{J=-1/2}$ are expected to be hidden by operator mixing, much in the same
way as discussed in the previous section~\ref{sec23}.
One can compute the leading order anomalous dimensions of the $\D-J=2$ operators
in equation~\eqref{op}, perturbatively in ${\cal N}= 2$ quiver gauge
theory and confirm that the proposal for the twisted sector operators
reproduces the correct light-cone string energy spectrum.
In fact, in the setup I have outlined above one can proceed
with the computations essentially parallel to those of~\cite{Berenstein:2002jq}, see for
example~\cite{Kim:2002fp} for more details.

\subsection{Further directions}

So far I mainly considered closed strings in IIB string theory on the
plane wave background, their duality to ${\mc N}=4$ SYM in the BMN limit
and generalizations thereof. In this subsection I would like to discuss
two further interesting issues: D-branes on the plane wave and
string theory on more general pp-wave backgrounds.

\subsubsection{D-branes on the plane wave}\label{dbranes}

Since D-branes capture non-perturbative effects in string theory, their understanding in the plane wave background
is important. They can be studied by various means: in perturbative string theory they are defined as hypersurfaces
on which open strings end and hence can be analyzed by finding consistent boundary conditions for open strings; alternatively they
can be described using boundary states, that is coherent states in closed string theory. The boundary state imposes certain
gluing conditions on the closed string fields that arise through the presence of the D-brane. Interactions between two static D-branes
through the exchange of closed strings at tree level can then be computed by sandwiching the closed string propagator between two
boundary states. The same process can be re-interpreted as an open string one-loop diagram,
i.e.\ the open string partition function. This is open-closed duality, which has to be satisfied for a D-brane to be consistent.
Yet another way to describe D-branes is by considering their world-volume theory, consisting of a Dirac-Born-Infeld and a Wess-Zumino
term. Solutions to the resulting field equations describe the embedding of the D-brane into the target space. Finally, at low energies
D-branes arise as solitonic solutions to the supergravity equations of motion.

All of these different approaches have been used to obtain a rather detailed picture of supersymmetric D-branes in the plane wave
background via open strings in light-cone
gauge~\cite{Dabholkar:2002zc,Skenderis:2002wx,Skenderis:2002ps}, covariant open
strings~\cite{Bain:2002tq},
boundary states~\cite{Billo:2002ff,Bergman:2002hv,Gaberdiel:2002hh,Skenderis:2002ps}
and the open-closed consistency conditions~\cite{Bergman:2002hv,Gaberdiel:2002hh},
D-brane embeddings~\cite{Skenderis:2002vf} and supergravity solutions~\cite{Bain:2002nq}
(for a supergravity analysis of branes in the pp-wave space-time originating from the Penrose limit of $AdS_3\times S^3$, see 
e.g.~\cite{Kumar:2002ps,Alishahiha:2002rw,Biswas:2002yz}).
I will summarize these results below,
overviews over
many aspects on D-branes on the plane wave can be found in~\cite{Gaberdiel:2002vz,Skenderis:2003wx}. For a
discussion of open strings in the plane wave with a constant B-field turned on,
see~\cite{Chu:2002in}.

Let me start with the open string analysis. The covariant action for strings in the plane
wave~\cite{Metsaev:2001bj} is invariant under local $\kappa$-symmetry. For open strings additional
boundary terms arise under $\kappa$-variations and for supersymmetry preserving configurations
these have to be cancelled by imposing suitable boundary conditions. In~\cite{Bain:2002tq} this
analysis was performed for longitudinal D$p$-branes $(+,-,m,n)$,
i.e.\ branes whose world-volume is along $x^+$, $x^-$
and $m$ and $n$ denote the number of coordinates along the two transverse $\mathbb{R}^4$'s.
Branes with $p=3$, $5$, $7$ and $|m-n|=2$ are
half-supersymmetric\footnote{This means that half of the kinematical as well as half of the dynamical
supercharges are preserved. Kinematical (non-linearly realized) supercharges square to $P^+$, whereas
dynamical (linearly realized) supercharges square to the Hamiltonian plus additional
generators.} if they are located `at the origin',
whereas `outside the origin' only one quarter of the supercharges, namely half of the
kinematical ones, are preserved~\cite{Bain:2002tq};
these results agree with the analysis of
open strings in light-cone gauge performed previously in~\cite{Dabholkar:2002zc}, as well
as the supergravity analysis~\cite{Bain:2002nq} and D-brane embeddings~\cite{Skenderis:2002vf}.
Moreover, the
D1-brane $(+,-,0,0)$ at any position only preserves half of the dynamical supercharges~\cite{Bain:2002tq}.
As the plane wave is a homogeneous space it is rather counterintuitive
that the number of preserved supersymmetries may depend on the position of the brane.
In fact, a more precise statement is that these branes are flat in Brinkmann coordinates.
As the $P^I$ are time dependent in these
coordinates and do not simply generate translations along the $x^I$ (cf.\
equation~\eqref{generators}), a half-supersymmetric brane related to a
flat brane at the origin by a translation is curved~\cite{Bain:2002nq}. Hence flat branes
at different transverse positions do not fall in the same equivalence class with respect to
translations generated by the $P^I$, see
also~\cite{Skenderis:2002wx}.

In light-cone gauge boundary states can only describe instantonic
D$(p+1)$-branes~\cite{Green:1996um}. These are formally related to the
longitudinal branes discussed above by a double Wick rotation and will be denoted by $(m,n)$.
Boundary states in the plane wave
preserving half of both kinematical and dynamical supercharges were first constructed
in~\cite{Billo:2002ff} closely following the flat space
description of~\cite{Green:1996um}. Assume as in flat space that the D-brane preserves half of
the dynamical supersymmetries, i.e.
\begin{equation}\label{odyn}
\left(Q+i\eta M\wt{Q}\right)_{\da}|\!|\,(m,n),{\bf y}_t,\eta\ra\!\ra = 0\,,
\end{equation}
where $\eta=\pm1$ distinguishes a brane from an anti-brane, ${\bf y}_t$ is the transverse
position and
\begin{equation}
M_{\da\db} = \left(\prod_{I\in N}\g^I\right)_{\da\db}\,.
\end{equation}
Here $\g^I$ are the gamma-matrices of $SO(8)$ and the product is over the Neumann directions.
Together
with {\em standard} Neumann and
Dirichlet boundary conditions on the transverse bosons this implies that half of the
kinematical supersymmetries are preserved (see e.g.~\cite{Gaberdiel:2002vz})
\begin{equation}\label{okin}
\left(Q+i\eta M\wt{Q}\right)_{a}|\!|\,(m,n),{\bf y}_t,\eta\ra\!\ra = 0\,.
\end{equation}
Here $M_{ab}$ is analogous to $M_{\da\db}$. The structure of the
boundary state and consistency of the corresponding brane is crucially dependent on the
choice of $M$. It is useful to distinguish the cases $\Pi M\Pi M=\mp 1$, the resulting
branes will be sometimes called D$_-$- and D$_+$-branes, respectively. Boundary states for
D$_-$ were constructed in~\cite{Billo:2002ff,Bergman:2002hv}. The condition on $M$ is equivalent to
$|m-n|=2$ and thus leads to an analogous splitting of transverse coordinates as found from
the open string analysis~\cite{Dabholkar:2002zc}. The allowed values for $p$ are $p=1$, 3, 5 and
moreover, the condition~\eqref{odyn} is only satisfied if ${\bf y}_t=0$, otherwise only
half of the kinematical supercharges are preserved.
A detailed analysis and proof of the open-closed consistency
conditions was given in~\cite{Bergman:2002hv}. In flat space the cylinder diagram can be
expressed in terms of certain standard $\vt$-functions and open-closed duality arises as a
consequence of the properties of $\vt$-functions under modular transformations.
In the plane wave the
cylinder diagram involves deformed $\vt$-functions, where the deformation depends on the
mass parameter~\cite{Bergman:2002hv}. It has been proven in~\cite{Bergman:2002hv}
that these deformed $\vt$-functions satisfy certain transformation properties that assure that the
open-closed consistency conditions are precisely satisfied for the
half-supersymmetric branes. On the other hand, branes away from the origin, i.e.\
those preserving only half of the kinematical supercharges, appeared to violate
open-closed duality and hence be inconsistent. It is also worthwhile to note, that the
kinematical conditions~\eqref{okin} are not preserved as a function of time
$x^+$~\cite{Bergman:2002hv}. Indeed, the open string kinematical supercharge does not commute
with the Hamiltonian and hence is spectrum generating as is the case for closed strings.
The open string ground state is an unmatched boson~\cite{Dabholkar:2002zc} and it follows that
the open string partition function does not vanish~\cite{Bergman:2002hv}.

Boundary states for D$_+$ and the analysis of open-closed duality was considered
in~\cite{Gaberdiel:2002hh}; independently this class was studied in detail in~\cite{Skenderis:2002wx} from the open
string side. As mentioned above, these branes also arose in the supergravity
analysis~\cite{Skenderis:2002vf,Bain:2002nq} and from the covariant open string~\cite{Bain:2002tq}. In this case
the condition on $M$ is equivalent to $|m-n|=0$, $4$, however the coupling of $(0,4)$ and
$(4,0)$-branes to the background \RR flux induces a flux on the world-volume~\cite{Skenderis:2002vf}
and correspondingly the boundary conditions for bosons have to be modified. From the analysis
of~\cite{Gaberdiel:2002hh} it seems that the only consistent boundary state with standard bosonic boundary
conditions is the $(0,0)$ at any position, i.e.\ the D-instanton. Again, this is in
agreement with the open string analysis of~\cite{Skenderis:2002vf,Bain:2002nq,Bain:2002tq} where the
corresponding object, the D1-brane, is found to preserve half of the dynamical supersymmetries
at any position. In this case the kinematical conditions~\eqref{okin} are preserved as a
function of time $x^+$~\cite{Bergman:2002hv}, corresponding to a vanishing mass term for the open
string zero-modes. Hence in this case the ground states form a degenerate supermultiplet and
the open string partition function vanishes~\cite{Gaberdiel:2002hh}.

However, this might not be the full story yet~\cite{Skenderis:2002wx,Skenderis:2002ps}. The reason for this
is that the world-sheet theory being free, it possesses an countably infinite set of world-sheet
symmetries. These simply correspond to transformations shifting
the fields by a parameter satisfying the free field equations. For the open string such a
shift changes the action by a boundary term, so it is a symmetry if it satisfies
appropriate boundary conditions. As shown in~\cite{Skenderis:2002wx} the dynamical supercharges broken
by D$_-$-branes located outside the origin and the kinematical supercharges broken by the
D1-brane can be combined with world-sheet transformations that generate a {\em non-vanishing}
boundary term in such a way that the combined transformation {\em is} a symmetry of the open
string. Together with open string symmetries originating from
closed string symmetries compatible with the boundary conditions they generate
a superalgebra similar to that of the other half-supersymmetric branes~\cite{Skenderis:2002ps}.
An analysis of the boundary states for D$_-$-branes located outside the origin showed that
these do preserve a combination of eight dynamical and kinematical closed string supercharges
in addition to the eight standard kinematical ones. It would be interesting
to see whether these D$_-$-branes turn out to be consistent with open-closed duality.

The BMN correspondence can be extended to open
strings~\cite{Berenstein:2002zw,Lee:2002cu,Skenderis:2002vf,Balasubramanian:2002sa}. It was shown
in~\cite{Skenderis:2002vf}
that the D$_-$-branes located at the origin, descend from supersymmetric $AdS$ embeddings
in $AdS_5\times S^5$ through the Penrose limit; these originate from
the near-horizon limit of supersymmetric intersections of the D$p$-branes with a stack of
D3-branes. For example, in the near-horizon limit, a suitable D3-D5 system leads to a
D5 wrapping a $AdS_4\times S^2$ submanifold in $AdS_5\times S^5$. AdS/CFT is then supposed to
act twice and the holographic dual is SYM coupled to a three-dimensional defect.
The defect theory lives on the boundary of $AdS_4$ and as such is a CFT. The physics of
closed strings and 5-5 open strings is described by the bulk theory, whereas the boundary
theory captures 3-3, 3-5 and 5-3 strings~\cite{Karch:2000ct,Karch:2000gx,DeWolfe:2001pq}.
In particular, the 3-5 and 5-3 strings give rise to hypermultiplets in the fundamental
of the gauge group.
Applying the
Penrose limit results in the D5 $(+,-,3,1)$ brane at the origin. The dual
description is through the BMN limit of SYM coupled to the three-dimensional defect.
The closed string vacuum is dual to the trace of $Z$'s and intuitively one expects
the open string vacuum also to be dual to a large number of $Z$'s,
but instead of the trace with `quarks' at the end of the `string'. This is indeed the case,
the `quarks' are scalars in the hypermultiplet originating from 3-5 and 5-3 strings
and $\bar{q}Z^Jq$ represents the open string vacuum~\cite{Lee:2002cu}.
Open string excitations are then dual to insertions of defect fields and, for non-zero-modes,
in analogy with the insertion of phases for the closed string,
cosines and sines for Neumann and Dirichlet boundary conditions,
respectively~\cite{Lee:2002cu}. The D7 $(+,-,4,2)$ was discussed
in~\cite{Berenstein:2002zw}, this is more involved as orientifold planes have to be added to have a
consistent theory, but the basic idea remains the same. A further interesting example is the
giant graviton, i.e.\ a D3-brane wrapped on a $S^3$ in the $S^5$, which in the Penrose
limit gives rise to the $(+,-,0,2)$ brane. Here the open string fluctuations arise from
subdeterminant operators in SYM with large $R$-charge,
see~\cite{Balasubramanian:2002sa} for details.

\subsubsection{Strings on pp-waves and interacting field theories}\label{massive}

So far we have seen that we can get solvable string theories in light-cone gauge
turning on null, {\em constant} \RR field-strengths in a plane wave geometry.
As first discussed in~\cite{Maldacena:2002fy}, a large class of {\em interacting}
string models with world-sheet supersymmetry, can be engineered in more general pp-wave
geometries with {\em non-constant} fluxes and possibly transverse spaces with special
holonomy; for example
\begin{equation}
\begin{split}
ds^2 & = -2dx^+dx^-+H(x^i)\bigl(dx^+\bigr)^2+ds_8^2\,,\\
F_5 & = dx^+\wedge \varphi(x^i)\,,
\end{split}
\end{equation}
and all other background fields set to zero. It is convenient to split the
candidate Killing spinor $\e$ into two parts of opposite $SO(8)$ chiralities,
$\e=\e_++\e_-$. Analyzing the gravitino variation, one finds that $\e_+$ is independent of
all the coordinates; at lowest order in $\varphi$ this is the supernumerary spinor we
encountered before and gives rise to linearly realized supersymmetry on the
world-sheet in light-cone gauge. On the other hand, it is useful to split $\e_-$ into two
parts as well: one, independent of $x^+$ (and $x^-$) is determined through $\e_+$ by the
Killing equation, see~\cite{Maldacena:2002fy} for the explicit solution. This completes
the supernumerary Killing spinor for non-constant $\varphi$, however, as it is annihilated
by $\G^+$ it does not survive as part of the linearly realized supersymmetry in light-cone
gauge.
Depending on $\varphi$ one might also have a number of
kinematical supersymmetries; these correspond to the part of $\e_-$ depending only on $x^+$ and
solving the Killing equation with
$\e_+=0$; they imply that an even number of fermions (and hence also bosons) are free on the
world-sheet and decouple from the remaining interacting fields. Generically
there will be no kinematical supersymmetries.
If the transverse space is curved, space-time supersymmetry requires it to have special
holonomy. For example, for solutions with at least $N=(2,2)$ world-sheet supersymmetry the
most general possibility is a Calabi-Yau four-fold. The Killing spinor equation
determines the bosonic
potential $H$ in terms of $\varphi$ and imposes additional constraints on the allowed
four-forms. For $N=(2,2)$ the solution is parameterized in terms of
a holomorphic function $W$ and a real, harmonic Killing potential $U$.
Moreover, the Lie-derivative of $W$ along the holomorphic Killing vector $V_{\m}=i\N_{\m}U$
has to vanish~\cite{Maldacena:2002fy}. Explicitly, the general solution
leading to $N=(2,2)$ world-sheet theories in light-cone gauge is~\cite{Maldacena:2002fy}
\begin{align}
ds^2 & = -2dx^+dx^--32\bigl(|dW|^2+|V|^2\bigr)\bigl(dx^+\bigr)^2
+2g_{\m\bar{\n}}dz^{\m}d\bar{z}^{\bar{\n}}\,,\\
\varphi_{\m\n}&
\equiv\frac{1}{3!}\varphi_{\m\overline{\r\s\t}}{\e^{\overline{\r\s\t}}}_{\n}
=\N_{\m}\N_{\n}W\,,\qquad \varphi_{\overline{\m\n}}=\varphi_{\m\n}^*\,,\\
\varphi_{\m\bar{\n}}& \equiv\frac{1}{2}{\varphi_{\m\bar{\n}\r}}^{\r}
=\N_{\m}\N_{\bar{\n}}U\,.
\end{align}
Holomorphicity of $W$ follows because the $(1,3)$ forms in the ${\bf 10}$ of $SU(4)$
are co-closed, whereas $U$ is harmonic due to tracelessness of the $(2,2)$ forms in the
${\bf 15}$. To get interesting interacting world-sheet theories the transverse space needs
to be non-compact~\cite{Maldacena:2002fy}. As the geometry is that of a pp-wave, one can
still choose the light-cone gauge; the form of the resulting world-sheet theory is
dictated by supersymmetry~\cite{Alvarez-Gaume:1983ab}. Notice that, pp-wave string theories
do not lead to the most general 2d supersymmetric field theories: the target space is
always eight-dimensional of special holonomy and the Killing potential $U$ has to be
harmonic due to the
self-duality of $F_5$. Turning on an additional null \RR three-form leads to a second
Killing vector (commuting with the first one), and again the corresponding potential is
harmonic as a consequence of the variation of the dilatino~\cite{Kim:2002gi}.
In the case of $N=(1,1)$ the transverse space has Spin(7) holonomy,
one gets a real harmonic superpotential~\cite{Maldacena:2002fy} and,
if the \RR three-form is non-zero, one harmonic Killing potential~\cite{Kim:2002gi}.

This general class of pp-wave solutions of type IIB supergravity is
interesting for several reasons. They are
{\em exact} string solutions, i.e.\ they do not receive $\a'$ corrections.
In particular this is true for the plane wave background, see~\cite{Berkovits:2002zv} for a
proof based on the pure spinor approach for a covariant description of
strings in \RR backgrounds. In semi-light-cone gauge, conformal invariance of the GS superstring on the plane wave 
has also been studied in~\cite{Walton:2003nd}.
As shown in~\cite{Berkovits:2002vn}, for the pp-wave
space-times it is more advantageous to use the $U(4)$ formalism, where strings are governed
by exact {\em interacting} $N=2$ superconformal world-sheet theories. This proves the
exactness of this general class of solutions, see also~\cite{Berkovits:2002rd}
for an extension to a larger class of \RR backgrounds, some of which cannot be studied in
light-cone gauge. For an alternative argument, based
on space-time properties, essentially the existence of a covariantly
constant null vector, see~\cite{Russo:2002qj}.

Another interesting feature is the
possibility to choose the superpotential such that the world-sheet theory becomes
integrable~\cite{Maldacena:2002fy}; in that case one may hope to use known properties
of integrable models to learn about strings propagating in these backgrounds, see
also~\cite{Russo:2002qj,Bakas:2002kt} for further discussions and examples.

D-branes in these backgrounds have been analyzed in~\cite{Hikida:2002qk}, for example
for $N=(2,2)$ branes are supersymmetric if they wrap complex manifolds and the
superpotential (and Killing potential) are constant on the world-volume; one can also
have supersymmetric D5-branes wrapped on special Lagrangian submanifolds and
appropriate conditions on the potentials.
These results were derived in~\cite{Hikida:2002qk} in two ways, in the same spirit
I described in the previous section: by analyzing supersymmetry preserving boundary
conditions in the world-sheet theory and by finding supersymmetric embeddings in target
space. Interestingly, for the special case of the plane wave, the branes found
in~\cite{Hikida:2002qk} are `oblique', that is they are oriented in directions that couple
the two transverse ${\mathbb R}^4$'s; these however, generically preserve less supersymmetry than the
branes considered in the previous section. Recently `oblique' branes in the plane wave background have been 
analysed in detail in~\cite{Gaberdiel:2003sb}.

\section{String interactions in the plane wave background}\label{chapter4}

In the previous two sections I have among other things discussed and explained how free strings on the plane wave
background and its orbifold
arise in a double-scaling limit of ${\mc N}=4$ SYM and ${\mc N}=2$ quiver gauge theory, respectively.
A computation of the anomalous dimensions of BMN single-trace operators in interacting {\em planar} ${\mc N}=4$
SYM~\cite{Berenstein:2002jq,Gross:2002su,Santambrogio:2002sb} reproduces the mass spectrum of {\em free} string
theory~\cite{Metsaev:2001bj,Metsaev:2002re}.
It is obviously an interesting question how string interactions and the non-planar sector of
(interacting) gauge theory will fit into this picture. Before going into details let me
first make a few general remarks. The proposed duality between free string theory
and planar, interacting ${\mc N}=4$ SYM in the BMN limit
\begin{equation}\label{stgt}
\frac{1}{\m}H \cong \D-J
\end{equation}
should encompass interactions and non-planar effects, respectively. This follows from
the fact that the global symmetries of both sides of the duality are not expected to be
broken by quantum effects and hence the relation~\eqref{stgt} should hold to all orders in
the string coupling as a consequence of the AdS/CFT correspondence~\cite{Gross:2002mh}.
As the two operators act on different Hilbert spaces, this identity should be interpreted
with some care. One information encoded in~\eqref{stgt} is the identification of
eigenvalues of the two operators. This is a basis-independent statement, on both sides of the
duality we can choose any suitable basis, compute the matrix elements of the operator and
obtain the eigenvalues by diagonalization. Subsequently the corresponding eigenstates can
be identified (up to degeneracy ambiguities).
Recall once more the relations
\begin{equation}\label{dual}
\frac{1}{(\m\a'p^+)^2}=\lambda'\,,\qquad 4\pi g_{\text{s}}(\m\a'p^+)^2=g_2\,.
\end{equation}
As already stated, considering planar ($g_2=0$) gauge theory for small $\l'$ is equivalent to free
($g_{\text{s}}=0$) string theory for large $\m\a'p^+$. Now what happens if we take $g_2$ to be non-zero in the
{\em free} gauge theory? We see from~\eqref{dual} that this means to take $\m\a'p^+\to\infty$, $g_{\text{s}}\to0$,
such that $g_{\text{s}}(\m\a'p^+)^2$ is finite. As single- and multi-string states are orthogonal to
each other, whereas single-trace BMN operators start to mix with multi-trace ones at finite
$g_2$ in the free gauge
theory~\cite{Kristjansen:2002bb,Constable:2002hw,Beisert:2002bb,Gross:2002mh,Constable:2002vq},
the identification of states with operators is modified for finite $g_2$.
The fact that the required transformation is
not unique~\cite{Beisert:2002bb,Gross:2002mh,Constable:2002vq,Vaman:2002ka,Gomis:2002wi} can be
intuitively understood from string theory, because string states
become highly degenerate for $\m\a'p^+=~\infty$.
Taking into account string interactions is equivalent to considering non-planar,
{\em interacting} gauge theory.
Then the freedom of mixing is getting more constrained because the dual operators now
have to be eigenstates of the
interacting dilatation operator. The ambiguity is still present for protected operators or ones where the interaction does not
lift degeneracies present in the free theory.

As we are only able to obtain the free string spectrum in light-cone gauge,
we should ask how interactions can be studied in this picture.
In flat space, the usual strategy is the vertex operator approach and the difficulties 
associated with the fact that $x^-$ is quadratic in the
transverse coordinates are circumvented by using the ten-dimensional Lorentz invariance to set $p^+=0$ in general scattering
amplitudes. However, in the plane wave background transverse momentum is not a good quantum number due to the harmonic oscillator potential
confining the string to the origin of transverse space. Moreover ten-dimensional Lorentz invariance is broken by the non-zero \RR flux, in
particular there is no $J^{+-}$ generator.
This obstruction significantly hinders the
vertex operator approach to string interactions. There is only one other known way of studying string interactions in light-cone gauge,
namely light-cone string field theory pioneered by Mandelstam~\cite{Mandelstam:1973jk,Mandelstam:1974fq} for the bosonic string, see
also~\cite{Cremmer:1974jq,Cremmer:1975ej,Kaku:1974zz,Kaku:1974xu}, and
extended to the superstring in~\cite{Mandelstam:1974hk,Green:1983tc,Green:1983hw,Green:1984fu}. The construction of light-cone string field theory in the plane wave
geometry~\cite{Spradlin:2002ar,Spradlin:2002rv,Pankiewicz:2002gs,Pankiewicz:2002tg} and the derivation of the leading non-planar correction to the
anomalous dimension of BMN operators with two defects (cf.\ equation~\eqref{anom})
from string theory~\cite{Roiban:2002xr,Gomis:2003kj} is the main subject of this section and will be
discussed in detail in the following subsections. For a qualitative discussion of closed and
open string interactions from the gauge theory point of view see~\cite{Berenstein:2002sa}.

Further studies of string interactions and their comparison with gauge theory in the
BMN limit include~\cite{Chu:2002eu,Chu:2002wj,Chu:2003qd,Georgiou:2003aa}, where an alternative construction of the string field
theory vertex is pursued. Recently two inequivalent supersymmetric completions have been put forward in 
~\cite{DiVecchia:2003yp} and~\cite{Pankiewicz:2003kj}, respectively.
I will discuss
this issue in more detail in section~\ref{kinvertex}.
In~\cite{Kiem:2002xn,Kiem:2002pb} cubic interactions of IIB supergravity scalars
arising from the dilaton-axion sector and the chiral primary sector -- corresponding to
mixtures of the metric and the five-form -- were analyzed, the role of the bosonic prefactor
in string field theory on the plane wave was studied in~\cite{Lee:2002rm,Lee:2002vz}.
For an investigation of the S-matrix for strings in the plane wave see~\cite{Bak:2002ku}.
In~\cite{Huang:2002wf,Chu:2002pd} interactions of supergravity and string states were computed to
leading and subleading order in $\m\a' p^+$ and agreement with the planar three-point
functions of BMN operators was established. For an extension to non-planar
corrections and higher string interactions see~\cite{Huang:2002yt,Chu:2002qj}. Here the comparison was
based on the earlier
proposal of~\cite{Constable:2002hw} that the coefficient of the three-point function of BMN
operators is proportional to the matrix element of the cubic interaction in the plane wave.
With the work of~\cite{Gross:2002mh} (see also~\cite{Janik:2002bd}) this proposal has been replaced by the
more rigorous expression in equation~\eqref{stgt}.
Indeed, see also~\cite{Spradlin:2003bw} for a derivation of a vertex-correlator duality slightly different 
from~\cite{Constable:2002hw}.
By identifying {\em single} string states with mixtures of single and
multi-trace BMN operators -- defined such that the redefined single/multi-trace
operators are orthogonal in the non-planar, {\em free} gauge theory --
general matrix elements of the two sides in~\eqref{stgt} have been compared
in~\cite{Gross:2002mh,Vaman:2002ka,Pearson:2002zs,Gomis:2002wi,Georgiou:2003ab}. In~\cite{deMelloKoch:2002nq,deMelloKoch:2003pv} 
methods of collective field theory
have been employed to derive the string field theory vertex for supergravity and (certain) string states
from the matrix model truncation of SYM in the BMN limit.

The algebraic structure of the cubic interaction vertex, in particular
its expansion in powers of $\m\a' p^+$ was first examined in~\cite{Spradlin:2002rv,Klebanov:2002mp} and
subsequently studied in~\cite{Schwarz:2002bc,Pankiewicz:2002gs}. For comments on a non-trivial dependence of the string
coupling on $\m\a' p^+$ see~\cite{Gopakumar:2002dq}. Most notably,
closed expressions for all the quantities appearing in the interaction vertex as functions
of $\m\a' p^+$ were provided in~\cite{He:2002zu}.

This section is organized as follows. To make the presentation self-contained and to introduce necessary notation I briefly
review the free string on the plane wave in section~\ref{free}. In section~\ref{principles} I discuss the general features of
light-cone string field theory.
The construction of the kinematical and dynamical parts of the vertex and the (dynamical) supercharges in the number basis
is described in detail in sections~\ref{kinvertex} and~\ref{sec4}. The functional expressions for the dynamical generators are
given in section~\ref{sec54}. The results are applied in
section~\ref{ch6} to recover in light-cone string field theory the leading non-planar
correction to the anomalous dimension . Several technical details that are not included 
in this section are given in appendices~\ref{appA} and~\ref{appB}.

\subsection{Review of free string theory on the plane wave}\label{free}

In this subsection I briefly review some basic properties of free string theory on the plane wave
background~\cite{Metsaev:2001bj} and
introduce some notation. After fixing fermionic $\kappa$-symmetry and
world-sheet diffeomorphism and Weyl-symmetry in
light-cone gauge, the $r$-th free string propagating on the plane wave is described by
$x^I_r(\s_r)$ and $\vt^a_r(\s_r)$\footnote{$\vt_r$ are the non-vanishing
components of the $SO(9,1)$ spinor $S$ satisfying the light-cone gauge $\G^+S=0$.} in position space or
by $p^I_r(\s_r)$ and $\l^a_r(\s_r)$ in momentum space, where $I=1,\ldots,8$ is a transverse SO(8) vector index, $a=1,\ldots,8$ is a
SO(8) spinor index. I will often suppress these indices in what follows. The bosonic part of the light-cone action
is~\cite{Metsaev:2001bj}
\begin{equation}
S_{\text{B}(r)}=\frac{e(\a_r)}{4\pi\a'}\int\,d\t\int_0^{2\pi|\a_r|}
\,d\s_r\bigl[\dot{x}_r^2-x^{\prime\,2}_r-\m^2x_r^2\bigr]\,,
\end{equation}
where
\begin{equation}
\dot{x}_r\equiv\p_{\t}x_r\,,\qquad x'_r\equiv\p_{\s_r}x_r\,,\qquad
\a_r\equiv\a'p^+_r\,,\qquad e(\a_r)\equiv\frac{\a_r}{|\alpha_r|}\,.
\end{equation}
In a collision process $p^+_r$ will be negative for an incoming string and positive for an outgoing one.
The mode expansions of the fields $x_r^I(\s_r,\t)$ and $p_r^I(\s_r,\t)$ at $\t=0$ are
\begin{equation}
\begin{split}
x_r^I(\s_r)& = x_{0(r)}^I+\sqrt{2}\sum_{n=1}^{\infty}
\bigl(x_{n(r)}^I\cos\frac{n\s_r}{|\a_r|}+x_{-n(r)}^I\sin\frac{n\s_r}{|\a_r|}\bigr)\,,\\
p_r^I(\s_r) & =\frac{1}{2\pi|\a_r|}\bigl[p_{0(r)}^I+\sqrt{2}\sum_{n=1}^{\infty}
\bigl(p_{n(r)}^I\cos\frac{n\s_r}{|\a_r|}+p_{-n(r)}^I\sin\frac{n\s_r}{|\a_r|}\bigr)\bigr]\,.
\end{split}
\end{equation}
The Fourier modes can be re-expressed in terms of creation and annihilation operators as
\begin{equation}\label{xp}
x_{n(r)}^I=i\sqrt{\frac{\a'}{2\o_{n(r)}}}\bigl(a_{n(r)}^I-a_{n(r)}^{I\,\dg}\bigr)\,,\qquad
p_{n(r)}^I=\sqrt{\frac{\o_{n(r)}}{2\a'}}\bigl(a_{n(r)}^I+a_{n(r)}^{I\,\dg}\bigr)\,,
\end{equation}
where
\begin{equation}
\o_{n(r)} = \sqrt{n^2+\bigl(\m\a_r\bigr)^2}\,.
\end{equation}
Canonical quantization of the bosonic coordinates
\begin{equation}
[x_r^I(\s_r),p_s^J(\s_s)]=i\d^{IJ}\d_{rs}\d(\s_r-\s_s)
\end{equation}
yields the usual commutation relations
\begin{equation}
[a_{n(r)}^I,a_{m(s)}^{J\,\dg}]=\d^{IJ}\d_{nm}\d_{rs}\,.
\end{equation}
The fermionic part of the light-cone action in the plane wave is~\cite{Metsaev:2001bj}
\begin{equation}\label{ferlcaction}
S_{\text{F}(r)}=\frac{1}{8\pi}\int\,d\t\int_0^{2\pi|\a_r|}\,d\s_r
[i(\bar{\vt}_r\dot{\vt}_r+\vt_r\dot{\bar{\vt}}_r)
-\vt_r\vt'_r+\bar{\vt}_r\bar{\vt}'_r-2\m\bar{\vt}_r\Pi\vt_r]\,,
\end{equation}
where $\vt^a_r$ is a complex, positive chirality SO(8) spinor and
\begin{equation}
\Pi_{ab}\equiv(\g^1\g^2\g^3\g^4)_{ab}
\end{equation}
is symmetric, traceless and squares to one.\footnote{In comparison with section~\ref{chapter2}, here $\g^I$ are the gamma-matrices
of $SO(8)$. Throughout this chapter I use
the gamma matrix conventions of~\cite{Green:1983hw}. }
The matrix $\Pi$ breaks the transverse $SO(8)$ symmetry of the metric to
$SO(4)\times SO(4)$ and induces a projection of $SO(8)$ spinors to subspaces
of positive (negative) chirality under both $SO(4)$'s.
The mode expansion of $\vt^a_r$ and its conjugate momentum $i\l^a_r\equiv i\frac{1}{4\pi}\bar{\vt}^a_r$ at $\t=0$ is
\begin{equation}
\begin{split}
\vt^a_r(\s_r) & =\vt^a_{0(r)}+\sqrt{2}\sum_{n=1}^{\infty}
\bigl(\vt^a_{n(r)}\cos\frac{n\s_r}{|\a_r|}+\vt^a_{-n(r)}\sin\frac{n\s_r}{|\a_r|}\bigr)\,,\\
\l^a_r(\s_r) & =\frac{1}{2\pi|\a_r|}\bigl[\l^a_{0(r)}+\sqrt{2}\sum_{n=1}^{\infty}
\bigl(\l^a_{n(r)}\cos\frac{n\s_r}{|\a_r|}+\l^a_{-n(r)}\sin\frac{n\s_r}{|\a_r|}\bigr)\bigr]\,.
\end{split}
\end{equation}
The Fourier modes satisfy
\begin{equation}
\l_{n(r)}^a=\frac{|\a_r|}{2}\bar{\vt}^a_{n(r)}\,,
\end{equation}
and, due to the canonical anti-commutation relations for the fermionic coordinates
\begin{equation}
\{\vt^a_r(\s_r),\l^b_s(\s_s)\}=\d^{ab}\d_{rs}\d(\s_r-\s_s)\,,
\end{equation}
they obey the following anti-commutation rules
\begin{equation}
\{\vt^a_{n(r)},\l^b_{m(s)}\}=\d^{ab}\d_{nm}\d_{rs}\,.
\end{equation}
It is convenient to define a new set of fermionic operators~\cite{Spradlin:2002ar}
\begin{equation}
\vt_{n(r)}=\frac{c_{n(r)}}{\sqrt{|\a_r|}}\left[(1+\r_{n(r)}\Pi)b_{n(r)}+e(\a_r)e(n)(1-\r_{n(r)}\Pi)b_{-n(r)}^{\dg}\right]\,,
\end{equation}
which explicitly break the $SO(8)$ symmetry to $SO(4)\times SO(4)$. Here
\begin{equation}
\r_{n(r)}=\r_{-n(r)}=\frac{\o_{n(r)}-|n|}{\m\a_r}\,,\qquad
c_{n(r)}=c_{-n(r)}=\frac{1}{\sqrt{1+\r_{n(r)}^2}}\,.
\end{equation}
These modes satisfy
\begin{equation}
\{b^a_{n(r)},b^{b\,\dg}_{m(s)}\}=\d^{ab}\d_{nm}\d_{rs}\,.
\end{equation}
The free string light-cone Hamiltonian is
\begin{equation}
H_{2(r)}=\frac{1}{\a_r}\sum_{n\in\Zop}\o_{n(r)}\bigl(a_{n(r)}^{\dg}a_{n(r)}+b_{n(r)}^{\dg}b_{n(r)}\bigr)\,.
\end{equation}
In the above the zero-point energies cancel between bosons and fermions. Since the Hamiltonian only depends on two dimensionful
quantities $\m$ and $\a_r$, $\a'$ and $p^+_r$ should not be thought of as separate parameters.

The single string Hilbert space is built out of creation operators acting on the vacuum $|v\ra_r$ defined by
\begin{equation}
a_{n(r)}|v\ra_r=0\,,\qquad b_{n(r)}|v\ra_r=0\,,\qquad n\in{\mathbb Z}\,.
\end{equation}
Physical states have to satisfy the level-matching constraint
\begin{equation}
\sum_{n\in{\mathbb Z}}n\bigl(a_{n(r)}^{\dg}a_{n(r)}+b_{n(r)}^{\dg}b_{n(r)}\bigr)=0\,,
\end{equation}
which expresses the fact that there is
no physical significance to the choice of origin for $\s_r$.

The isometries of the plane wave background are generated by $H$, $P^+$, $P^I$, $J^{+I}$, $J^{ij}$ and $J^{i'j'}$. The latter two are
angular momentum generators of the transverse $SO(4)\times SO(4)$ symmetry of the plane wave. The 32 supersymmetries are generated
by $Q^+$, $\bar{Q}^+$ and $Q^-$, $\bar{Q}^-$.
The former correspond to inhomogeneous shift symmetries on the world-sheet ('non-linearly realized' supersymmetries), whereas the
latter generate the linearly realized world-sheet supersymmetries. In
sigma models the isometries of the target space-time result in conserved currents on the world-sheet. These have been obtained
in~\cite{Metsaev:2001bj} by the standard Noether method. I will need the following expressions (at $\t=0$)
\begin{equation}\label{PI}
P^I_{(r)} = \int_0^{2\pi|\a_r|}d\s_r\,p^I_r\,,\qquad
J^{I+}_{(r)} = \frac{e(\a_r)}{2\pi\a'} \int_0^{2\pi|\a_r|}d\s_r\,x^I_r\,.
\end{equation}
Conservation of (angular) momentum at the time of interaction ($\t=0$) will then be achieved by
{\em local} conservation of $\sum p^I_r(\s_r)$ and $\sum e(\a_r)x^I_r(\s_r)$, see equation~\eqref{kinb} below. The supercharges are
\begin{align}
\label{Q+}
Q^+_{(r)} & = \sqrt{\frac{2}{\a'}}\int_0^{2\pi|\a_r|}d\s_r\,\sqrt{2}\l_r\,,\\
\label{qfield}
Q^-_{(r)} & =\sqrt{\frac{2}{\a'}}\int_0^{2\pi|\a_r|}d\s_r\,\left[2\pi\a'e(\a_r)p_r\g\l_r-ix'_r\g\bar{\l}_r-i\m x_r\g\Pi\l_r\right]\,,
\end{align}
and $\bar{Q}^{\pm}_{(r)}=e(\a_r)\bigl[Q_{(r)}^{\pm}\bigr]^{\dg}$. Conservation of the non-linearly realized supercharges by
the interaction is
established by local conservation of $\sum\l_r(\s_r)$ and $\sum e(\a_r)\vt_r(\s_r)$, cf.\ equation~\eqref{kinf}.
Expanding $Q^-$ in modes one finds
\begin{equation}\label{q-mode}
\begin{split}
Q^-_{(r)} & =\frac{e(\a_r)}{\sqrt{|\a_r|}}\g
\Bigl(\sqrt{\m}\left[a_{0(r)}(1+e(\a_r)\Pi)+a_{0(r)}^{\dg}(1-e(\a_r)\Pi)\right]\l_{0(r)}\\
&+\sum_{n\neq 0}\sqrt{|n|}\left[a_{n(r)}P_{n(r)}^{-1}b_{n(r)}^{\dg}
+e(\a_r)e(n)a_{n(r)}^{\dg}P_{n(r)}b_{-n(r)}\right]\Bigr)\,,
\end{split}
\end{equation}
where
\begin{equation}
P_{n(r)}=\frac{1}{\sqrt{1-\rho_{n(r)}^2}}(1-\rho_{n(r)}\Pi)\,.
\end{equation}

\subsection{Principles of light-cone string field theory}\label{principles}

\begin{figure}[htb]
\begin{center}
\vspace{1cm}
\includegraphics{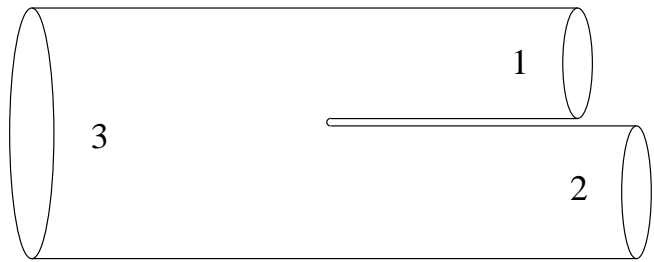}
\end{center}
\caption{The world-sheet of the three string interaction vertex.}
\label{fig1}
\end{figure}

The basic object in string field theory is an operator $\Psi$ that, roughly speaking, creates or annihilates strings and is acting on
a Hilbert space ${\mc H}$.\footnote{${\mc H}$ is the direct sum of $m$-string Hilbert spaces ${\mc H}_m$, the latter being the direct
product of the single-string Hilbert space ${\mc H}_1$.} In light-cone string field theory
$\Psi$ is a functional of the light-cone time $x^+$, the string length $|\a|$ and the momentum densities $p^I(\s)$ and
$\l^a(\s)$ specifying the configuration of the created/annihilated string. Observables of the free theory are expressed in terms of
$\Psi$, for example for the free light-cone Hamiltonian
\begin{equation}
H_2= \frac{1}{2}\int\,d|\a|{\mc D}^8p(\s){\mc D}^8\l(\s)\Psi^{\dg}
\left(\frac{\a'^2}{4}p^2-\frac{\m^2\a^2}{4}\frac{\d^2}{\d p^2}+\m|\a|\frac{\a'}{2}\l\Pi\frac{\d}{\d\l}\right)\Psi\,.
\end{equation}
To add interactions to the theory we have to ask the following question: what are the guiding principles in the
construction of the interaction? For the bosonic string the answer is very intuitive and
geometric~\cite{Mandelstam:1973jk,Mandelstam:1974fq}, the interaction
should couple the string world-sheets in a continuous way. For example, the interaction vertex for the scattering of three strings
depicted in Figure~\ref{fig1} is constructed with a Delta-functional enforcing world-sheet continuity.
The functional approach~\cite{Mandelstam:1973jk,Mandelstam:1974fq,Kaku:1974zz,Kaku:1974xu} can be extended to the
superstring~\cite{Mandelstam:1974hk,Green:1983tc,Green:1983hw,Green:1984fu}. Here the situation is slightly
more complicated, but the basic principle governing the construction of interactions is very simple: the superalgebra has to
be realized in the full interacting theory. It is easy to understand why this complicates matters, as the supercharges that square
to the Hamiltonian have to receive corrections as well when adding interactions. This is the essential difference to the bosonic
string and modifies the form of the vertex~\cite{Mandelstam:1974hk,Green:1983tc}. In a way the picture remains quite geometric, but in addition to a
Delta-functional enforcing continuity in superspace, one has to insert local operators at the interaction
point~\cite{Mandelstam:1974hk,Green:1983tc}.
These operators represent functional generalizations of derivative couplings.

To be more precise, consider the plane wave geometry and the behavior of the various generators of its superalgebra~\cite{Blau:2001ne}
when interactions are
taken into account. In fact, one can distinguish two different sets of generators.
The first set consists of the kinematical generators
\begin{equation}
P^+\,,\,P^I\,,\,J^{+I}\,,\,J^{ij}\,,\,J^{i'j'}\,,\,Q^+\,,\,\bar{Q}^+\,,
\end{equation}
which are not corrected by interactions, in other words the symmetries they generate are not
affected by adding higher order terms to the action. Hence these generators remain quadratic in the
string field $\Psi$ in the interacting field theory and act diagonally on ${\mc H}$.
On the other hand, as alluded to above, the dynamical generators
\begin{equation}
H\,,\,Q^-\,,\,\bar{Q}^-\,,
\end{equation}
do receive corrections in the presence of interactions
and couple different numbers of strings. The requirement that the
superalgebra is satisfied in the interacting theory, now gives rise to two kinds of constraints: {\em kinematical} constraints arising
from the (anti)commutation relations of kinematical with dynamical generators and {\em dynamical} constraints arising from the
(anti)commutation relations of dynamical generators alone. As I will explain below, the former will lead to the continuity conditions
in superspace, whereas the latter require the insertion of the interaction point operators. In practice these constraints will be
solved in perturbation theory, for example $H$, the full Hamiltonian of the interacting theory, has an expansion in the string coupling
\begin{equation}\label{hexp}
H=H_2+g_{\text{s}}H_3+\cdots\,,
\end{equation}
and $H_3$ leads to the three-string interaction depicted in Figure~\ref{fig1}.
To illustrate the procedure, consider the commutator
\begin{equation}
[H,P^I] = -i\m^2J^{+I}\,,
\end{equation}
which is of course different from the one in flat space. In the plane wave geometry transverse momentum is not a good quantum number due to
the confining harmonic oscillator potential. However, expansion in $g_{\text{s}}$ implies the same kinematical constraint as in flat space
\begin{equation}\label{kinematical1}
[H_3,P^I]=0\,,
\end{equation}
and, therefore, the interaction is translationally invariant.
In fact, the relation~\eqref{kinematical1} is also valid for all higher order interactions and as it is identical to the one in flat
space many of the techniques developed in~\cite{Green:1983tc,Green:1983hw} may be used in the plane wave case as well. In momentum space
the conservation of transverse momentum by the interaction will be implemented by a Delta-functional (cf.~\eqref{PI})
\begin{equation}
\D^8\left[\sum_{r=1}^3p_r(\s)\right]\,,
\end{equation}
for a precise definition of this functional see Appendix~\ref{appA}, equation~\eqref{Delta}.
Here the coordinate $\s$ of the three-string world-sheet is related to the coordinates $\s_r$ of the $r$-th string as
\begin{equation}
\begin{split}
\s_1 & =\s \qquad\quad\qquad -\pi\a_1\le\s\le\pi\a_1\,, \\
\s_2 & =\begin{cases} \s-\pi\a_1 & \quad\pi\a_1\le\s\le\pi(\a_1+\a_2)\,, \\
\s+\pi\a_1 & \quad-\pi(\a_1+\a_2)\le\s\le-\pi\a_1\,, \end{cases} \\
\s_3 & =-\s \qquad\quad\qquad -\pi(\a_1+\a_2)\le\s\le \pi(\a_1+\a_2)
\end{split}
\end{equation}
and $\a_1+\a_2+\a_3=0$, $\a_3<0$, i.e.~the process where the incoming string
splits into two strings. The joining of two strings into one is the adjoint of this process, see also section~\ref{ch6}.
In general, when I write an expression like $p_r(\s)$ it is understood that the function has support only for $\s$ within
the range that coincides with that of the $r$-th string. So, for example $p_r(\s)$ actually denotes $p_r(\s)=p_r(\s_r)\Theta_r(\s)$,
where
\begin{equation}
\Theta_1(\s)=\theta(\pi\a_1-|\s|)\,,\qquad \Theta_2(\s)=\theta(|\s|-\pi\a_1)\,,\qquad \Theta_3(\s)=1\,.
\end{equation}
Analogously from
\begin{equation}\label{q+}
[H,Q^+]=-\m\Pi Q^+\quad\Longrightarrow\quad[H_3,Q^+]=0\,,
\end{equation}
and, since light-cone momentum is a good quantum number, $[H,P^+]=0$,
one concludes that the cubic interaction contains (cf.~\eqref{Q+})
\begin{equation}
\D^8\left[\sum_{r=1}^3\l_r(\s)\right]\d\left(\sum_{r=1}^3\a_r\right)\,.
\end{equation}
Most interesting is the supersymmetry algebra
\begin{equation}\label{susy}
\{Q^-_{\da},\bar{Q}^-_{\db}\} = 2\d_{\da\db}H
-i\m\bigl(\g_{ij}\Pi\bigr)_{\da\db}J^{ij}+i\m\bigl(\g_{i'j'}\Pi\bigr)_{\da\db}J^{i'j'}\,,
\end{equation}
which also differs from the one in flat space. Expanding the supercharges
$Q^-_{\da}=Q^-_{2\,\da}+g_{\text{s}}Q^-_{3\,\da}+\cdots$,
and analogously for $\bar{Q}^-$, the
dynamical constraint following from~\eqref{susy} at ${\mc O}(g_{\text{s}})$
\begin{equation}
\{Q^-_{2\,\da},\bar{Q}^-_{3\,\db}\}+\{Q^-_{3\,\da},\bar{Q}^-_{2\,\db}\} = 2\d_{\da\db}H_3\,,
\end{equation}
is again the same as in flat space.
This constraint will be solved by inserting
a prefactor $h_3(\a_r,p_r(\s),\l_r(\s))$ into the ansatz for $H_3$ and analogously for $Q_3^-$ and $\bar{Q}_3^-$.
As I have already mentioned, the prefactors are operators inserted at the interaction point as required by locality, see also
section~\ref{sec54}. In summary, the structure of the superalgebra implies that the cubic interaction
can formally be written in the form
\begin{equation}\label{H3}
H_3 = \int d\m_3\,h_3(\a_r,p_r(\s),\l_r(\s))\Psi(1)\Psi(2)\Psi(3)\,,
\end{equation}
where $\Psi(r)$ is the string field for the $r$-th string, $h_3$ is the prefactor determined by the dynamical constraints and
the measure is
\begin{equation}\label{measure}
d\m_3\equiv \prod_{r=1}^3d\a_rD^8\l_r(\s)D^8p_r(\s)
\d\bigl(\sum_s\a_s\bigr)\D^8\bigl[\sum_s\l_s(\s)\bigr]\D^8\bigl[\sum_sp_s(\s)\bigr]\,.
\end{equation}
The expressions for $Q_3^-$ and $\bar{Q}_3^-$ are similar with different prefactors but the same measure $d\m_3$.

To give a precise meaning to the above functional expressions and in particular, to solve the dynamical constraints, it is
essential to do computations in the number basis~\cite{Cremmer:1974jq,Cremmer:1975ej}. For simplicity consider the bosonic part, also
the dependence on (and integration over) $\a_r$ will be suppressed in what follows.
The bosonic part of the string field $\Psi$ can be expanded in the number basis as
\begin{equation}
\Psi=\sum_{\{m_k\}}\phi_{m_k}\prod_{k\in{\mathbb Z}}\psi_{m_k}(p_k)\,,
\end{equation}
where $\phi_{m_k}$ is an operator that creates/annihilates a number basis state $|{m_k}\ra$
and $\psi_{m_k}(p_k)$ is the $m_k$-th oscillator wave function in momentum space. Substituting this into~\eqref{H3} yields
the cubic coupling $C(m_{k(1)},m_{k(2)},m_{k(3)})$ of three fields $\phi_{m_{k(r)}}$. It is convenient to
express $H_3$ not as an operator mapping ${\mc H}_1\to{\mc H}_2$ (or the adjoint process) but as a state in the 3-string Hilbert space via
\begin{equation}
C(m_{k(1)},m_{k(2)},m_{k(3)}) = \la m_{k(1)}|\la m_{k(2)}|\la m_{k(3)}|H_3\ra\,.
\end{equation}
Analogously the operators  $Q_3^-$ and $\bar{Q}_3^-$ will be identified with states  $|Q_3^-\ra$ and $|\bar{Q}_3^-\ra$ in ${\mc H}_3$.
Then we can write
\begin{equation}\label{defprefactor}
|H_3\ra\equiv\hat{h}_3|V\ra\,,
\end{equation}
where $\hat{h}_3$ is the prefactor (operator) and the {\em kinematical part of the vertex} $|V\ra$, common to
all the dynamical generators, is
\begin{equation}\label{vertex}
|V\ra\equiv|E_a\ra|E_b\ra\d\left(\sum_{r=1}^3\a_r\right)\,,\qquad
|E_a\ra\equiv\prod_{r=1}^3\int\,{\mc D}p_r\,\D^8\bigl[\sum_{s=1}^3p_s(\s)\bigr]|p_r\ra\,,
\end{equation}
and a similar expression for the fermionic contribution $|E_b\ra$. Here $|p\ra$ is the momentum eigenstate
\begin{equation}\label{boswv}
\begin{split}
|p\ra & =\prod_{k\in\mathbb{Z}}|p_{k}\ra=
\sum_{\{m_k\}}\prod_{k\in{\mathbb Z}}\psi_{m_{k}}(p_{k})|m_{k}\ra \\
& = \prod_{k\in\mathbb{Z}}\Bigl(\frac{\o_{k}\pi}{\a'}\Bigr)^{-1/4}
\exp\left(-\frac{\a'}{2\o_{k}}p_{k}^2+\sqrt{\frac{2\a'}{\o_{k}}}a_{k}^{\dg}p_{k}
-\frac{1}{2}a_{k}^{\dg}a_{k}^{\dg}\right)|0\ra\,,
\end{split}
\end{equation}
and $|0\ra$ is annihilated by $a_n$. Using ~\eqref{xp} one can check that this is indeed a momentum eigenstate.
It is not too difficult to derive the analogous expression for the fermionic contribution, but I will not need it in what follows.

\subsection{The kinematical part of the vertex}\label{kinvertex}

In the previous subsection I have explained the general ideas underlying light-cone string field theory and presented formal
expressions for the cubic corrections to the dynamical generators of the plane wave superalgebra. In particular we have seen
that the solution to the kinematical constraints can be constructed as a functional integral, which is common to all the
dynamical generators, cf.~\eqref{vertex}. To obtain the full solution we still
need to determine the explicit form of the prefactors and for
this it is necessary to explicitly compute the functional integral in the number basis.

The bosonic contribution $|E_a\ra$ to the exponential part of the three-string interaction
vertex has to satisfy the kinematic constraints~\cite{Green:1983tc,Green:1983hw}
\begin{equation}\label{kinb}
\sum_{r=1}^3p_r(\s)|E_a\ra=0\,,\qquad \sum_{r=1}^3e(\a_r)x_r(\s)|E_a\ra=0\,.
\end{equation}
These are the same as in flat space and arise from the commutation relations of $H$ with
$P^I$ and $J^{+I}$. They guarantee momentum conservation and continuity of the string world-sheet in the interaction.
The solution of the constraints in~\eqref{kinb} can be constructed as
the functional integral (cf.~\eqref{vertex})
\begin{equation}\label{ea}
\begin{split}
|E_a\ra & = \prod_{r=1}^3\int\,{\mc D}p_r\,\D^8\bigl[\sum_{s=1}^3p_s(\s)\bigr]|p_r\ra \\
& = \prod_{r=1}^3\prod_{n\in{\mathbb Z}}\int\,dp_{n(r)}\d^8\bigl[\sum_{s=1}^3\bigl(X^{(s)}p_s\bigr)_n\bigr]|p_{n(r)}\ra\,.
\end{split}
\end{equation}
In the second equality the precise definition of the Delta-functional in terms of an infinite
product of delta-functions for the individual Fourier modes of its argument was used, see appendix~\ref{appA}, equations~\eqref{Delta}--\eqref{AE7}
for details and the explicit expressions of the $X^{(r)}$.
As the resulting integrals are Gaussian (cf.~\eqref{boswv}) the evaluation is straightforward and the result
is~\cite{Spradlin:2002ar}
\begin{equation}\label{bv}
|E_a\ra\sim\exp\left(\frac{1}{2}
\sum_{r,s=1}^3\sum_{m,n\in{\mathbb Z}}a^{\dg}_{m(r)}\bar{N}^{rs}_{mn}a^{\dg}_{n(s)}
\right)|0\ra_{123}\,,
\end{equation}
where $|0\ra_{123}=|0\ra_1\otimes|0\ra_2\otimes|0\ra_3$ is annihilated by $a_{n(r)}$, $n\in{\mathbb Z}$.
Apart from the zero-mode part, the determinant factor coming
from the functional integral will be cancelled by the fermionic determinant.
In~\eqref{bv} the~non-vanishing elements of the so called bosonic Neumann matrices $\bar{N}^{rs}_{mn}$ for $m$, $n>0$
are~\cite{Spradlin:2002ar}
\begin{align}
\label{mn} \bar{N}^{rs}_{mn} & =\d^{rs}\d_{mn}
-2\sqrt{\frac{\o_{m(r)}\o_{n(s)}}{mn}}\left(A^{(r)\,T}\G^{-1}A^{(s)}\right)_{mn}\,,\\
\label{m0} \bar{N}^{rs}_{m0} & =
-\sqrt{2\m\a_s\o_{m(r)}}\e^{st}\a_t\bar{N}^r_m\,,\qquad s\in\{1,2\}\,,\\
\label{00a} \bar{N}^{rs}_{00} & =
(1-4\m\a K)\left(\d^{rs}+\frac{\sqrt{\a_r\a_s}}{\a_3}\right)\,,\qquad  r,s\in\{1,2\}\,,\\
\label{00b} \bar{N}^{r3}_{00} & =
-\sqrt{-\frac{\a_r}{\a_3}}\,,\qquad
r\in\{1,2\}\,.
\end{align}
Here
\begin{equation}
\a\equiv\a_1\a_2\a_3\,,\qquad
\G\equiv\sum_{r=1}^3 A^{(r)}U_{(r)}A^{(r)\,T}\,,
\end{equation}
where
\begin{equation}\label{rem}
U_{(r)}\equiv C^{-1}\bigl(C_{(r)}-\m\a_r\bigr)\,,\qquad C_{mn}\equiv m\d_{mn}\,,\quad
\bigl(C_{(r)}\bigr)_{mn}\equiv\o_{m(r)}\d_{mn}\,.
\end{equation}
The matrices $A^{(r)}$ are related to the $X^{(r)}$ in a simple way, see equation~\eqref{Ar}. The terms in
$\bar{N}^{rs}_{00}$ and
$\bar{N}^{r3}_{00}$ that are not proportional to $\m$
give the pure supergravity contribution to the Neumann matrices. The part of $\bar{N}^{rs}_{00}$ that is proportional to $\m$
is induced by positive string modes of $p_3$. I also defined
\begin{equation}
\bar{N}^r\equiv -C^{-1/2}A^{(r)\,T}\G^{-1}B\,,\qquad
K\equiv-\frac{1}{4}B^T\G^{-1}B\,.
\end{equation}
An explicit expression for the vector $B$ is given in~\eqref{B}. The quantities $\G$, $\bar{N}^r$ and $K$ manifestly reduce to
their flat space counterparts, defined in~\cite{Green:1983tc,Green:1983hw}, as $\m\to 0$. The only non-vanishing matrix elements with negative indices are
$\bar{N}^{rs}_{-m,-n}$. They are related to $\bar{N}^{rs}_{mn}$ via~\cite{Spradlin:2002ar}
\begin{equation}\label{neg}
\bar{N}^{rs}_{-m,-n}=-\left(U_{(r)}\bar{N}^{rs}U_{(s)}\right)_{mn}\,,\qquad m,\,n>0\,.
\end{equation}
As such the above expressions are already quite useful, though still formal in the sense that I did not present their explicit
expressions as functions of $\m$, $\a_r$. As the inverse of the infinite-dimensional matrix $\G$ appears in the expressions
for the Neumann matrices this is a formidable problem. In flat space the results were
known~\cite{Mandelstam:1973jk,Mandelstam:1974fq} due to the
identity\footnote{Notice that in comparison with~\cite{Green:1983tc} we have $\bar{N}^{rs}_{\text{here}}=C^{1/2}\bar{N}^{rs}_{\text{there}}C^{1/2}$.}
\begin{equation}\label{nnflat}
\bar{N}^{rs}_{mn}=-\a\frac{(mn)^{3/2}}{\a_rn+\a_sm}\bar{N}^r_m\bar{N}^s_n\,,
\end{equation}
and the explicit expressions
\begin{equation}\label{nrkflat}
\bar{N}^r_m=\frac{1}{\a_r}f_m\left(-\frac{\a_{r+1}}{\a_r}\right)e^{m\t_0/\a_r}\,,\qquad
K=-\frac{\t_o}{2\a}\,,
\end{equation}
where $\a_4\equiv\a_1$ is understood and
\begin{equation}
f_m(\g) = \frac{\G(m\g)}{m!\G\bigl(m(\g-1)+1\bigr)}\,,\qquad \t_0 = \sum_{r=1}^3\a_r\ln|\a_r|\,.
\end{equation}
The generalization of equation~\eqref{nnflat} to the plane wave background
is~\cite{Schwarz:2002bc,Pankiewicz:2002gs}
\begin{equation}\label{nnpp}
\begin{split}
\bar{N}^{rs}_{mn} & =-(1-4\m\a K)^{-1}\frac{\a}{\a_r\o_{n(s)}+\a_s\o_{m(r)}}\times\\
& \times\left[U_{(r)}^{-1}C_{(r)}^{1/2}C\bar{N}^r\right]_m\left[U_{(s)}^{-1}C_{(s)}^{1/2}C\bar{N}^s\right]_n\,,
\end{split}
\end{equation}
and reduces to equation~\eqref{nnflat} as $\m\to0$. This factorization theorem can also be used to verify
directly~\cite{Pankiewicz:2002gs} that $|E_a\ra$ satisfies the kinematic constraints in equation~\eqref{kinb}, see also
appendix~\ref{A4}. It will
also prove essential throughout the next section. The remaining problem of deriving explicit expressions for $K$ and $\bar{N}^r$ as in
equation~\eqref{nrkflat} has been solved in~\cite{He:2002zu}, however as I will not need these results in the remainder of this 
section I shall not give them here and refer the reader to~\cite{He:2002zu}.

Analogously to the bosonic case, the fermionic exponential part of the interaction vertex has to
satisfy~\cite{Green:1983tc,Green:1983hw}
\begin{equation}\label{kinf}
\sum_{r=1}^3\l_r(\s)|E_b\ra=0\,,\qquad
\sum_{r=1}^3e(\a_r)\vt_r(\s)|E_b\ra=0\,.
\end{equation}
These constraints arise from the commutation relations of $H$ with $Q^+$ and $\bar{Q}^+$, cf.~equation~\eqref{q+}.
As in the bosonic case its solution could be obtained by constructing the fermionic analogue of the wavefunction~\eqref{boswv} and then
performing the resulting integrals over the non-zero-modes. The pure zero-mode contribution has to be treated separately.
Instead of using the functional integral the exponential can be obtained (up to the normalization) by making a suitable
ansatz and imposing the constraints~\eqref{kinf}~\cite{Green:1983tc,Green:1983hw}. The solution
is~\cite{Pankiewicz:2002gs} (cf.\ appendix~\ref{A4} for the details;
the notation is defined below)
\begin{equation}\label{fv}
|E_b\ra \sim
\exp\left[\sum_{r,s=1}^3\sum_{m,n=1}^{\infty}b^{\dg}_{-m(r)}Q^{rs}_{mn}b^{\dg}_{n(s)}
-\sqrt{2}\L\sum_{r=1}^3\sum_{m=1}^{\infty}Q^r_mb^{\dg}_{-m(r)}\right]|E_b^0\ra\,,
\end{equation}
where
\begin{equation}
\L\equiv\a_1\l_{0(2)}-\a_2\l_{0(1)}
\end{equation}
and $|E_b^0\ra$ is the pure zero-mode part of the fermionic vertex
\begin{equation}\label{zero}
|E^0_b\ra=\prod_{a=1}^8\left[\sum_{r=1}^3\l_{0(r)}^a\right]|0\ra_{123}
\end{equation}
and satisfies $\sum_{r=1}^3\l_{0(r)}|E^0_b\ra=0$ and $\sum_{r=1}^3\a_r\vt_{0(r)}|E^0_b\ra=0$. Notice that $|0\ra_r$ is not the
plane wave vacuum defined to be annihilated by the $b_{0(r)}$. Rather, it satisfies $\vt_{0(r)}|0\ra_r=0$ and
$H_{2(r)}|0\ra_r=4\m e(\a_r)|0\ra_r$. In the limit $\m\to 0$ it coincides with the $SO(8)$ invariant 
flat space state that generates the massless multiplet
by acting with $\l_{0(r)}^a$ on it. The fermionic Neumann matrices can be expressed in terms of the bosonic ones
as~\cite{Pankiewicz:2002gs}
\begin{align}
\label{qmn}
Q^{rs}_{mn} & =e(\a_r)\sqrt{\left|\frac{\a_s}{\a_r}\right|}
\bigl[P_{(r)}^{-1}U_{(r)}C^{1/2}\bar{N}^{rs}C^{-1/2}U_{(s)}P_{(s)}^{-1}\bigr]_{mn}\,,\\
\label{qm}
Q^r_n & =\frac{e(\a_r)}{\sqrt{|\a_r|}}(1-4\m\a
K)^{-1}(1-2\m\a K(1+\Pi))\bigl[P_{(r)}C_{(r)}^{1/2}C^{1/2}\bar{N}^r\bigr]_n\,.
\end{align}
Let me comment on the choice of zero-mode vertex in equation~\eqref{zero}. 
Instead of constructing the vertex on $|0\ra_r$ (`the $SO(8)$ formulation'), it was
proposed in~\cite{Chu:2002eu} to use a different zero-mode vertex built on the
plane wave vacuum $|v\ra_r$ which is $SO(4)\times SO(4)$ invariant and annihilated by all the 
$b_{0(r)}$ (`the $SO(4)\times SO(4)$ formulation'). This also modifies the
non-zero-mode part of $|E_b\ra$, a complete solution to the kinematic constraints was
given in~\cite{Chu:2002wj,Pankiewicz:2002gs}. The motivation for this proposal originally was twofold.
First, it was shown in~\cite{Gursoy:2002yy} that the torus anomalous dimension of BMN operators
with mixed scalar/vector impurities is the same as that for scalar/scalar impurities.
This was in disagreement with the proposal of~\cite{Constable:2002hw} that the coefficient of the
three-point function of BMN operators is proportional to the matrix element of the cubic
interaction in the plane wave, which due to the structure of the string field theory vertex 
would predict vanishing anomalous dimension for these
class of operators at the torus level. 
One possible resolution of this discrepancy was to think about a modification of the
string vertex. Another possibility is of course to abandon the proposal
of~\cite{Constable:2002hw} which was not derived from first principles. In fact, I will show in
section~\ref{ch6} that using the identification in equation~\eqref{stgt},
the anomalous dimension of BMN operators transforming as $({\bf 4},{\bf 4})$ under
$SO(4)\times SO(4)$ {\em is} reproduced in string theory using the vertex with
fermionic zero-mode part as in~\eqref{zero}. A second (related) reason was
based on the fact that the plane wave has a discrete $\mathbb{Z}_2$ symmetry that exchanges
the two transverse $\mathbb{R}^4$'s. This discrete symmetry should be preserved by the
interaction.
It was shown in~\cite{Chu:2002eu} that the  ${\mathbb Z}_2$ parity of $|v\ra$ is opposite to the
one of $|0\ra$. Then it followed that we have to assign positive parity to $|0\ra$ in order to preserve the full
transverse symmetry in the $SO(8)$ formulation.
This seems strange, as $|v\ra$ has negative parity although it is the ground state of the theory. 
However, the spectrum of type IIB string theory on the plane wave was analyzed in detail
in~\cite{Metsaev:2002re}, in particular the precise correspondence between the lowest lying string
states and the fluctuation modes of supergravity on the plane wave was established.
It turns out that the state $|0\ra$ corresponds to the complex scalar arising from the
dilaton-axion system, whereas the state $|v\ra$ corresponds to a complex scalar
being a mixture of the trace of the graviton and the \RR potential
on one of the ${\mathbb R}^4$'s, that is the chiral primary sector. As dilaton and axion are
scalars under $SO(8)$ and
the discrete $\mathbb{Z}_2$ is just a particular $SO(8)$ transformation, we see that
the assignment of positive parity to $|0\ra$ appears to be correct. Moreover, analysis
of the interaction Hamiltonian for the chiral primary sector shows that invariance of the
Hamiltonian under the ${\mathbb Z}_2$ requires the chiral primaries to have negative
parity~\cite{Kiem:2002pb}. Finally, the implications of this assignment on matrix elements of the cubic vertex 
were successfully tested from the gauge theory side in~\cite{Georgiou:2003ab}. 

In~\cite{Pankiewicz:2003kj} the solution of the kinematical constraints in the $SO(4)\times SO(4)$ formulation
was extended to include the required prefactors for the three-string interaction vertex and dynamical supercharges. 
In particular, contrary to previous expectations, evidence was presented that the two vertices constructed on $|v\ra$ 
and on $|0\ra$, respectively,  are in fact one and the same, see~\cite{Pankiewicz:2003kj} for details. 
In what follows, I will 
keep on working with the $SO(8)$ formulation, though for computations involving fermionic oscillators the $SO(4)\times SO(4)$ one
is better suited. Finally, let me remark that if the plane wave ground state $|v\ra$ is odd under the $\mathbb{Z}_2$, then 
the supersymmetric extension proposed in~\cite{DiVecchia:2003yp} which is of the form $\p_{\t}|V\ra_{SO(4)\times SO(4)}$ 
is not $\mathbb{Z}_2$ invariant. 

\subsection{The complete ${\mc O}(g_{\text{s}})$ superstring vertex}\label{sec4}

In the previous subsection I reviewed the exponential part of the vertex, which solves the kinematic constraints.
The remaining dynamic constraints are much more restrictive and are solved by introducing
prefactors~\cite{Green:1983tc,Green:1983hw}, polynomial in creation
operators, in front of $|V\ra$ (cf.~\eqref{defprefactor}). 
Within the functional formalism, the prefactors can be re-interpreted as insertions of local operators at
the interaction point~\cite{Mandelstam:1974hk,Green:1983tc}. In this section I present expressions for the dynamical generators in
the number basis and prove that they satisfy the superalgebra at order
${\mc O}(g_{\text{s}})$~\cite{Spradlin:2002ar,Pankiewicz:2002tg}. The functional form of the
leading order corrections to the dynamical generators~\cite{Spradlin:2002ar,Pankiewicz:2002gs,Pankiewicz:2002tg} 
will be discussed in section~\ref{sec54}.

Define the linear combinations of the free supercharges ($\eta=e^{i\pi/4}$)
\begin{equation}
\sqrt{2}\eta\,Q\equiv Q^-+i\bar{Q}^-\,,\qquad\sqrt{2}\bar{\eta}\,\wt{Q}=Q^--i\bar{Q}^-
\end{equation}
which, on the subspace of physical states satisfying the level-matching condition, satisfy
\begin{equation}
\begin{split}
\{Q_{\dot{a}},Q_{\dot{b}}\}& = \{\wt{Q}_{\dot{a}},\wt{Q}_{\dot{b}}\}
=2\d_{\dot{a}\dot{b}}H\,,\\
\{Q_{\dot{a}},\wt{Q}_{\dot{b}}\} & = 
-\m\bigl(\g_{ij}\Pi\bigr)_{\dot{a}\dot{b}}J^{ij}
+\m\bigl(\g_{i'j'}\Pi\bigr)_{\dot{a}\dot{b}}J^{i'j'}
\,.
\end{split}
\end{equation}
Since $J^{ij}$ and $J^{i'j'}$ are not corrected by the interaction, it follows that at order ${\mc O}(g_{\text{s}})$ the dynamical
generators have to satisfy
\begin{align}
\label{dyn1}
\sum_{r=1}^3Q_{\dot{a}(r)}|Q_{3\,\dot{b}}\ra+\sum_{r=1}^3Q_{\dot{b}(r)}|Q_{3\,\dot{a}}\ra
& = 2\d_{\dot{a}\dot{b}}|H_3\ra\,,\\
\label{dyn2}
\sum_{r=1}^3\wt{Q}_{\dot{a}(r)}|\wt{Q}_{3\,\dot{b}}\ra+\sum_{r=1}^3\wt{Q}_{\dot{b}(r)}|\wt{Q}_{3\,\dot{a}}\ra
& = 2\d_{\dot{a}\dot{b}}|H_3\ra\,,\\
\label{dyn3}
\sum_{r=1}^3Q_{\dot{a}(r)}|\wt{Q}_{3\,\dot{b}}\ra+\sum_{r=1}^3\wt{Q}_{\dot{b}(r)}|Q_{3\,\dot{a}}\ra   & = 0\,.
\end{align}
In order to derive equations that determine the full expressions for the dynamical generators one has to compute (anti)commutators of
the free supercharges $Q_{\da(r)}$ and $\wt{Q}_{\da(r)}$ with the prefactors appearing in $|Q_{3\,\dot{a}}\ra$ and
$|\wt{Q}_{3\,\dot{a}}\ra$. Moreover, the action of the supercharges on $|V\ra$ has to be known. Here the factorization
theorem~\eqref{nnpp} for the bosonic Neumann matrices and the relation between the bosonic and fermionic Neumann matrices given in
equations~\eqref{qmn} and~\eqref{qm} prove to be essential.

\subsubsection{The bosonic constituents of the prefactors}

An important constraint on the prefactors (that I will collectively denote by ${\mc P}$) is that they must respect
the local conservation laws ensured by $|E_a\ra$ and $|E_b\ra$. For the bosonic part this means that it must commute
with~\cite{Green:1983tc,Green:1983hw}
\begin{equation}\label{bospre}
\bigl[\,\sum_{r=1}^3p_r(\s),{\mc P}\bigr]=0=\bigl[\,\sum_{r=1}^3e(\a_r)x_r(\s),{\mc P}\bigr]\,.
\end{equation}
Consider first an expression of the form
\begin{equation}
K_0+K_+=\sum_{r=1}^3\sum_{m=0}^{\infty}F_{m(r)}a^{\dg}_{m(r)}\,.
\end{equation}
The Fourier transform of \eqref{bospre} leads to the equations~\cite{Spradlin:2002rv}
\begin{equation}\label{kplus}
\sum_{r=1}^3\bigl[X^{(r)}C_{(r)}^{1/2}F_{(r)}\bigr]_m=0=\sum_{r=1}^3\a_r\bigl[X^{(r)}C_{(r)}^{-1/2}F_{(r)}\bigr]_m\,.
\end{equation}
Here the components $m=0$ and $m>0$ decouple from each other.
It is convenient to write the solution for $m=0$ in
a form which makes the flat space limit manifest~\cite{Pankiewicz:2002gs}
\begin{equation}
K_0=(1-4\m\a K)^{1/2}\left(\mathbb{P}-i\m\frac{\a}{\a'}\mathbb{R}\right)\,.
\end{equation}
Here
\begin{equation}
\mathbb{P}\equiv\a_1p_{0(2)}-\a_2p_{0(1)}\,,\qquad
\a_3\mathbb{R}\equiv x_{0(1)}-x_{0(2)}\,,\qquad [\mathbb{R},\mathbb{P}]=i\,,
\end{equation}
that is (no sum on $r$)
\begin{equation}
F_{0(r)} = -(1-4\m\a K)^{1/2}\sqrt{\frac{2}{\a'}}\e^{rs}\sqrt{\m\a_r}\a_s\,,\qquad F_{0(3)}=0\,.
\end{equation}
The overall normalization of $K_0$ is of course not determined by~\eqref{kplus}. The
inclusion of the overall factor $(1-4\m\a K)^{1/2}$ will be convenient in what follows.
For $m>0$ we have
\begin{equation}\label{kplus2}
\sum_{r=1}^3\bigl[A^{(r)}C^{-1/2}C_{(r)}^{1/2}F_{(r)}\bigr]_m=\frac{1}{\sqrt{\a'}}\m\a B_m=
\sum_{r=1}^3\m\a_r\bigl[A^{(r)}C^{-1/2}C_{(r)}^{-1/2}F_{(r)}\bigr]\,.
\end{equation}
These equations can be solved using the identities~\eqref{id2} and~\eqref{ups1} given in
appendix~\ref{appA}. One finds~\cite{Spradlin:2002rv,Pankiewicz:2002gs}
\begin{equation}\label{f+}
F_{m(r)}=-\frac{\a}{\sqrt{\a'}\a_r}(1-4\m\a K)^{-1/2}
\bigl[U_{(r)}^{-1}C_{(r)}^{1/2}C\bar{N}^r\bigr]_m\,.
\end{equation}
In the limit $\m\to0$
\begin{equation}
\lim_{\m\to0}\bigl(K_0+K_+\bigr)=
\mathbb{P}-\frac{\a}{\sqrt{\a'}}\sum_{r=1}^3\sum_{m=1}^{\infty}\frac{1}{\a_r}
\bigl[C\bar{N}^r\bigr]_m\sqrt{m}a_{m(r)}^{\dg}
\end{equation}
coincides with the flat space result of~\cite{Green:1983hw}.
Now take into account the negatively moded creation oscillators, i.e.\ consider
\begin{equation}
K_-=\sum_{r=1}^3\sum_{m=1}^{\infty}F_{-m(r)}a^{\dg}_{-m(r)}\,.
\end{equation}
This leads to the equations
\begin{equation}
\sum_{r=1}^3\frac{1}{\a_r}\bigl[A^{(r)}C^{1/2}C_{(r)}^{1/2}F_{(r)}\bigr]_{-m}=0=
\sum_{r=1}^3\bigl[A^{(r)}C^{1/2}C_{(r)}^{-1/2}F_{(r)}\bigr]_{-m}\,.
\end{equation}
Comparing the second equation with the difference of the two equations in~\eqref{kplus2} it follows
\begin{equation}\label{f-+1}
F_{-m(r)}\sim U_{m(r)}F_{m(r)}\,.
\end{equation}
However, if one substitutes this into the first equation one actually sees that the sum is
divergent~\cite{Green:1983tc,Green:1983hw,Spradlin:2002rv}. This phenomenon
already appears in flat space and it is known~\cite{Green:1983tc} that the function of $\s$ responsible for the divergence is
$\d(\s-\pi\a_1)-\d(\s+~\pi\a_1)$. However, since $\pm\pi\a_1$ are actually identified this divergence is merely an artifact of our
parametrization. I will argue in section~\ref{sec42} that the appropriate relative normalization
is~\cite{Pankiewicz:2002gs}
\begin{equation}\label{f-+2}
F_{-m(r)}=iU_{m(r)}F_{m(r)}\,.
\end{equation}

\subsubsection{The fermionic constituents of the prefactors}

The fermionic constituents of the prefactors have to satisfy the conditions
\begin{equation}\label{ferpre}
\bigl\{\,\sum_{r=1}^3\l_r(\s),{\mc P}\bigr\}=0=\bigl\{\,\sum_{r=1}^3e(\a_r)\vt_r(\s),{\mc P}\bigr\}\,.
\end{equation}
Consider
\begin{equation}
Y=\sum_{r=1}^2G_{0(r)}\l_{0(r)}+\sum_{r=1}^3\sum_{m=1}^{\infty}G_{m(r)}b^{\dg}_{m(r)}\,.
\end{equation}
For the zero-modes we can set the coefficient of, say, $\l_{0(3)}$ to zero due to the property of the fermionic supergravity vertex that
$\sum_{r=1}^3\l_{0(r)}|E_b^0\ra=0$ . The Fourier transform of~\eqref{ferpre} leads to the equations
\begin{align}
\sum_{r=1}^3\frac{1}{\sqrt{|\a_r|}}\bigl[A^{(r)}CC_{(r)}^{-1/2}P_{(r)}G_{(r)}\bigr]_m & =0\,,\\
\sum_{r=1}^3e(\a_r)\sqrt{|\a_r|}\bigl[C^{1/2}A^{(r)}C_{(r)}^{-1/2}P_{(r)}^{-1}G_{(r)}\bigr]_m & =\sum_{r=1}^3\a_rX_{m0}^{(r)}G_{0(r)}\,.
\end{align}
The components $m=0$ and $m>0$ decouple from each other. For $m=0$ the solution is
\begin{equation}\label{y}
Y=(1-4\m\a K)^{-1/2}(1-2\m\a K(1+\Pi))\sqrt{\frac{2}{\a'}}\L+\cdots
\end{equation}
As in the previous subsection the normalization is not determined and is chosen for
further convenience. For $m>0$ we can rewrite the second equation as
\begin{equation}
\sum_{r=1}^3e(\a_r)\sqrt{|\a_r|}\bigl[A^{(r)}C_{(r)}^{-1/2}P_{(r)}^{-1}G_{(r)}\bigr]_m=\frac{\a}{\sqrt{\a'}}B_m\,.
\end{equation}
Then the solution can be expressed in terms of $F_{(r)}$ as~\cite{Pankiewicz:2002gs}
\begin{equation}\label{gf}
G_{(r)}=\sqrt{|\a_r|}P_{(r)}^{-1}U_{(r)}C^{-1/2}F_{(r)}\,.
\end{equation}
As $\m\to0$ we have
\begin{equation}
\lim_{\m\to0}Y=\sqrt{\frac{2}{\a'}}\L+
\sum_{r=1}^3\sum_{m=1}^{\infty}\frac{F_{m(r)}}{\sqrt{m}}\sqrt{|\a_r|}b_{m(r)}^{\dg}\,.
\end{equation}
Taking into account that $\sqrt{|\a_r|}b_{m(r)}^{\dg}\longleftrightarrow Q_{-m(r)}^I$
in the notation of~\cite{Green:1983hw} this is exactly the flat space expression.
We will see below that as in flat space~\cite{Green:1983tc,Green:1983hw}, it turns out that the prefactors
do not involve negatively moded fermionic creation oscillators.

\subsubsection{The dynamical generators at order ${\mc O}(g_{\text{s}})$}\label{sec42}

Below I present the results~\cite{Pankiewicz:2002tg} necessary to verify the dynamical
constraints in equations~\eqref{dyn1} and~\eqref{dyn2}, given the
ansatz~\eqref{h}-\eqref{tq} for the cubic vertex and dynamical supercharges. Computational details are relegated to
appendix~\ref{appB}. We need
\begin{equation}\label{qk}
\sqrt{2}\eta\sum_{r=1}^3[Q_{(r)},\wt{K}^I]\,|V\ra=
\sqrt{2}\bar{\eta}\sum_{r=1}^3[\wt{Q}_{(r)},K^I]\,|V\ra=\m\g^I(1+\Pi)Y|V\ra\,,
\end{equation}
where
\begin{equation}
K^I\equiv K_0^I+K_+^I+K_-^I\,,\qquad
\wt{K}^I\equiv K^I_0+K_+^I-K_-^I
\end{equation}
and
\begin{equation}\label{qy}
\begin{split}
\sqrt{2}\eta\sum_{r=1}^3\{Q_{(r)},Y\}\wt{K}^I|V\ra & =
i\g^JK^J\wt{K}^I|V\ra-i\m\frac{\a}{\a'}(1-4\m\a K)\g^I(1-\Pi)|V\ra\,,\\
\sqrt{2}\bar{\eta}\sum_{r=1}^3\{\wt{Q}_{(r)},Y\}K^I|V\ra & =
-i\g^J\wt{K}^JK^I|V\ra+i\m\frac{\a}{\a'}(1-4\m\a K)\g^I(1-\Pi)|V\ra\,.
\end{split}
\end{equation}
Notice that the above identities are only valid when both sides of the equation act on
$|V\ra$. The action of the supercharges on $|V\ra$ is
\begin{equation}
\begin{split}
\label{qv}
\sqrt{2}\eta\sum_{r=1}^3Q_{(r)}|V\ra & =
-\frac{\a'}{\a}K^I\g^IY|V\ra\,,\\
\sqrt{2}\bar{\eta}\sum_{r=1}^3\wt{Q}_{(r)}|V\ra & =
-\frac{\a'}{\a}\wt{K}^I\g^IY|V\ra\,.
\end{split}
\end{equation}
The latter two equations actually lead to the insight that one has to consider the combinations $K^I$ and $\wt{K}^I$, as they are
solely determined by the kinematical part of the vertex and the quadratic pieces of the dynamical supercharges. In this way it is then
possible to fix the relative normalization as has been done in equation~\eqref{f-+2}~\cite{Pankiewicz:2002gs}. The results summarized in
equations~\eqref{qk}-\eqref{qv} motivate the following ansatz for the explicit form of the dynamical supercharges and the three-string
interaction vertex~\cite{Pankiewicz:2002tg,Spradlin:2002ar}
\begin{align}
\label{h}
|H_3\ra & =\left(\wt{K}^IK^J-\m\frac{\a}{\a'}\d^{IJ}\right)v_{IJ}(Y)|V\ra\,,\\
\label{q}
|Q_{3\,\dot{a}}\ra & = {\wt K}^Is_{\dot{a}}^I(Y)|V\ra\,,\\
\label{tq}
|\wt{Q}_{3\,\dot{a}}\ra & = K^I\tilde{s}_{\dot{a}}^I(Y)|V\ra\,.
\end{align}
Substituting the above ansatz into~\eqref{dyn1} and~\eqref{dyn2} and using~\eqref{qk}-\eqref{qv},
one gets the following equations for $v^{IJ}$, $s_{\da}^I$ and
$\tilde{s}_{\da}^I$\,\footnote{Here $(\da\db)$ denotes symmetrization in $\da$, $\db$.}
\begin{equation}\label{flat}
\d_{\dot{a}\dot{b}}v^{IJ} =
\frac{i}{\sqrt{2}}\frac{\a'}{\a}\g^J_{a(\dot{a}}D^as^I_{\dot{b})}\,,\qquad
\d_{\dot{a}\dot{b}}v^{IJ} =
-\frac{i}{\sqrt{2}}\frac{\a'}{\a}\g^I_{a(\dot{a}}\bar{D}^a\tilde{s}^J_{\dot{b})}\,,
\end{equation}
which originate from terms proportional to $\wt{K}_IK_J$ and $K_I\wt{K}_J$
and are identical to the flat space equations of~\cite{Green:1983hw}.
Two additional equations, arising from terms proportional to $\m\d_{IJ}$, are
\begin{equation}\label{ppwave1}
\begin{split}
-\d_{\dot{a}\dot{b}}v^{II} & =\frac{i}{\sqrt{2}}\frac{\a'}{\a}
\g^I_{a(\dot{a}}\Bigl(D^a+i\bigl[\Pi\bar{D}\bigr]^a\Bigr)s^I_{\dot{b})}\,,\\
-\d_{\dot{a}\dot{b}}v^{II} & = -\frac{i}{\sqrt{2}}\frac{\a'}{\a}
\g^I_{a(\dot{a}}\Bigl(\bar{D}^a-i\bigl[\Pi D\bigr]^a\Bigr)\tilde{s}^I_{\dot{b})}\,.
\end{split}
\end{equation}
As in flat space~\cite{Green:1983hw} one defines
\begin{equation}
D^a\equiv\eta Y^a+\bar{\eta}\frac{\a}{\a'}\frac{\p}{\p Y_a}\,,\qquad
\bar{D}^a\equiv\bar{\eta} Y^a+\eta\frac{\a}{\a'}\frac{\p}{\p Y_a}\,.
\end{equation}
Recall first the solution of the flat space equations~\eqref{flat}~\cite{Green:1983hw}. One introduces the following functions of $Y^a$
\begin{align}
w^{IJ} & = \d^{IJ} +\left(\frac{\a'}{\a}\right)^2\frac{1}{4!}t^{IJ}_{abcd}Y^aY^bY^cY^d
+\left(\frac{\a'}{\a}\right)^4\frac{1}{8!}\d^{IJ}\e_{abcdefgh}Y^a\cdots Y^h\,,\\
iy^{IJ} & = \frac{\a'}{\a}\frac{1}{2!}\g^{IJ}_{ab}Y^aY^b+\left(\frac{\a'}{\a}\right)^3\frac{1}{2\cdot 6!}
\g^{IJ}_{ab}{\e^{ab}}_{cdefgh}Y^c\cdots Y^h\,,\\
s^I_{1\,\dot{a}} & = 2\g^I_{a\dot{a}}Y^a+\left(\frac{\a'}{\a}\right)^2\frac{2}{6!}
u^I_{abc\dot{a}}{\e^{abc}}_{defgh}Y^d\cdots Y^h\,,\\
s^I_{2\,\dot{a}} & = -\frac{\a'}{\a}\frac{2}{3!}u^I_{abc\dot{a}}Y^aY^bY^c+
\left(\frac{\a'}{\a}\right)^3\frac{2}{7!}\g^I_{a\dot{a}}{\e^{a}}_{bcdefgh}Y^b\cdots Y^h\,.
\end{align}
Here
\begin{equation}
t^{IJ}_{abcd}\equiv\g^{IK}_{[ab}\g^{JK}_{cd]}\,,\qquad u^I_{abc\dot{a}}\equiv -\g^{IJ}_{[ab}\g^J_{c]\dot{a}}\,.
\end{equation}
$t^{IJ}_{abcd}$ is traceless and symmetric in $I$, $J$, hence $w^{IJ}$ is a symmetric tensor of $SO(8)$, whereas $y^{IJ}$ is
antisymmetric.
Apart from the coefficients, in flat space the structure of the individual terms is
completely fixed by the $SO(8)$ symmetry.
The solution of equations~\eqref{flat} is~\cite{Green:1983hw}
\begin{equation}\label{sol}
v^{IJ}\equiv w^{IJ}+y^{IJ}\,,\quad
s^I_{\dot{a}}\equiv-\frac{2}{\a'}\frac{i}{\sqrt{2}}
\bigl(\eta s^I_{1\,\dot{a}}+\bar{\eta}s^I_{2\,\dot{a}}\bigr)\,,\quad
\tilde{s}^I_{\dot{a}}\equiv \frac{2}{\a'}\frac{i}{\sqrt{2}}
\bigl(\bar{\eta} s^I_{1\,\dot{a}}+\eta s^I_{2\,\dot{a}}\bigr)\,.
\end{equation}

Next consider the additional equations~\eqref{ppwave1}. Using the flat space solution, these can be rewritten as
\begin{equation}
0=\g^I_{a(\dot{a}}\bigl[\Pi\bar{D}\bigr]^as^I_{\dot{b})}\,\qquad
0=\g^I_{a(\dot{a}}\bigl[\Pi D\bigr]^a\tilde{s}^I_{\dot{b})}\,.
\end{equation}
The proof that these equations are also satisfied by~\eqref{sol} is given in appendix~\ref{appB}.

The proof~\cite{Pankiewicz:2002tg} of equation~\eqref{dyn3} is more involved and provides an important consistency check of the
ansatz~\eqref{h}-\eqref{tq}. It leads to the equations (cf. appendix~\ref{B3})
\begin{align}
\label{qq1}
&\d^{IJ}m_{\da\db}-\frac{1}{\sqrt{2}}\frac{\a'}{\a}\g^{(I}_{a\da}D^a\tilde{s}^{J)}_{\db}=0\,,\\
&\d^{IJ}m_{\da\db}-\frac{1}{\sqrt{2}}\frac{\a'}{\a}\g^{(I}_{a\db}\bar{D}^as^{J)}_{\da}=0\,,\\
\label{qq3}
&\sqrt{2}\bigl(\g^I_{a\da}\eta\tilde{s}^I_{\db}-\g^{I}_{a\db}\bar{\eta}s^I_{\da}\bigr)-4im_{\da\db}Y_a=0\,,\\
\label{ppwave2}
&\bigl(\g^I_{a\da}\bar{D}_b\tilde{s}^I_{\db}+\g^I_{a\db}D_bs^I_{\da}\bigr)(1-\Pi)^{ab}=0\,.
\end{align}
Here
\begin{align}
m_{\dot{a}\dot{b}} & =\d_{\dot{a}\dot{b}}+\frac{i}{4}\frac{\a'}{2\a}\g^{IJ}_{\dot{a}\dot{b}}\g^{IJ}_{ab}Y^aY^b
-\frac{1}{4\cdot 4!}\left(\frac{\a'}{2\a}\right)^2\g^{IJKL}_{\dot{a}\dot{b}}t^{IJKL}_{abcd}Y^aY^bY^cY^d
\nonumber\\
&-\frac{i}{6!}\left(\frac{\a'}{2\a}\right)^3\g^{IJ}_{\dot{a}\dot{b}}\g^{IJ}_{ab}{\e^{ab}}_{cdefgh}Y^c\cdots
Y^h-\frac{2}{7!}\left(\frac{\a'}{2\a}\right)^4\d_{\dot{a}\dot{b}}\e_{abcdefgh}Y^a\cdots Y^h
\end{align}
and
\begin{equation}
t^{IJKL}_{abcd}\equiv\g^{[IJ}_{[ab}\g^{KL]}_{cd]}\,.
\end{equation}
The first three equations are identical to those in flat space
and have been proven in~\cite{Green:1983hw}. The additional equation~\eqref{ppwave2} is proved in 
appendix~\ref{B2}.

The dynamical constraints do not 
fix the overall normalization of the dynamical generators which can depend on $\m$ and the $\a_r$'s.
In flat space, the fact that the $J^{-I}$ generator of the Lorentz algebra is also dynamical imposes further constraints
on the other dynamical generators and apart from trivial rescaling uniquely fixes their
normalization~\cite{Green:1984fu}.
As the $J^{-I}$ generator is not part of the plane wave superalgebra this procedure cannot be applied to our setup.
A comparison with a supergravity calculation fixes the normalization for small $\m$ to be
$\sim(\a'\m^2)/(\a_3^4)$~\cite{Kiem:2002xn},
whereas a comparison with the dual field theory implies that for large $\m$ it is
$\sim\a'/\a^2$~\cite{Vaman:2002ka,Gomis:2002wi,Pankiewicz:2002tg}.
It was conjectured in~\cite{Pankiewicz:2002tg} that the normalization valid for all $\m$ is
\begin{equation}\label{norm}
16\pi\a'\m^2\a_3^{-4}(1-4\m\a K)^2\,,
\end{equation}
which has the correct small- and large-$\m$ behavior~\cite{Schwarz:2002bc}.
On the other hand, the non-trivial normalization of $Y$ (cf.\ equation~\eqref{y}) and the fact that the terms $\wt{K}^IK^J$ and $\m\d^{IJ}$ in
equation~\eqref{h} involve different powers of $1-4 \m\a K$ is fixed by requiring the closure of the superalgebra
at ${\mc O}(g_{\text{s}})$. In order to obtain the supergravity expressions for the dynamical generators from
equations~\eqref{h}-\eqref{tq}, one should set $K$ to zero, as it originates from massive string modes, cf.\ the remark below
equation~\eqref{rem}. 



\subsection{Functional expressions}\label{sec54}

The functional expressions for the cubic corrections to the dynamical generators can be provided by defining the operator analogues
for the constituents of the prefactor. These operators depend on $p_r(\s)$, $x'_r(\s)$ and $\l_r(\s)$ and since $p_r(\s)$ and
$\l_r(\s)$ correspond to functional derivatives with respect to $x_r(\s)$ and $\vt_r(\s)$ the only physically sensible value of $\s$ to
choose is the interaction point $\s=\pm\pi\a_1$. As operators at this point are singular the prefactor must be carefully defined
in the limit $\s\to|\pi\a_1|$~\cite{Green:1983tc}. Rewriting the operators in the number basis
one obtains expressions containing both creation and annihilation operators of the various oscillators.
Eliminating the annihilation operators by (anti)commuting them through the exponential factors of the vertex one recovers the
number basis expressions for the constituents of the
prefactors~\cite{Green:1983tc,Green:1983hw,Pankiewicz:2002gs}.

As in flat space~\cite{Green:1983tc,Green:1983hw} consider the following operators
\begin{equation}\label{p}
P(\s) \equiv -2\pi\sqrt{-\a}(\pi\a_1-\s)^{1/2}\bigl(p_1(\s)+p_1(-\s)\bigr)\,,
\end{equation}
\begin{equation}\label{dx}
\p X(\s)\equiv 4\pi\frac{\sqrt{-\a}}{\a'}(\pi\a_1-\s)^{1/2}\bigl(x'_1(\s)+x'_1(-\s)\bigr)\,,
\end{equation}
\begin{equation}\label{y2}
Y(\s) \equiv -2\pi\frac{\sqrt{-2\a}}{\sqrt{\a'}}(\pi\a_1-\s)^{1/2}\bigl(\l_1(\s)+\l_1(-\s)\bigr)\,.
\end{equation}
One also defines $P|V\ra\equiv\lim\limits_{\s\to\pi\a_1}P(\s)|V\ra$ and analogously for $\p X$.
Acting on the exponential part of the vertex and taking the limit $\s\to\pi\a_1$ we
have~\cite{Pankiewicz:2002gs}
\begin{align}
\lim_{\s\to\pi\a_1}K^I(\s)|V\ra &\equiv\left(P^I+\frac{1}{4\pi}\p X^I\right)|V\ra=K^I|V\ra\,,\\
\lim_{\s\to\pi\a_1}\wt{K}^I(\s)|V\ra &\equiv\left(P^I-\frac{1}{4\pi}\p X^I\right)|V\ra=\wt{K}^I|V\ra\,,\\
\lim_{\s\to\pi\a_1}Y(\s)|V\ra  & = Y|V\ra\,.
\end{align}
Here I prove only the last equation, for more details see~\cite{Pankiewicz:2002gs}. Substituting the mode expansion for $\l_1(\s)$ 
into~\eqref{y2} one gets
\begin{align}
\lim_{\s\to\pi\a_1}Y(\s)|V\ra & =-\sqrt{\frac{2}{\a'}}\sqrt{\frac{-2\a}{\a_1}}\lim_{\e\to0}\e^{1/2}
\sum_{n=1}^{\infty}(-1)^n\cos(n\e/\a_1)\times\nonumber\\
&\times\left[\sqrt{2}\L Q^1_n+\sum_{r=1}^3\sum_{m=1}^{\infty}Q^{1r}_{nm}b^{\dg}_{m(r)}\right]|V\ra\,.
\end{align}
Now the singular behavior of the sum as $\e\to0$ can be traced to the way it diverges as $n\to\infty$. Therefore to take the limit
$\e\to0$ we can approximate the summand for large $n$ and using the factorization theorem~\eqref{nnpp} one
finds~\cite{Pankiewicz:2002gs}
\begin{equation}
\lim_{\s\to\pi\a_1}Y(\s)|V\ra = f(\m)(1-4\m\a K)^{-1/2}Y|V\ra\,,
\end{equation}
where 
\begin{equation}
f(\m)\equiv-2\frac{\sqrt{-\a}}{\a_1}\lim_{e\to0}\e^{1/2}\sum_{n=1}^{\infty}(-1)^nn\cos(n\e/\a_1)\bar{N}^1_n\,.
\end{equation}
The identity
\begin{equation}\label{fmu}
f(\m)=(1-4\m\a K)^{1/2}
\end{equation}
was conjectured to hold on general grounds (the closure of the superalgebra) in~\cite{Pankiewicz:2002tg} and shown to be true
in~\cite{He:2002zu}.
This concludes the proof of equation~\eqref{y2}.

So up to the overall normalization one can write the
functional equivalent of equations~\eqref{h}, \eqref{q} and~\eqref{tq} as
\begin{align}
H_3 & = \lim_{\s\to\pi\a_1}\int d\m_3
\Bigl(\wt{K}^I(\s)K^J(\s)-\m\frac{\a}{\a'}\d^{IJ}\Bigr)v_{IJ}(Y(\s))\Psi(1)\Psi(2)\Psi(3)\,,\\
Q_{3\,\da} & = \lim_{\s\to\pi\a_1}\int d\m_3\wt{K}^I(\s)s^I_{\da}(Y(\s))\Psi(1)\Psi(2)\Psi(3)\,,\\
\wt{Q}_{3\,\da} & = \lim_{\s\to\pi\a_1}\int d\m_3K^I(\s)\tilde{s}^I_{\da}(Y(\s))\Psi(1)\Psi(2)\Psi(3)\,,
\end{align}
where $d\m_3$ is the functional expression leading to the kinematical part of the vertex, cf.\ equation~\eqref{measure}.

Finally, I would like to point out the following subtlety.
One can check for example that
\begin{equation}
\sqrt{2}\bar{\eta}\sum_{r=1}^3[\wt{Q}_{(r)},\lim_{\s\to\pi\a_1}K^I(\s)]\,|V\ra=\m\g^I\Pi Y|V\ra\,.
\end{equation}
However, this is not equal to the commutator of $\sum_r\wt{Q}_{(r)}$ with $K^I$. Using equation~\eqref{qv} and
\begin{equation}
[\lim_{\s\to\pi\a_1}K^I(\s),\wt{K}^J]|V\ra=-\frac{\m\a}{\a'}(1-4\m\a K)^{-1/2}\d^{IJ}|V\ra\,,
\end{equation}
leads to~\cite{Pankiewicz:2002tg}
\begin{equation}
\sqrt{}2\bar{\eta}\sum_{r=1}^3[\wt{Q}_{(r)},K^I]|V\ra=\m\g^I(1+\Pi)Y|V\ra\,,
\end{equation}
which is equivalent to equation~\eqref{qk} of section~\ref{sec4}. It is this appearance of the matrix $1+\Pi$ as opposed to
just $\Pi$, that is responsible for the term proportional to $\m\d^{IJ}$ in the cubic interaction vertex.

\subsection{Anomalous dimension from string theory}\label{ch6}

In this section I discuss how the result for the anomalous dimension in
equation~\eqref{anom} can be recovered in string theory.
This has been done for the symmetric-traceless ${\bf 9}$ and antisymmetric
${\bf 6}={\bf 3}+\bar{{\bf 3}}$ of
either one of the $SO(4)$'s in~\cite{Roiban:2002xr} and for the trace ${\bf 1}$
in~\cite{Gomis:2003kj}.
Here I review this work and also include the states in 
$({\bf 4},{\bf 4})_{\pm}$\footnote{We define the states in  $({\bf 4},{\bf 4})_{\pm}$
as $\frac{1}{2}\bigl(\a_{n(3)}^{i\,\dag}\a_{-n(3)}^{j'\,\dag}\pm\a_{-n(3)}^{i\,\dag}\a_{n(3)}^{j'\,\dag}\bigr)|v\ra_3$.
The change of basis $\a_n=\frac{1}{\sqrt{2}}\bigl(a_{|n|}+ie(n)a_{-|n|}\bigr)$
for $n\neq0$ is convenient and
an analogous transformation will be made for the fermions.}
of $SO(4)\times SO(4)$ in the analysis. These
correspond to BMN operators with mixed scalar/vector impurities and
superconformal symmetry of the gauge theory implies that they
have the same anomalous dimension as the other representations~\cite{Beisert:2002tn}.

To compute the mass shift of the single string state
due to interactions
\begin{equation}\label{nij}
|n\ra\equiv\a^{I\,\dg}_{n(3)}\a^{J\,\dg}_{-n(3)}|v\ra_3\,,
\end{equation}
non-degenerate perturbation theory was used
in~\cite{Roiban:2002xr,Gomis:2003kj}.
In principle one should use degenerate perturbation theory as the
single string state can mix with multi-string states having the same energy. The same caveat
holds for the computation in gauge theory and we will ignore this complication here, see however~\cite{Freedman:2003bh}.
At lowest order the eigenvalue correction comes from two contributions; one-loop diagram
and contact term
\begin{equation}\label{de1}
\d E_n^{(2)}\la n|n\ra = g_2^2\sum_{1,2}
\left[\frac{1}{2}\frac{\left|\la n|H_3|1,2\ra\right|^2}{E_n^{(0)}-E_{1,2}^{(0)}}
+\frac{1}{8}\sum_{\da}\left|\la n|Q_{3\,\da}|1,2\ra\right|^2\right]\,.
\end{equation}
Factors different from $g_2$ in the normalization (cf.~\eqref{norm}) are absorbed in the definition of $H_3$ 
and $Q_3$, the extra factor of $1/2$ in the first term is due to the reflection symmetry of
the one-loop diagram.
The sum
over $1$,~$2$ is over physical double-string states, that is those obeying the level-matching condition and for the case at hand
$Q_3^2$ is the only relevant contribution to the quartic coupling. 
As the generators are hermitian we take the absolute value squared of the matrix elements.
In fact, time-reversal in the plane wave background consists of the transformation
\begin{equation}\label{tr}
x^+ \to -x^+\,,\qquad x^- \to - x^-\,,\qquad \m \to -\m\,,
\end{equation}
in particular the reversal of $\m$ is needed due to the presence of the \RR flux. Previously I have always assumed
that $\m$ is non-negative and $\a_3<0$, $\a_1$, $\a_2>0$. This is, say, the process where a single string {\em splits} into two strings.
One can show that for the process in which
two strings {\em join} to form a single string, i.e.\ $\a_1$, $\a_2<0$ and $\a_3>0$, one should make the additional replacements
\begin{equation}
\m \to -\m\,,\qquad \Pi \to -\Pi
\end{equation}
in equations~\eqref{h}-\eqref{tq} and~\eqref{norm}. This is in agreement with equation~\eqref{tr}. Notice that the
transformation of $\Pi$ is needed to leave the fermionic mass term invariant, cf.~\eqref{ferlcaction}.
From the formal expressions for the Neumann matrices it is not manifest that the cubic corrections to the dynamical generators are
hermitian as they have to be. However, from the explicit expressions for the Neumann
matrices~\cite{He:2002zu} one can see that
all the quantities are in fact invariant under the time-reversal.
The string states obey the delta-function normalization
$\la n|n'\ra = {\mc N}|\a_3|\d(\a_3-\a_4)$, where
${\mc N}=\frac{1}{2}(1+\d^{ij})$ for the ${\bf 9}$, ${\mc N}=\frac{1}{4}$ for the ${\bf 1}$ and
${\mc N}=\frac{1}{2}$ otherwise. The sum over double-string states includes a double integral
over light-cone momenta, one integral is trivial due to the string-length conservation of the
cubic interaction and the factor of $|\a_3|\d(\a_3-\a_4)$ can be cancelled on both sides of
equation~\eqref{de1}.
The remaining sum is then the usual completeness relation for harmonic
oscillators projected on physical states and we have ($\b\equiv\a_1/\a_3$)
\begin{equation}\label{de2}
{\mc N}\d E_n^{(2)}=-g_2^2\int_{-1}^0\frac{d\b}{\b(\b+1)}
\sum_{\text{modes}}\left[\frac{1}{2}
\frac{\left|\la n|H_3|1,2\ra\right|^2}{E_n^{(0)}-E_{1,2}^{(0)}}
+\frac{1}{8}\sum_{\da}\left|\la n|Q_{3\,\da}|1,2\ra\right|^2\right]\,.
\end{equation}
The measure arises due to the fact that string states are delta-function normalized.

It is important to note that in gauge theory the dilatation operator was diagonalized
within the subspace of two-impurity BMN operators in perturbation theory in
the 't Hooft coupling $\l$
and then extrapolated to $\l$, $J\to\infty$.
But it is not obvious that the large $J$ limit of the perturbation series in $\l$ has to
agree order by order with the perturbation series in $\l'$, see for
example~\cite{Klebanov:2002mp}.
Indeed there is evidence from string theory that this is not the case:
for large $\m$ the denominator of the first term in equation~\eqref{de2}
is of order ${\mc O}(\m^{-1})$ in the
impurity conserving channel, whereas it is of order ${\mc O}(\m)$ in the
impurity non-conserving one. However, as already noticed in~\cite{Spradlin:2002rv},
matrix elements where the number of impurities changes by two are of
order ${\mc O}(1)$ and, therefore potentially can contribute to the mass-shift at leading
order, that is ${\mc O}(\m g_2^2\l')$. Notice that impurity non-conserving matrix
elements being of order one, means actually ${\mc O}(\m g_2\sqrt{\l'})$ and as the overall
factor of $\m$ is simply for dimensional reasons and should not be counted when
translating to gauge theory (cf.\ equation~\eqref{stgt}) implies contributions
$\sim g_2\sqrt{\l'}$ to matrix elements of the dilatation operator.
It was observed in~\cite{Roiban:2002xr} that the contribution of
the impurity non-conserving channel to~\eqref{de2} is linearly divergent.
This is due to the fact that the large $\m$ limit does not commute
with the infinite sums over mode numbers;
for finite $\m$ the divergence is regularized. So a linear divergence reflects
a contribution $\sim \m g_2^2\l'(-\m\a_3)=\m g_2^2\sqrt{\l'}$ and hence of order
$g_2^2\sqrt{\l'}$ to the anomalous dimension. This constitutes a non-perturbative, `stringy'
effect.
It remains a very interesting challenge to
investigate the contribution of the impurity non-conserving channel in detail.
However, to reproduce the result~\eqref{anom}
for the anomalous dimensions of two-impurity BMN operators in string theory one is led to
a truncation of equation~\eqref{de2} to the impurity conserving channel~\cite{Roiban:2002xr}.
This analysis will be performed below.

\subsubsection{Contribution of one-loop diagrams}

The matrix element $\la n|H_3|1,2\ra$ in the impurity conserving channel
is non-zero only if the double-string state contains either two bosonic or two fermionic
oscillators. The relevant projection operator is
\begin{equation*}
\begin{split}
\sum_{K,L}
\a_{0(1)}^{\dg\,K}\a_{0(2)}^{\dg\,L}|v\ra\la v|\a_{0(2)}^{L}\a_{0(1)}^{K}
&+\frac{1}{2}\sum_{k\in\Zop}\sum_{r,K,L}
\a_{k(r)}^{\dg\,K}\a_{-k(r)}^{\dg\,L}|v\ra\la v|\a_{-k(r)}^{L}\a_{k(r)}^{K}\\
+\sum_{a,b}
\b_{0(1)}^{\dg\,a}\b_{0(2)}^{\dg\,b}|v\ra\la v|\b_{0(2)}^{b}\b_{0(1)}^{a}
&+\frac{1}{2}\sum_{k\in\Zop}\sum_{r,a,b}
\b_{k(r)}^{\dg\,a}\b_{-k(r)}^{\dg\,b}|v\ra\la v|\b_{-k(r)}^{b}\b_{k(r)}^{a}\,.
\end{split}
\end{equation*}
For the first case the fermionic contribution to the matrix elements is simple to determine.
Using a $\g$-matrix representation in which $\Pi=\text{diag}({\bf 1}_4,-{\bf 1}_4)$,
the plane wave vacua ${}_r\la v|$ are related to ${}_r\la 0|$ (up to an irrelevant phase) via
\begin{equation}\label{vo}
{}_r\la v|={}_r\la 0|\left(\frac{\a_r}{2}\right)^2\prod_{a=5}^8\vt^a_{0(r)}\,,\qquad
{}_3\la v|=-{}_3\la 0|\left(\frac{\a_3}{2}\right)^2\prod_{a=1}^4\vt^a_{0(r)}\,.
\end{equation}
Directions $1,\ldots,4$ and $5,\ldots,8$ correspond to positive and negative chirality under
$SO(4)\times SO(4)$, respectively.
Eight of the zero-modes in equation~\eqref{vo},
namely $\vt_{0(3)}^a$, $a=1,\ldots,4$ and, say, $\vt_{0(2)}^a$, $a=5,\ldots,8$ are saturated
by $|E_b^0\ra$, so to give a non-zero contribution the remaining four zero-modes must be
contracted with the ${\mc O}(Y^4)$ term in $v_{MN}(Y)$. Hence, the fermionic contribution is
\begin{equation}
\left(\frac{\a'}{\a}\right)^2\frac{1}{4!}t^{MN}_{abcd}
{}_{123}\la v|Y^{abcd}|E^0_b\ra =
-\left(\frac{\a_3}{2}\right)^4(1-4\m\a K)^{-2}t^{MN}_{5678}\,.
\end{equation}
One can show that $t^{MN}_{5678}=(\d^{mn},-\d^{m'n'})$ in the $\g$-matrix basis used here.
The bosonic part of the matrix element is not difficult to evaluate and I will not go into details.
Using the large $\m$ expansions for the bosonic Neumann
matrices~\cite{Schwarz:2002bc,He:2002zu} one finds, for example for $(I,J)=(i,j)$,
\begin{equation}
\begin{split}
\la n|H_3|\a_{0(r)}^{\dg\,k}\a_{0(s)}^{\dg\,l}|v\ra_{12}& \sim
\m\l'\frac{\sin^2 n\pi\b}{2\pi^2}\left(\d^{rs}+\frac{\sqrt{\a_r\a_s}}{\a_3}\right)S^{ijkl}\,,\\
\la n|H_3|\a_{k(r)}^{\dg\,K}\a_{-k(r)}^{\dg\,L}|v\ra_{12} & \sim
\m\l'\b(\b+1)\frac{\a_3}{\a_r}\frac{\sin^2 n\pi\b}{2\pi^2}S^{ijkl}\,,
\end{split}
\end{equation}
and the analogous expression for $(I,J)=(i',j')$ with an (inessential) overall minus sign.
Here
\begin{equation}
S^{ijkl}\equiv T^{ijkl}+\frac{1}{4}\d^{ij}T^{kl}\,,\quad
T^{ijkl}=\d^{ik}\d^{jl}+\d^{jk}\d^{il}-\frac{1}{2}\d^{ij}\d^{kl}\,,\quad
T^{kl} = -2\d^{kl}
\end{equation}
can be split into a symmetric-traceless and a trace part.
There is no contribution to the ${\bf 6}$ nor to $({\bf 4},{\bf 4})_{\pm}$.
The sum over $k$ and the integral over $\b$ can
be done and the complete contribution of the impurity conserving channel with
bosonic excitations at one-loop is
\begin{equation}
\frac{\m g_2^2\l'}{4\pi^2}\frac{15}{16\pi^2n^2}
\begin{cases} \frac{1}{4}\sum_{k,l}T^{ijkl}T^{ijkl}=
\frac{1}{2}(1+\frac{1}{2}\d^{ij}) \\
\frac{1}{64}\sum_{k,l}T^{kl}T^{kl}=\frac{1}{4} \end{cases}\,.
\end{equation}
The factors of $\frac{1}{2}(1+\frac{1}{2}\d^{ij})$ and $\frac{1}{4}$
equal the normalization ${\mc N}$ of the string states.
Thus the contribution to the ${\bf 9}$ and ${\bf 1}$ is in both cases~\cite{Roiban:2002xr,Gomis:2003kj}
\begin{equation}\label{9-1}
\frac{\m g_2^2\l'}{4\pi^2}\frac{15}{16\pi^2n^2}\,.
\end{equation}
The second case with two fermionic oscillators in the double-string
was not analyzed in~\cite{Roiban:2002xr,Gomis:2003kj}. For example, one
has to evaluate the tensor $t^{MN}_{abcd}$ for spinor indices belonging to different
chiralities of $SO(4)\times SO(4)$. Then $t^{MN}_{abcd}$ is non-zero only if $M$ and $N$ are
not in the same $SO(4)$.
The resulting contribution is the same as
in equation~\eqref{9-1} for the representation $({\bf 4},{\bf 4})_+$.

\subsubsection{Contribution of contact terms}

To have a non-zero contribution from $Q_3^2$ the intermediate states need
to have an odd number of bosonic oscillators and an odd number of fermionic oscillators. Thus
the simplest contribution comes from the impurity conserving channel.
In this case the projector is
\begin{equation*}
\sum_{K,a}
\a_{0(1)}^{\dg\,K}\b_{0(2)}^{\dg\,a}|v\ra\la v|\b_{0(2)}^{a}\a_{0(1)}^{K}
+(1\leftrightarrow 2)
+\sum_{k\in\Zop}\sum_{r,K,a}
\a_{k(r)}^{\dg\,K}\b_{-k(r)}^{\dg\,a}|v\ra\la v|\b_{-k(r)}^{a}\a_{k(r)}^{K}\,.
\end{equation*}
At leading order in $\m$ one finds that
for the bosonic part of the matrix element the zero-modes
contribute only to the antisymmetric representations,
whereas the
non-zero-modes contribute to all representations. For the fermionic part of the matrix
element a simple counting of zero-modes shows that only terms of order ${\mc O}(Y^3)$ and
${\mc O}(Y^5)$ in $v_{MN}(Y)$ can contribute. One also needs to evaluate the tensor
$u^I_{abc\da}$ and the large $\m$ expansion of the fermionic Neumann matrices, which
due to the relation to the bosonic Neumann matrices~\cite{Pankiewicz:2002gs} can be
inferred from the latter. The final result is
\begin{equation}
\frac{1}{2}\frac{\m g_2^2\l'}{4\pi^2}\left(\frac{1}{12}+\frac{35}{32n^2\pi^2}\right)\,,
\end{equation}
for the antisymmetric ${\bf 6}$ and $({\bf 4},{\bf 4})_-$ and
\begin{equation}
\frac{\m g_2^2\l'}{4\pi^2}\left(\frac{1}{12}+\frac{5}{32n^2\pi^2}\right)
\begin{cases}
\frac{1}{2}\bigl(1+\frac{1}{2}\d^{IJ}\bigr)\\
\frac{1}{4}
\end{cases}\,,
\end{equation}
for the ${\bf 1}$, ${\bf 9}$ and $({\bf 4},{\bf 4})_+$.
Summing the contributions of one-loop and contact diagrams
we see that
{\em all} (bosonic) two-impurity irreducible representations of
$SO(4)\times SO(4)$ get the same contribution to the mass-shift from the
impurity-conserving channels
\begin{equation}
\d E_n^{(2)}=\frac{\m g_2^2\l'}{4\pi^2}\left(\frac{1}{12}+\frac{35}{32n^2\pi^2}\right)\,.
\end{equation}
This is in exact agreement with the gauge theory
result of~\cite{Beisert:2002bb,Constable:2002vq}, cf.~\eqref{anom}.

\section{Summary}\label{chapter5}

The realization of BMN that the
Penrose limit of $AdS_5\times S^5$ and the knowledge of
the full string spectrum on the plane wave, allowed
to study AdS/CFT -- albeit in a special limit --
beyond the supergravity approximation, has
ignited a lot of activity. 
The purpose of this work was to give an overview over various developments
that have taken place.

In section~\ref{chapter2} I gave an introduction to the BMN correspondence. Several aspects of this duality were discussed
in some detail both from the string theory as well as the gauge theory point of view. 

Extensions of the BMN duality to less trivial backgrounds have been the topic of section~\ref{chapter3}. 
Having first considered several illustrative examples, 
we studied supersymmetric ${\mathbb Z}_k$ orbifolds of the plane
wave space-time and showed that free string theory in the orbifolded plane wave is dual to a subsector of planar
${\mc N}=2$ $[U(N)]^k$ quiver gauge theory. In particular, we gave an explicit identification of gauge theory operators and
string states both in the untwisted and twisted sectors.
As interesting examples of further aspects of string theory on pp-wave space-times, I discussed
D-branes on the plane wave and string theory on pp-waves with non-constant \RR fluxes and curved transverse spaces.
 
To investigate the BMN correspondence beyond the free string/planar gauge theory level,
string interactions and the non-planar gauge theory sector have to be taken into account.
In section~\ref{chapter4} string interactions in the plane wave background were studied in the framework of light-cone
string field theory. At first order in the string coupling, interactions in this setup are encoded in a cubic vertex. 
We analyzed in detail the construction of this vertex as well as the dynamical supercharges
and presented their complete expressions both in the oscillator as well as the continuum
basis. We proved that these satisfy the plane wave superalgebra to first order in the
string coupling. In the process, several results that had been known in flat space light-cone string field 
theory, e.g.\ a factorization theorem for the bosonic Neumann matrices, were generalized to the plane wave 
space-time. We used the vertex and supercharges to compute the leading order mass shift of certain string states
in a truncation to the impurity-conserving channel. The result exactly agreed with the leading non-planar correction to 
the anomalous dimension of the dual operators in ${\mc N}=4$ SYM. 

There are a number of interesting problems we have encountered:
for example, it would be nice to extend the computation of the mass shift
for the simplest string states in section~\ref{ch6} beyond the contribution of
the impurity-conserving channel. As I have explained, in the large $\m$ limit this
presumably translates to non-perturbative effects in the dual gauge theory.
Indeed, a non-vanishing contribution of order $g_2^2\sqrt{\l'}$ to the anomalous dimension
would only constitute the leading term in a power series in fractional powers of $\l'$;
verifying the presence of such a contribution could eventually lead to better understanding
the nature of the BMN limit in ${\mc N}=4$ SYM.
One should be aware, however, that even a computation of the leading order `stringy'
effect along the lines of section~\ref{ch6} seems unfeasible, as infinitely many
intermediate states have to be taken into account. So the way out seems to be to perform a
full-fledged one-loop/contact term computation. Again, this is difficult, as one has to compute the
inverse of infinite-dimensional matrices (involving e.g.\ the product of two
Neumann matrices) exactly, before taking the large $\m$ limit.
Nevertheless, some progress might be achieved along the lines of~\cite{He:2002zu} using the
techniques developed there.


It is natural to extend the research on light-cone string field theory
to include open strings, i.e.\ D-branes on the plane wave. In particular, as
explained in section~\ref{dbranes}, D$_-$-branes outside the origin preserve dynamical
supercharges which involve certain world-sheet symmetries~\cite{Skenderis:2002wx}.
One way to understand the consistency of these branes in the presence of interactions is to construct the
corresponding cubic open string interaction vertex: for D$_-$ branes at the origin this has been
done in~\cite{Chandrasekhar:2003fq,Stefanski:2003zc}.  In fact, recent analysis of the world-volume supersymmetries 
of M2-branes in the KG space suggests that these additional dynamical supercharges are not respected by string interactions, 
see~\cite{Freedman:2003kb} for details. 
Of course, open/closed string interactions are interesting as well given the
expected duality to the BMN limit of ${\mc N}=4$ SYM coupled to defect conformal field
theories. Here studies have been initiated in~\cite{Gomis:2003kb}.

As we have seen, the light-cone GS action is well-suited to obtain the spectrum of string
theories in simple backgrounds with \RR flux. Although the construction of the cubic
interaction vertex is technically quite involved, it is a viable possibility to
study simple tree- and -- at least in the approximation described in section~\ref{ch6} --
one-loop interactions.
However, as discussed in~\cite{Berkovits:2002zv},
even for studying higher point tree-amplitudes in flat space
this approach is not useful, as the vertex explicitly depends on the interaction point.
Moreover, it is difficult to describe physical states with vanishing $p^+$ in the
light-cone formalism. These caveats become even
more problematic for backgrounds without the full Lorentz isometry, such
as the plane wave.
It appears to be a worthwhile prospect to use the $U(4)$ formalism as advocated
in~\cite{Berkovits:2002vn,Berkovits:2002rd} to overcome some of these drawbacks. In this
approach strings on the plane wave are described by an exact {\em interacting} $N=2$ superconformal
field theory and standard CFT techniques may be used
for computations. One can also naturally describe strings in the more general pp-wave
geometries of section~\ref{massive} in this setup, which makes this approach
potentially even more interesting.


\section*{Acknowledgments}

I would like to thank N. Kim, S.-J. Rey, B. Stefa\'nski and S. Theisen for the good collaboration on which part 
of this work is based. I am grateful to G. Arutyunov, N. Beisert and J. Plefka for useful discussions.   
In particular I would like to thank S. Theisen for encouragement and advice during my thesis and A. Klemm for his help and support.   
This work was supported by GIF, the German-Israeli foundation for Scientific Research, and 
from the European Community's Human Potential Programme under contracts HPRN-CT-2000-00131 and HPMT-CT-2001-00296.

\appendix

\section{The kinematical part of the vertex}\label{appA}

\subsection{The Delta-functional}\label{A1}

The precise definition of the Delta-functional is
\begin{equation}\label{Delta}
\D^8\bigl[\sum_{r=1}^3p_r(\s)]\equiv \prod_{m\ge0}\d^8\left(\int_{-\pi|\a_3|}^{\pi|\a_3|}d\s\,e^{im\s/|\a_3|}\sum_{r=1}^3p_r(\s)\right)\,.
\end{equation}
The pure zero-mode contribution decouples from the Delta-functional, so
\begin{equation}\label{AE2}
\D^8\bigl[\sum_{r=1}^3p_r(\s)\bigr]=\d^8\bigl(\sum_{r=1}^3p_{0(r)}\bigr)\prod_{m=1}^{\infty}
\d^8\left(\int_{-\pi|\a_3|}^{\pi|\a_3|}\,d\s e^{im\s/|\a_3|}\sum_{r=1}^3p_r(\s)\right)\,.
\end{equation}
We need the following integrals for $m>0$, $n\ge0$ ($\b\equiv\a_1/\a_3$)
\begin{equation}
\begin{split}
\frac{1}{\pi\a_1}\int_{-\pi\a_1}^{\pi\a_1}\,d\s\cos\frac{m\s}{\a_3}\cos\frac{n\s}{\a_1}
& =(-1)^n\frac{2m\b}{\pi}\frac{\sin m\pi\b}{m^2\b^2-n^2}\equiv X^{(1)}_{mn}\,,\\
\frac{1}{\pi\a_1}\int_{-\pi\a_1}^{\pi\a_1}\,d\s\sin\frac{m\s}{\a_3}\sin\frac{n\s}{\a_1}
& =\frac{n}{m\b}X^{(1)}_{mn}\,,
\end{split}
\end{equation}
and
\begin{equation}
\begin{split}
\frac{2}{\pi\a_2}\int_{\pi\a_1}^{-\pi\a_3}\!\!\!d\s\cos\frac{m\s}{\a_3}\cos\frac{n}{\a_2}(\s-\pi\a_1)
& =\frac{2m(\b+1)}{\pi}\frac{\sin m\pi\b}{m^2(\b+1)^2-n^2}\equiv X^{(2)}_{mn},\\
\frac{2}{\pi\a_2}\int_{\pi\a_1}^{-\pi\a_3}\!\!\!d\s\sin\frac{m\s}{\a_3}\sin\frac{n}{\a_2}(\s-\pi\a_1)
& =-\frac{n}{m(\b+1)}X^{(2)}_{mn}\,.
\end{split}
\end{equation}
Then the delta-functions over the non-zero-modes contribute
\begin{equation}
\prod_{m=1}^{\infty}\d^8\left(\frac{1}{\sqrt{2}}\sum_{r=1}^3\left[\sum_{n=1}^{\infty}X^{(r)}_{mn}
\bigl(p_{n(r)}-i\frac{\a_3}{\a_r}\frac{n}{m}p_{-n(r)}\bigr)+\frac{1}{\sqrt{2}}X^{(r)}_{m0}p_{0(r)}\right]\right)
\end{equation}
and I have defined $X^{(3)}_{mn}=\d_{mn}$. We see that negative and non-negative modes decouple from each other.
We can extend the range of $m,n$ to ${\mathbb Z}$ by introducing
\begin{equation}\label{xr}
X^{(r)}_{mn}\equiv\begin{cases}
X^{(r)}_{mn}\,,\qquad m\,,n>0 \\
\frac{\a_3}{\a_r}\frac{n}{m}X^{(r)}_{-m,-n}\,,\qquad m\,,n<0 \\
\frac{1}{\sqrt{2}}X^{(r)}_{m0}\,,\qquad m>0\,,r\in\{1,2\}\\ 1\,,\qquad m=0=n \\ 0\,,\qquad\text{otherwise}
\end{cases}
\end{equation}
Then the Delta-functional takes the form
\begin{equation}\label{AE7}
\D\bigl[\sum_{r=1}^3p_r(\s)\bigr]\sim\prod_{m\in{\mathbb Z}}
\d\left(\sum_{r=1}^3\sum_{n\in{\mathbb Z}}X^{(r)}_{mn}p_{n(r)}\right)\,.
\end{equation}
Here I ignored factors of $\sqrt{2}$ which can be absorbed in the measure.
It is convenient to introduce the matrices for $m$, $n>0$
\begin{equation}\label{Ar}
\begin{split}
C_{mn} & = m\d_{mn}\,,\\
A^{(1)}_{mn} & = (-1)^n\frac{2\sqrt{mn}\b}{\pi}\frac{\sin m\pi\b}{m^2\b^2-n^2}
=\bigl(C^{-1/2}X^{(1)}C^{1/2}\bigr)_{mn}\,,\\
A^{(2)}_{mn} & = \frac{2\sqrt{mn}(\b+1)}{\pi}\frac{\sin m\pi\b}{m^2(\b+1)^2-n^2}
=\bigl(C^{-1/2}X^{(2)}C^{1/2}\bigr)_{mn}\,,\\
A^{(3)}_{mn} & = \d_{mn}
\end{split}
\end{equation}
and the vector $(m>0)$
\begin{equation}\label{B}
B_m=-\frac{2}{\pi}\frac{\a_3}{\a_1\a_2}m^{-3/2}\sin m\pi\b
\end{equation}
related to $X^{(r)}_{m0}$ via
\begin{equation}
X^{(r)}_{m0}=-\e^{rs}\a_s\bigl(C^{1/2}B\bigr)_m\,.
\end{equation}
These satisfy the following very useful identities~\cite{Green:1983tc}
\begin{equation}\label{id1}
-\frac{\a_3}{\a_r}\,CA^{(r)T}C^{-1}A^{(s)}=\d^{rs}{\bf 1}\,,\quad
-\frac{\a_r}{\a_3}\,C^{-1}A^{(r)T}CA^{(s)}=\d^{rs}{\bf 1}\,,\quad
A^{(r)T}CB=0
\end{equation}
valid for $r,s\in\{1,2\}$ and
\begin{equation}\label{id2}
\sum_{r=1}^3\frac{1}{\a_r}A^{(r)}CA^{(r)\,T}=0\,,\qquad \sum_{r=1}^3\a_rA^{(r)}C^{-1}A^{(r)\,T}=\frac{\a}{2}BB^T\,.
\end{equation}
In terms of the big matrices $X^{(r)}_{mn}$, $m,n\in{\mathbb Z}$ the relations~\eqref{id1} and~\eqref{id2} can be
written in the compact form
\begin{equation}
\bigl(X^{(r)T}X^{(s)}\bigr)_{mn}=-\frac{\a_3}{\a_r}\d^{rs}\d_{mn}\,,\quad r,s\in\{1,2\}\,,\quad
\sum_{r=1}^3\a_r\bigl(X^{(r)}X^{(r)T}\bigr)_{mn}=0\,.
\end{equation}

\subsection{Structure of the bosonic Neumann matrices}\label{A3}

Evaluating the Gaussian integrals in equation~\eqref{ea} one finds the following
expressions for the bosonic Neumann matrices~\cite{Spradlin:2002ar}
\begin{equation}
\bar{N}^{rs}_{mn}=\d^{rs}\d_{mn}-2\bigl(C_{(r)}^{1/2}X^{(r)T}\G_a^{-1}X^{(s)}C_{(s)}^{1/2}\bigr)_{mn}\,,
\qquad \G_a=\sum_{r=1}^3 X^{(r)}C_{(r)}X^{(r)T}\,.
\end{equation}
From the structure of the $X^{(r)}$ given in equation~\eqref{xr} it follows that $\G_a$ is
block diagonal and using the identities~\eqref{id2} one can write the blocks as~\cite{Spradlin:2002ar}
\begin{equation}
\bigl[\G_a\bigr]_{mn}=
\begin{cases}
\bigl(C^{1/2}\G C^{1/2}\bigr)_{mn}\,, & m\,,n>0\,, \\
-2\m\a_3\,, & m=0=n\,, \\
\bigl(C^{1/2}\G_-C^{1/2}\bigr)_{-m,-n}\,, & m\,,n>0\,,
\end{cases}
\end{equation}
where
\begin{equation}
\G_-\equiv\sum_{r=1}^3 A_-^{(r)}U^{-1}_{(r)}A_-^{(r)\,T}\,,\qquad A_-^{(r)}=\frac{\a_3}{\a_r}C^{-1}A^{(r)}C\,.
\end{equation}
The matrix $\G$ (which reduces to the flat space $\G$ of~\cite{Green:1983tc,Green:1983hw} for $\m\to0$) exists and is invertible, whereas
$\G_-$ is ill-defined since the above sum is divergent. Nevertheless it is possible to derive a well-defined identity for
$\G_-^{-1}$~\cite{Spradlin:2002ar}
\begin{equation}\label{gamma-}
\G_-^{-1}=U_{(3)}\bigl(1-\G^{-1}U_{(3)}\bigr)\,.
\end{equation}
Since $\G_-^{-1}$ is related to $\G^{-1}$ it is possible to relate the Neumann matrices with positive and negative indices. This
results in equation~\eqref{neg}.
To derive the factorization theorem~\eqref{nnpp}~\cite{Schwarz:2002bc,Pankiewicz:2002gs} introduce
\begin{equation}
\Upsilon\equiv\sum_{r=1}^3A^{(r)}U_{(r)}^{-1}A^{(r)\,T}=\G+\m\a BB^T\,,
\end{equation}
where I have used equation~\eqref{id2}. Its inverse is related to $\G^{-1}$ by
\begin{equation}\label{ups1}
\Upsilon^{-1}=\G^{-1}-\frac{\m\a}{1-4\m\a K}\left(\G^{-1}B\right)\left(\G^{-1}B\right)^T\,.
\end{equation}
For $r$, $s\in\{1,2\}$ one can derive the following relations
\begin{align}
A^{(r)\,T}C^{-1}U_{(3)}\G^{-1} & =
A^{(r)\,T}C^{-1}+\frac{\a_r}{\a_3}C^{-1}U_{(r)}A^{(r)\,T}\G^{-1}\,,\\
\Upsilon^{-1}U_{(3)}^{-1}C^{-1}A^{(r)} & =
C^{-1}A^{(r)}+\frac{\a_r}{\a_3}\Upsilon^{-1}A^{(r)}U_{(r)}^{-1}C^{-1}\,,\\
2C^{-1} & = \G^{-1}U_{(3)}C^{-1}+C^{-1}U_{(3)}\G^{-1}+ \Upsilon^{-1}U_{(3)}^{-1}C^{-1}
+C^{-1}U_{(3)}^{-1}\Upsilon^{-1}\nonumber\\
&-\a_1\a_2\Upsilon^{-1}B\left(\G^{-1}B\right)^T\,.
\end{align}
Using equations~\eqref{ups1} and~\eqref{id1} results in the factorization theorem~\eqref{nnpp}.

\subsection{The kinematical constraints at ${\mc O}(g_{\text{s}})$}\label{A4}

\subsubsection{The bosonic part}

The bosonic constraints the exponential part of the vertex has to satisfy are
\begin{equation}
\sum_{r=1}^3\sum_{n\in{\mathbb Z}}X_{mn}^{(r)}p_{n(r)}|V\ra=0\,,\qquad
\sum_{r=1}^3\sum_{n\in{\mathbb Z}}\a_rX_{mn}^{(r)}x_{n(r)}|V\ra=0\,.
\end{equation}
For $m=0$ this leads to
\begin{equation}
\sum_{r=1}^3p_{0(r)}|V\ra=0\,,\qquad \sum_{r=1}^3\a_rx_{0(r)}|V\ra=0\,.
\end{equation}
Substituting \eqref{xp} and commuting the annihilation operators through the exponential this requires
\begin{align}
&\sum_{r,s=1}^3\sqrt{|\a_r|}\bigl[\bigl(\bar{N}^{rs}_{00}+\d^{rs}\bigr)a_{0(s)}^{\dg}
+\sum_{n=1}^{\infty}\bar{N}^{rs}_{0n}a_{n(s)}^{\dg}\bigr]|V\ra=0\,,\\
&\sum_{r,s=1}^3e(\a_r)\sqrt{|\a_r|}\bigl[\bigl(\bar{N}^{rs}_{00}-\d^{rs}\bigr)a_{0(s)}^{\dg}
+\sum_{n=1}^{\infty}\bar{N}^{rs}_{0n}a_{n(s)}^{\dg}\bigr]|V\ra=0\,.
\end{align}
Using the expressions given for $\bar{N}^{rs}_{0n}$ and $\bar{N}^{rs}_{00}$ in \eqref{m0}, \eqref{00a}
and \eqref{00b} one can check that the above equations are satisfied trivially,
i.e.\ without further use of additional non-trivial identities.
For $m>0$ we find the following constraints
\begin{align}
\label{bos1}
B+\sum_{r=1}^3A^{(r)}C^{1/2}U_{(r)}\bar{N}^r & = 0\,,\\
\label{bos2}
A^{(s)}C_{(s)}^{-1/2}U_{(s)}^{-1}+\sum_{r=1}^3A^{(r)}C_{(r)}^{-1/2}U_{(r)}C^{1/2}\bar{N}^{rs}C^{-1/2} & = 0\,,\\
\label{bos3}
-\a_sA^{(s)}C_{(s)}^{-1/2}+\sum_{r=1}^3\a_rA^{(r)}C_{(r)}^{-1/2}C^{-1/2}\bar{N}^{rs}C^{1/2} & =
\a B\bigl[C_{(s)}^{1/2}C^{1/2}\bar{N}^s\bigr]^T\,.
\end{align}
Equation~\eqref{bos1} is satisfied by the definition for $\bar{N}^r$. Equations~\eqref{bos2}
and~\eqref{bos3} are proved by substituting the expression for $\bar{N}^{rs}$ given in~\eqref{mn}.
For $m<0$ there is one additional constraint
\begin{equation}
A^{(s)}C_{(s)}^{-1/2}U_{(s)}^{-1}-
\a_s\sum_{r=1}^3\frac{1}{\a_r}A^{(r)}C_{(r)}^{1/2}U_{(r)}C^{1/2}\bar{N}^{rs}C^{-1/2}C_{(s)}^{-1}=0
\end{equation}
which is verified by subtracting it from equation~\eqref{bos1} and using~\eqref{nnpp}.
Here I used the identity
\begin{equation}
\sum_{r=1}^3\a_rA^{(r)}C^{-1/2}\bar{N}^r=2\a K B\,.
\end{equation}

\subsubsection{The fermionic part}

The fermionic constraints the exponential part of the vertex has to satisfy are
\begin{equation}
\sum_{r=1}^3\sum_{n\in{\mathbb Z}}X_{mn}^{(r)}\l_{n(r)}|V\ra=0\,,\qquad
\sum_{r=1}^3\sum_{n\in{\mathbb Z}}\a_rX_{mn}^{(r)}\vt_{n(r)}|V\ra=0\,.
\end{equation}
For $m=0$ this leads to
\begin{equation}
\sum_{r=1}^3\l_{0(r)}|V\ra=0\,,\qquad \sum_{r=1}^3\a_r\vt_{0(r)}|V\ra=0\,.
\end{equation}
These equations are satisfied by construction of the zero-mode part of $|V\ra$.
For $m>0$ we get
\begin{align}
\label{ferm1}
B+\sum_{r=1}^3e(\a_r)\sqrt{|\a_r|}A^{(r)}C_{(r)}^{-1/2}P_{(r)}Q^r & = 0\,,\\
\label{ferm2}
\sqrt{|\a_s|}A^{(s)}C_{(s)}^{-1/2}P_{(s)}^{-1}+
\sum_{r=1}^3e(\a_r)\sqrt{|\a_r|}A^{(r)}C_{(r)}^{-1/2}P_{(r)}Q^{rs} & = 0\,,\\
\label{ferm3}
-\sqrt{|\a_s|}A^{(s)}C_{(s)}^{-1/2}P_{(s)}+\frac{1}{\a_s}
\sum_{r=1}^3|\a_r|^{3/2}A^{(r)}C_{(r)}^{-1/2}P_{(r)}^{-1}C^{-1}Q^{rs}C & = \a BQ^{s\,T}\,,
\end{align}
whereas for $m<0$ the constraints are
\begin{align}
\label{ferm4}
\sum_{r=1}^3\frac{1}{\sqrt{|\a_r|}}A^{(r)}CC_{(r)}^{-1/2}P_{(r)}^{-1}Q^r & = 0\,,\\
\label{ferm5}
A^{(s)}CC_{(s)}^{-1/2}P_{(s)}-e(\a_s)\sqrt{|\a_s|}\sum_{r=1}^3\frac{1}{\sqrt{|\a_r|}}
A^{(r)}CC_{(r)}^{-1/2}P_{(r)}^{-1}Q^{rs} & = 0\,.
\end{align}
Now equations \eqref{ferm1} and \eqref{ferm4} uniquely determine
\begin{equation}
Q^r=\frac{e(\a_r)}{\sqrt{|\a_r|}}(1-4\m\a K)^{-1}(1-2\m\a K(1+\Pi))P_{(r)}C_{(r)}^{1/2}C^{1/2}\bar{N}^r\,.
\end{equation}
Furthermore comparing equations \eqref{ferm2} and \eqref{bos2} we see that
\begin{equation}
Q^{rs}=e(\a_r)\sqrt{\left|\frac{\a_s}{\a_r}\right|}
P_{(r)}^{-1}U_{(r)}C^{1/2}\bar{N}^{rs}C^{-1/2}U_{(s)}P_{(s)}^{-1}
\end{equation}
solves \eqref{ferm2}. Using
\begin{equation}
P^{-2}_{(r)}U_{(r)}\bar{N}^{rs}U_{(s)}P_{(s)}^{-2}=\bar{N}^{rs}
+\m\a(1-4\m\a K)^{-1}C_{(r)}^{1/2}\bar{N}^r\bigl[C_{(s)}^{1/2}\bar{N}^s\bigr]^T(1-\Pi)
\end{equation}
establishes \eqref{ferm3} by virtue of \eqref{bos3}. Finally, equation \eqref{ferm5} is satisfied due to the identity
\begin{equation}
A^{(s)}C_{(s)}^{-1/2}-\a_s\sum_{r=1}^3\frac{1}{\a_r}A^{(r)}C_{(r)}^{-1/2}C^{3/2}\bar{N}^{rs}C^{-3/2}=0
\end{equation}
which can be proved using the expression for $\bar{N}^{rs}$ given in \eqref{mn}. This
concludes the determination of the fermionic contribution to the kinematical part of the vertex.

\section{The dynamical constraints}\label{appB}

\subsection{More detailed computations}\label{B1}

Here I give the details leading to equations~\eqref{qk},~\eqref{qy} and~\eqref{qv}.
The following identities prove very useful ($\a_3\Theta\equiv\vt_{0(1)}-\vt_{0(2)}$)
\begin{align}
\label{r}
\mathbb{R}|V\ra & =i\sqrt{\a'}\left[2K\sqrt{\a'}
\left(\mathbb{P}-i\frac{\m\a}{\a'}\mathbb{R}\right)+
\sum_{r,n>0}C_{n(r)}^{1/2}\bar{N}^r_na^{\dg}_{n(r)}\right]|V\ra\,,\\
\label{theta}
\Theta|V\ra & =-\sqrt{2}\sum_{r,n}Q^r_nb_{-n(r)}^{\dg}|V\ra\,.
\end{align}
Using the mode expansions of $Q^-_{(r)}$, $\bar{Q}^-_{(r)}$,
$K_0+K_+$, $K_-$ and $Y$ one finds
\begin{align}
\sum_{r=1}^3\{Q^-_{(r)},Y\} & =
-\g\sum_{r=1}^3\frac{1}{\sqrt{|\a_r|}}\sum_{n=1}^{\infty}\bigl[P_{(r)}C^{1/2}G_{(r)}\bigr]_na_{-n(r)}^{\dg}\,,\\
\sum_{r=1}^3\{\bar{Q}^-_{(r)},Y\} & =
(1-4\m\a K)^{-1/2}(1-2\m\a K(1-\Pi))\left(\mathbb{P}\g-i\frac{\m\a}{\a'}\mathbb{R}\g\Pi\right)
\nonumber\\
&+\g\sum_{r=1}^3\frac{1}{\sqrt{|\a_r|}}\sum_{n=1}^{\infty}\bigl[P_{(r)}^{-1}C^{1/2}G_{(r)}\bigr]_na_{n(r)}^{\dg}\,,
\end{align}
\begin{align}
\sum_{r=1}^3[Q^-_{(r)},K_0+K_+] & = \m\g(1+\Pi)(1-4\m\a K)^{1/2}\sqrt{\frac{2}{\a'}}\L
\nonumber\\
&+\g\sum_{r=1}^3\frac{e(\a_r)}{\sqrt{|\a_r|}}\sum_{n=1}^{\infty}\bigl[P_{(r)}^{-1}C^{1/2}F_{(r)}\bigr]_nb_{n(r)}^{\dg}\,,\\
\sum_{r=1}^3[Q^-_{(r)},K_-] & = i\g\sum_{r=1}^3\frac{e(\a_r)}{\sqrt{|\a_r|}}
\sum_{n=1}^{\infty}\bigl[P_{(r)}^{-1}C^{1/2}U_{(r)}F_{(r)}\bigr]_nb_{-n(r)}^{\dg}\,,\\
\sum_{r=1}^3[\bar{Q}^-_{(r)},K_0+K_+] & = -\frac{\m\a}{\sqrt{2\a'}}\g(1-\Pi)(1-4\m\a K)^{1/2}\Theta
\nonumber\\
&+\g\sum_{r=1}^3\frac{e(\a_r)}{\sqrt{|\a_r|}}\sum_{n=1}^{\infty}\bigl[P_{(r)}C^{1/2}F_{(r)}\bigr]_nb_{-n(r)}^{\dg}\,,\\
\sum_{r=1}^3[\bar{Q}^-_{(r)},K_-] & = -i\g\sum_{r=1}^3\frac{e(\a_r)}{\sqrt{|\a_r|}}
\sum_{n=1}^{\infty}\bigl[P_{(r)}C^{1/2}U_{(r)}F_{(r)}\bigr]_nb_{n(r)}^{\dg}\,.
\end{align}
Using~\eqref{gf},~\eqref{r} and~\eqref{theta} leads to
equations~\eqref{qk} and~\eqref{qy}. The action of the supercharges on $|V\ra$ given in equation~\eqref{qv} can
be proven similarly. One needs
\begin{equation}
\begin{split}
\bar{N}^{rs}_{nm}+e(\a_s)\left(\frac{m}{n}\left|\frac{\a_r}{\a_s}\right|\right)^{3/2}
P_{n(r)}P_{m(s)}Q^{rs}_{nm} & = -\frac{\a}{\a_s}(1-4\m\a K)^{-1}\times\\
& \times\bigl[C_{(r)}^{1/2}\bar{N}^r\bigr]_n\bigl[U_{(s)}^{-1}C_{(s)}^{1/2}C\bar{N}^s\bigr]_m\,,\\
\bar{N}^{rs}_{-n,-m}+e(\a_r)\left(\frac{m}{n}\left|\frac{\a_r}{\a_s}\right|\right)^{1/2}P_{n(r)}P_{m(s)}Q^{rs}_{nm} &
= 0\,,
\end{split}
\end{equation}
\begin{equation}
\begin{split}
\bar{N}^{rs}_{nm}-e(\a_r)\left(\frac{m}{n}\left|\frac{\a_r}{\a_s}\right|\right)^{1/2}P_{n(r)}^{-1}P_{m(s)}^{-1}Q^{rs}_{nm} & =
-\m\a(1-4\m\a K)^{-1}\times \\
&\times(1-\Pi)\bigl[C_{(r)}^{1/2}\bar{N}^r\bigr]_n\bigl[C_{(s)}^{1/2}\bar{N}^s\bigr]_m,\\
\bar{N}^{rs}_{-n,-m}-e(\a_s)\left(\frac{m}{n}\left|\frac{\a_r}{\a_s}\right|\right)^{3/2}P_{n(r)}^{-1}P_{m(s)}^{-1}Q^{rs}_{nm} & =
\frac{\a}{\a_s}(1-4\m\a K)^{-1}\times\\
&\times\bigl[P^{-2}_{(r)}C_{(r)}^{1/2}\bar{N}^r\bigr]_n\bigl[C_{(s)}^{1/2}C\bar{N}^s\bigr]_m
\end{split}
\end{equation}
which follow from~\eqref{nnpp} and~\eqref{qmn}.

\subsection{Proof of the dynamical constraints}\label{B2}

In this appendix I prove that
\begin{align}
\label{1}
\g^I_{a(\dot{a}}\bigl[\Pi\bar{D}\bigr]^as^I_{\dot{b})} &=0\,,\\
\label{2}
\g^I_{a(\dot{a}}\bigl[\Pi D\bigr]^a\wt{s}^I_{\dot{b})} & =0\,,\\
\label{3} \Bigl(\g^I_{a\dot{a}}\bar{D}_b\wt{s}^I_{\dot{b}}+
\g^I_{a\dot{b}}D_bs^I_{\dot{a}}\Bigr)(1-\Pi)^{ab} & =0\,.
\end{align}
Equations~\eqref{1} and~\eqref{2} are equivalent to
\begin{align}
\label{1e}
\Bigl(\g^I_{a(\da}Y_bs^I_{1\,\db)}+\frac{\a}{\a'}\g^I_{a(\da}\frac{\p}{\p Y^b}s^I_{2\,\db)}\Bigr)\Pi^{ab} & =0\,,\\
\label{2e}
\Bigl(\g^I_{a(\da}Y_bs^I_{2\,\db)}-\frac{\a}{\a'}\g^I_{a(\da}\frac{\p}{\p Y^b}s^I_{1\,\db)}\Bigr)\Pi^{ab} & =0\,,
\end{align}
The first equation has terms of order ${\mc O}(Y^2)$ and ${\mc O}(Y^6)$, whereas the second one
has terms of order ${\mc O}(Y^0)$, ${\mc O}(Y^4)$ and ${\mc O}(Y^8)$.
There are two contributions to the order ${\mc O}(Y^2)$ in equation~\eqref{1e} , both vanish separately. The first one is
\begin{equation}
\g^I_{a(\da}Y_bs_{1\,\db)}^I\Pi^{ab}=
2\g^I_{a(\da}\g^I_{c\db)}Y^bY^c\Pi^{ab}=
-2\d_{\da\db}\Pi_{ab}Y^aY^b=0\,,
\end{equation}
whereas the second one is
\begin{align}
&\frac{\a}{\a'}\g^I_{a(\da}\frac{\p}{\p Y^b}s_{2\,\db)}^I\Pi^{ab}=
-\g^I_{a(\da}u^I_{bcd\db)}Y^cY^d\Pi^{ab}=\nonumber\\
&\frac{1}{16}\left(\g^{IJ}\g^{KL}\right)_{(\da\db)}
\g^{IJ}_{a[b}\g^{KL}_{cd]}\Pi^{ab}Y^cY^d=
\frac{1}{24}\left(\g^{IJ}\g^{KL}\right)_{(\da\db)}
\left(\g^{IJ}\Pi\g^{KL}\right)_{cd}Y^cY^d=0\,.
\end{align}
Here I have used equations~\eqref{gu} and~\eqref{klpij}. From the Fourier
identities~\cite{Green:1983hw}
\begin{equation}\label{fourier}
\begin{split}
s_{1\,\da}(\phi) & = \left(\frac{\a}{\a'}\right)^4\int
d^8Y\,s_{2\,\da}^I(Y)
e^{\frac{\a'}{\a}\phi Y}\,,\\
s_{2\,\da}(\phi) & = \left(\frac{\a}{\a'}\right)^4\int
d^8Y\,s_{1\,\da}^I(Y) e^{\frac{\a'}{\a}\phi Y}\,,
\end{split}
\end{equation}
it follows that the terms of order ${\mc O}(Y^6)$ vanish as well. This proves equation \eqref{1e}.
The ${\mc O}(Y^0)$ term in equation~\eqref{2e} is
\begin{equation}
\g^I_{a(\da}\g^I_{b\db)}\Pi^{ab}=\d_{\da\db}\tr(\Pi)=0\,,
\end{equation}
and the order ${\mc O}(Y^8)$ term vanishes by~\eqref{fourier}. The terms of order ${\mc O}(Y^4)$ in equation~\eqref{2e} are
\begin{align}
&{\Pi^a}_b\g^I_{a(\da}u^I_{cde\db)}
\left(Y^bY^cY^dY^e+\frac{1}{24}{\e^{cdeb}}_{ghij}Y^gY^hY^iY^j\right)\nonumber \\
&=-\frac{1}{16}{\Pi^a}_b\left(\g^{IJ}\g^{KL}\right)_{(\dot{a}\dot{b})}
\g^{IJ}_{a[c}\g^{KL}_{de]}\left(Y^bY^cY^dY^e+\frac{1}{24}{\e^{cdeb}}_{ghij}Y^gY^hY^iY^j\right)
\nonumber\\
&=-\frac{1}{16}{\Pi^a}_b\left(\g^{IJKL}_{\da\db}-2\d_{\da\db}\d^{IK}\d^{JL}\right)
\g^{IJ}_{a[c}\g^{KL}_{de]}\left(Y^bY^cY^dY^e\right.\nonumber\\
&\left.+\frac{1}{24}{\e^{cdeb}}_{ghij}Y^gY^hY^iY^j\right)
\nonumber\\
&=-\frac{1}{16}{\Pi^a}_b\g^{IJKL}_{\da\db}t^{IJKL}_{acde}
\left(Y^bY^cY^dY^e+\frac{1}{24}{\e^{cdeb}}_{ghij}Y^gY^hY^iY^j\right)=0\,.
\end{align}
In the last step I used that $\Pi$ is symmetric and traceless and
\begin{equation}
t^{IJKL}_{abcd}=-\frac{1}{24}{\e_{abcd}}^{efgh}t^{IJKL}_{efgh}\,.
\end{equation}
This proves equation~\eqref{2e}. Finally, equation~\eqref{3} is equivalent to
\begin{align}
\label{sym3} &
\Bigl(\g^I_{a(\da}Y_bs_{1\,\db)}^I-\frac{\a}{\a'}\g^I_{a(\da}\frac{\p}{\p
Y^b}s_{2\,\db)}^I
\Bigr)(1-\Pi)^{ab}=0\,,\\
\label{asym3} &
\Bigl(\g^I_{a[\da}Y_bs_{2\,\db]}^I+\frac{\a}{\a'}\g^I_{a[\da}\frac{\p}{\p
Y^b}s_{1\,\db]}^I \Bigr)(1-\Pi)^{ab}=0\,.
\end{align}
The first equation is symmetric in $\da$, $\db$ and contains terms of order ${\mc O}(Y^2)$ and ${\mc O}(Y^6)$. These vanish for the
same reason as those in equation~\eqref{1e}. The second equation is antisymmetric in $\da$, $\db$ and contains terms of order
${\mc O}(Y^0)$, ${\mc O}(Y^4)$ and ${\mc O}(Y^8)$. The ${\mc O}(Y^0)$ contribution to equation~\eqref{asym3} is
\begin{equation}
\g^I_{a[\da}\g^I_{b\db]}(1-\Pi)^{ab}=\frac{1}{4}\g^{IJ}_{\da\db}\g^I_{ab}(1-\Pi)^{ab}=0\,.
\end{equation}
From equation~\eqref{fourier} it follows that the term of order ${\mc O}(Y^8)$ vanishes as well. Finally, there are two
contributions to the terms of order ${\mc O}(Y^4)$, both of them vanish separately. The first one is
\begin{align}
&\frac{\a}{\a'}\g^I_{a[\da}Y_bs_{2\,\db]}^I(1-\Pi)^{ab}=
-\frac{1}{3}\g^I_{a[\da}u^I_{cde\db]}{(1-\Pi)^a}_bY^bY^cY^dY^e=\nonumber\\
&\frac{1}{12}\left(\g^{IJ}_{\da\db}\d_{a[c}\g^{IJ}_{de]}
+\frac{1}{4}\left(\g^{IJ}\g^{KL}\right)_{[\da\db]}\g^{IJ}_{a[c}\g^{KL}_{de]}\right)
{(1-\Pi)^a}_bY^bY^cY^dY^e=\nonumber\\
&\frac{1}{12}\g^{IJ}_{\da\db}\g^{IK}_{a[c}\g^{KJ}_{de]}{(1-\Pi)^a}_bY^bY^cY^dY^e=
\frac{1}{6}\g^{IJ}_{\da\db}\g^{IJ}_{bc}(1-\Pi)_{de}Y^bY^cY^dY^e=0\,.
\end{align}
In the last step I have used equation~\eqref{ij}. The second contribution of order ${\mc O}(Y^4)$ then vanishes by
equation~\eqref{fourier}. This concludes the proof of equation~\eqref{asym3}.

\noindent Apart from symmetry and tracelessness of $\Pi$ I have used the following identities
\begin{align}
\g^{IJ}_{ab}&=-\g^{IJ}_{ba}\,,\\
\g^I_{a\da}\g^I_{b\db}&=\d_{ab}\d_{\da\db}+\frac{1}{4}\g^{IJ}_{ab}\g^{IJ}_{\da\db}\,,\\
\left(\g^{IJ}\g^{KL}\right)_{ab}&=\g^{IJKL}_{ab}+\d^{IL}\g^{JK}_{ab}+\d^{JK}\g^{IL}_{ab}
\nonumber \\
&-\d^{IK}\g^{JL}_{ab}-\d^{JL}\g^{IK}_{ab}+\left(\d^{JK}\d^{IL}-\d^{JL}\d^{IK}\right)\d_{ab}
\,,\\
\label{gu}
\g^I_{a\da}u^I_{bcd\db}&=-\frac{1}{4}\g^{IJ}_{\da\db}\d_{a[b}\g^{IJ}_{cd]}
-\frac{1}{16}\left(\g^{IJ}\g^{KL}\right)_{\da\db}\g^{IJ}_{a[b}\g^{KL}_{cd]}\,,\\
\label{ij}
\g^{IK}_{a[b}\g^{JK}_{cd]} & =t^{IJ}_{abcd}-2\d_{a[b}\g^{IJ}_{cd]}\,,\\
\label{ijij}
\g^{IJ}_{ab}\g^{IJ}_{cd} & = 8\bigl(\d_{ac}\d_{bd}-\d_{ad}\d_{bc}\bigr)\,,\\
\label{klpij}
\gamma^{IJKL}_{\da\db}\left(\g^{KL}\Pi\g^{IJ}\right)_{[ab]}&=0 \,.
\end{align}

\subsection{$\{Q,\wt{Q}\}$ at order ${\mc O}(g_{\text{s}})$}\label{B3}

Here I demonstrate that equation~\eqref{dyn3} leads to the constraints~\eqref{qq1}-\eqref{ppwave2} given in section~\ref{sec4}.
To this end, I adopt a trick introduced in~\cite{Green:1983hw}. Namely, associate
the world-sheet coordinate dependence with the oscillators as
\begin{equation}
\begin{pmatrix} a_{n(r)} \\ a_{-n(r)} \end{pmatrix}
\longrightarrow e^{-i\o_{n(r)}\t/\a_r}
\begin{pmatrix}
\cos\frac{n\s_r}{\a_r} & -\sin\frac{n\s_r}{\a_r} \\
\sin\frac{n\s_r}{\a_r} & \cos\frac{n\s_r}{\a_r}
\end{pmatrix}
\begin{pmatrix} a_{n(r)} \\ a_{-n(r)} \end{pmatrix}\,,
\end{equation}
and analogously for the fermionic oscillators. Then integrate the constraint equation~\eqref{dyn3}
over the $\s_r$. In dealing with the resulting expressions one can integrate by parts since the integrand is periodic.
In addition to the identities in equations~\eqref{qy},\footnote{In fact, here we need the analogue of equation~\eqref{qy} with
$K^I\leftrightarrow\wt{K}^I$.}and~\eqref{qv} we have to calculate the commutator of
$\sum_rQ_{(r)}$ with $K^I$
and its tilded counterpart. One gets
\begin{equation}
\begin{split}
\sqrt{2}\eta &\sum_{r=1}^3[Q_{(r)},K^I]\,|V\ra=-2i\g^I
\left[\dot{Y}+Y'+\frac{i}{2}\m(1-\Pi)\left(Y-2Y_0\right)\right]|V\ra\,,\\
\sqrt{2}\bar{\eta} &\sum_{r=1}^3[\wt{Q}_{(r)},\wt{K}^I]\,|V\ra=-2i\g^I
\left[\dot{Y}-Y'+\frac{i}{2}\m(1-\Pi)\left(Y-2Y_0\right)\right]|V\ra\,.
\end{split}
\end{equation}
Here $Y_0$ is the zero-mode part of $Y$, I suppressed the $\t$, $\s_r$ dependence and
\begin{equation}
\dot{Y}\equiv\p_{\t}Y\,,\qquad Y'\equiv\sum_{r=1}^3\p_{\s_r}Y\,.
\end{equation}
The fact that the above equations have a term which only depends on the zero-mode $Y_0$ is important. Combining the various
contributions to equation~\eqref{dyn3}, removing the $\s_r$ derivatives from $Y$ by partial integration and using the
further identity~\cite{Green:1983hw}
\begin{equation}
\Bigl(\g^I_{a\dot{a}}\eta\tilde{s}^I_{\dot{b}}+\g^I_{a\dot{b}}\bar{\eta}s^I_{\dot{a}}\Bigr){Y'}^a
=-\frac{2^{3/2}\a}{\a'}m'_{\dot{a}\dot{b}}
\end{equation}
and
\begin{equation}
\sum_{r=1}^3\p_{\s_r}|V\ra=-\frac{i}{4}\frac{\a'}{\a}\Bigl(
\bigl(K^2-\wt{K}^2\bigr)+4\bigl(Y\dot{Y}+i\m(1-\Pi)YY_0\bigr)\Bigr)|V\ra\,,
\end{equation}
we find that equation~\eqref{dyn3} is equivalent to
\begin{align}
&\Bigl( \Bigl[
\sqrt{2}\bigl(\g^I_{a\dot{a}}\eta\tilde{s}^I_{\dot{b}}-\g^{I}_{a\dot{b}}\bar{\eta}s^I_{\dot{a}}\bigr)
-4im_{\dot{a}\dot{b}}Y_a\Bigr]\Bigl(\dot{Y}^a-\dot{Y}^a_0\Bigr)
-\frac{\m}{\sqrt{2}}\Bigl(\g^I_{a\dot{a}}\bar{D}_b\tilde{s}^I_{\dot{b}}+\g^I_{a\dot{b}}D_bs^I_{\dot{a}}\Bigr)
\nonumber\\
&(1-\Pi)^{ab}-iK^IK^J\Bigl[\d^{IJ}m_{\dot{a}\dot{b}}-\frac{\a'}{\sqrt{2}\a}\g^J_{a\dot{a}}D^a\tilde{s}^I_{\dot{b}}\Bigr]
+i\wt{K}^I\wt{K}^J\Bigl[\d^{IJ}m_{\dot{a}\dot{b}}\nonumber\\&-\frac{\a'}{\sqrt{2}\a}\g^J_{a\dot{b}}
\bar{D}^as^I_{\dot{a}}\Bigr]\Bigr)|V\ra=0\,.
\end{align}
This results in equations~\eqref{qq1}-\eqref{ppwave2}.


\end{document}